\documentclass[fleqn,usenatbib]{mnras}

\usepackage{newtxtext,newtxmath}

\usepackage[T1]{fontenc}
\usepackage{ae,aecompl}

\usepackage{graphicx}	
\usepackage{amsmath}	
\usepackage{caption}
\usepackage[position=top]{subfig}
\usepackage{multirow}
\usepackage{tablefootnote}
\usepackage{threeparttable}
\usepackage[capitalise]{cleveref}

\title[Pallene's Long-Term Dynamical Evolution]{Long-term Dynamical Evolution of Pallene (Saturn XXXIII) and its Diffuse, Dusty Ring}

\author[M. A. Mu\~noz-Guti\'errez et al.]
{Marco A. Mu\~noz-Guti\'errez,$^{1}$\thanks{E-mail: mmunoz@asiaa.sinica.edu.tw (MAM)}
A.~P. {Granados Contreras},$^{1}$
Gustavo Madeira,$^{2,3}$
\newauthor Joseph A. A'Hearn,$^{4}$
and Silvia Giuliatti Winter$^{2}$
\\
$^{1}$Institute of Astronomy and Astrophysics, Academia Sinica, 11F of AS/NTU Astronomy-Mathematics Building, No.1, Sec. 4, \\
Roosevelt Rd, Taipei 10617, Taiwan, R.O.C.\\
$^{2}$Grupo de Din\^{a}mica Orbital e Planetologia, S\~ao Paulo State University (UNESP), 333 Av.~Dr.~Ariberto Pereira da Cunha, Guaratinguet\'a-SP, 12516-410, Brazil\\
$^{3}$Universit\'e de Paris, Institut de Physique du Globe de Paris, CNRS, F-75005 Paris, France\\
$^{4}$Department of Physics, University of Idaho, 875 Perimeter Drive, Moscow, Idaho 83844, USA}

\date{Accepted XXX. Received YYY; in original form ZZZ}

\pubyear{2021}

\begin{document}
\label{firstpage}
\pagerange{\pageref{firstpage}--\pageref{lastpage}}
\maketitle

\begin{abstract}

The distinctive set of Saturnian small satellites, Aegaeon, Methone, Anthe, and Pallene, constitutes an excellent laboratory to understand the evolution of systems immersed in co-orbital dusty rings/arcs, subjected to perturbations from larger satellites and non-gravitational forces. In this work, we carried out a comprehensive numerical exploration of the long-term evolution of Pallene and its ring. Through frequency map analysis, we characterised the current dynamical state around Pallene. A simple tidal evolution model serves to set a time frame for the current orbital configuration of the system. With detailed short and long-term N-body simulations we determine whether Pallene is currently in resonance with one or more of six of Saturn's major moons. We analysed a myriad of resonant arguments extracted from the direct and indirect parts of the disturbing function, finding that Pallene is not in mean motion resonance from the present up to 5~Myr into the future; nonetheless, some resonant arguments exhibit intervals of libration and circulation at different timescales and moon pairings. We studied the dynamical evolution of micrometric particles forming the ring, considering gravitational and non-gravitational forces. Non-gravitational forces are responsible for particles vertical excursions and outward migration. By estimating the satellite's mass production rate, we find that Pallene could be responsible for keeping its ring in steady-state only if it is mainly composed of large micrometre-sized particles. If mainly composed of particles with a few micrometres for which Pallene is the only source, the ring will spread out, both radially and vertically, until it finally disappears.

\end{abstract}



\section{Introduction}

Pallene (Saturn XXXIII) is a satellite of only 2.23~km in radius \citep{Thomas13}, orbiting Saturn at an average distance of $\sim212\,283$~km, with an eccentricity of $\sim0.004$, and a relatively large inclination of $\sim0.18^\circ$ \citep{Spitale06,Jacobson08}.

This small Saturnian moon was first observed in a single photograph of the \emph{Voyager~2} spacecraft. It was reported, together with a preliminary orbital and physical characterisation, by \citet{Synnott86}. Pallene was then rediscovered in 2005 by the \emph{Cassini} Imaging Science team \citep{Porco05} and positively identified as the S/1981 S14 object from \emph{Voyager~2}.

Pallene is one of three small moons located between the orbits of Mimas and Enceladus, called collectively as the Alkyonides. Despite the presence of a vast number of resonances in the region, accentuated by the commensurabilities of Mimas with Tethys and Enceladus with Dione \citep[see e.g.][]{Sinclair72,Greenberg73,Peale76,Peale99}, Pallene doesn't seem to be in a mean motion resonance (MMR), unlike Methone and Anthe, trapped in 14:15 and 10:11 corotation eccentricity resonances with Mimas, respectively \citep{Cooper08,Hedman09,Moutamid14}. Nonetheless, Pallene is migrating away from Saturn via tidal evolution, though at a slower rate than Mimas. Thus the orbits of Pallene and Mimas are converging, which at some point either in the past or in the future should have resulted or should result in Pallene being captured into resonance with Mimas. A simple tidal evolution model \citep{MurrayDermott} suggests that the most recent first-order resonance that Pallene might have escaped from, perhaps around 40 Myr ago, is the 4:5 resonance with Mimas. Pallene's current eccentricity or inclination could be signs of this or another past resonance.

After Pallene's rediscovery, \citet{Spitale06} determined the orbital parameters of Pallene with high accuracy using images from \emph{Cassini} and \emph{Voyager~2}. \citeauthor{Spitale06} suggested that Pallene could be in an inner third-order resonance with Enceladus (i.e., $\phi = 19\lambda_\mathrm{Enc} - 16\lambda_\mathrm{Pal} -\varpi_\mathrm{Pal} - 2\Omega_\mathrm{Pal}$). However, recent short-term numerical integrations have shown that this resonance's libration angle actually circulates \citep[e.g. Fig.~1 in][]{Munoz17}.

Furthermore, synchronous periodic eccentricity and inclination oscillations were found while exploring a longer-term dynamical evolution of Pallene \citep[of up to $10^5$ yr,][]{Callegari10,Munoz17}, which could be the result of a resonant perturbation produced by either Mimas or Enceladus. Moreover, \citet{Callegari10} identifies a possible argument for the proposed quasi-resonance involving the apsidal and nodal longitudes of Pallene and Mimas, given by $\phi=\varpi_\mathrm{Pal}-\varpi_\mathrm{Mim}+\Omega_\mathrm{Pal}-\Omega_\mathrm{Mim}$. Nonetheless, \citet{Munoz17} found that this argument also circulates with a period of $\sim 4762.2$ yr, though, interestingly, with the same period of the observed oscillations of Pallene's eccentricity and inclination. 

Pallene shares its orbit with a diffuse ring of micrometre-sized dust, first reported by \citet{Hedman09}. The constant resupply of ring material is expected to come from impact debris, expelled from the satellite's surface by collisions between interplanetary dust particles (IDPs) and the moon. A similar mechanism has been proposed and explored in order to explain the existence of Aegaeon's ring arc inside the G~ring \citep{Hedman10,Madeira18}, the ring arcs of Methone and Anthe \citep{Sun17,Madeira20}, as well as the Neptunian rings and arcs \citep{Gaslac20,Giuliatti20}.

In this work we carry out a comprehensive study of the long-term dynamics of Pallene, as well as of the possible origin and dynamical evolution of its diffuse dusty ring, formed by micrometre-sized particles subject to gravitational and non-gravitational forces. We organise this paper as follows: in \cref{sec:methods}, we describe the different set-ups of our numerical simulations, performed to address various aspects of our study; we characterise the current dynamical environment of Pallene and its ring through frequency map analysis in \cref{sec:dynconx}. In \cref{sec:dymevol}, we first estimate the time span in which the current orbital configuration of the Saturnian system would remain approximately unchanged by using a simple tidal evolution model; then, with detailed short- and long-term simulations, we re-evaluate at different timescales all possible libration angles between Pallene and the six major Saturnian satellites considered in our study. Finally, a characterisation of the evolution of Pallene's ring is carried out in \cref{sec:ring}, where all the relevant non-gravitational forces that affect small particles are considered. We summarise our work and present our main conclusions in \cref{sec:summary}.

\section{Methods and Simulations}
\label{sec:methods}

\begin{table}
\centering
   \begin{threeparttable}
   \caption{Saturn's physical parameters.}
   \label{tab:phySat}
   \begin{tabular}{@{}llc}
        \hline \hline
        Parameter & Value & Reference \\
        \hline
        $R_\mathrm{S}$ [km] & $60\,330$ & \citet{Kliore80} \\
        $GM_\mathrm{S}$ [km$^3$ s$^{-2}$] & 3.793120749865220E+07 & \textit{gm\_de431.tpc} \tnote{\emph{a}} \\
        $J_2$ & 1.6290573E-02 & \citet{Iess19}\\
        $J_4$ & -9.35314E-04 & \citet{Iess19} \\
        $J_6$ & 8.6340E-05 & \citet{Iess19} \\
        $\Omega_\mathrm{S}$ [rad s$^{-1}$]& 1.65269E-04 & \citet{Helled15} \\
        \hline
   \end{tabular}
   \begin{tablenotes}
      \footnotesize
      \item[\emph{a}]{Available at \url{https://naif.jpl.nasa.gov/pub/naif/generic_kernels/pck/gm_de431.tpc}}
   \end{tablenotes}
   \end{threeparttable}
\end{table}

\begin{table}
\centering
    \begin{threeparttable}
    \caption{Summary of physical parameters of the six large moons in our system.}
    \label{tab:phyMoons}
    \begin{tabular}{@{}lccc}
        \hline \hline
        Name & $G M_m$\tnote{\emph{a}} & $\rho_m$ & $R_m$\tnote{\emph{b}} \\ 
         & [km$^3$ s$^{-2}$] & [g cm$^{-3}$] & [km] \\
        \hline
        Mimas & 2.503522884661795E+00 & 1.152 & 198.2 \\
        Enceladus & 7.211292085479989E+00 & 1.606 & 252.6 \\
        Tethys & 4.121117207701302E+01 & 0.956 & 537.5 \\
        Dione & 7.311635322923193E+01 & 1.469 & 561.4 \\
        Rhea & 1.539422045545342E+02 & 1.233 & 763.8 \\
        Titan & 8.978138845307376E+03 & 1.880 & 2574.7 \\
        \hline
    \end{tabular}
    \begin{tablenotes}
      \footnotesize
      \item[\emph{a}]{$G M_m$ values are taken from the planetary constant kernel \textit{gm\_de431.tpc}.}
      \item[\emph{b}]{Radius values, $R_m$, are taken from the planetary constant kernel \textit{pck00010.tpc} \citep[available at \url{https://naif.jpl.nasa.gov/pub/naif/generic_kernels/pck/pck00010.tpc},][]{Archinal11}.}
    \end{tablenotes}
    \end{threeparttable}
\end{table}

We carried out extensive and detailed numerical simulations of the evolution of the dynamical system formed by Pallene and six major Saturnian satellites, those 
gravitationally relevant in our region of interest, namely: Mimas, Enceladus, Tethys, Dione, Rhea, and Titan. Throughout this work, we consider Saturn's oblateness and take into account zonal harmonic terms up to $J_6$ in all simulations. Our numerical integrations cover several time spans, in order to study different aspects of the dynamics of Pallene, its phase-space surroundings, as well as the evolution of its dust-ring. Our shortest simulation lasts 18 yr, while the longest simulation is $5\times10^6$~yr long. 

Unless otherwise stated, the physical parameters of Saturn and the six major moons used throughout this work are summarised in \cref{tab:phySat,tab:phyMoons}. We use a rotation rate for Saturn $\Omega_\mathrm{S}=1.65269 \times10^{-4}$~rad/s from \citet{Helled15}. As initial conditions for the major Saturnian moons, we use the satellite's Saturn-centric state vectors taken from the JPL Horizons ephemerides service\footnote{\url{https://ssd.jpl.nasa.gov/horizons.cgi}} on $JD = 2459305.5$, corresponding to April 1, 2021. We scale the satellite semi-major axes and masses to Pallene's semi-major axis and Saturn's mass, respectively. The system's gravitational constant is scaled accordingly, for which we use Pallene's average semi-major axis $\bar{a}_{\mathrm{Pal}} = 2.1228335\times 10^5$~km \citep[as found in][]{Munoz17} and the $G M_\mathrm{S}$ parameter given in \cref{tab:phySat}. Consequently, our gravitational constant for this system is $G = 29.59895344398\; \bar{a}_{\mathrm{Pal}}^3\, M_\mathrm{S}^{-1}\, d^{-2}$. 

Pallene's mass is derived from its size, which has been measured with small uncertainty, i.e., $R_\mathrm{Pal} = 2.23 \pm 0.07$~km, as well as from its reported range of bulk density, i.e. $0.19 \leq \rho_\mathrm{Pal} \leq 0.34$~g/cm$^3$ \citep{Thomas13}. We explore three different density values to cover the uncertainty reported by \citeauthor{Thomas13}, i.e., $\rho_\mathrm{Pal}=$~0.19, 0.25, and 0.34~g/cm$^3$. This means that for each simulation suite described in the following paragraphs, we run three versions, each with Pallene's gravitational parameter given by $GM_\mathrm{Pal}=$~5.89064055531E-07, 7.75084283594E-07, and 1.05411462568E-06~km$^3$/s$^2$, corresponding to the selected density values. At the end of each of our integrations, we convert the state vectors ($\vec{r}$ and $\vec{v}$) to geometric orbital elements \citep{Renner06}, which reduces the short-term oscillations of the osculating elements due to the oblateness of the central mass.

In order to place Pallene within its current dynamical context, our first objective is to characterise the dynamics of a broad region of the geometric semi-major axis-eccentricity ($a$-$e$) phase-space plane around Pallene. With this in mind, we performed two numerical simulations (lasting 18 and $10^4$~yr, respectively), including a total of $13\,025$ test particles covering a grid of the geometric $a$-$e$ plane in the vicinity of Pallene. For these integrations, we used the Bulirsch-Stoer integrator from the Mercury6 package \citep{Chambers99}, with a toleration accuracy parameter of $10^{-12}$ and an initial time-step of 0.1 days.

Secondly, to examine the big picture of Pallene's tidal evolution, we use a simple model based on \citet{MurrayDermott}, which assumes a linear tidal dissipation mechanism and a constant $Q$, independent of frequency. We only examine the tidal evolution of Pallene and the large moons in its vicinity, Mimas and Enceladus, in order to look at resonances that may have been crossed in the recent past, as well as to establish a time limit of the validity of the current orbital configuration of the system for the longer-term simulations.

Next, in order to determine the possible resonant behaviour of Pallene, we performed a set of N-body simulations, spanning from 50 up to $5\times10^6$~yr of integration time. In this instance, the test particles are not included, and there are only seven bodies orbiting Saturn. The N-body simulations are performed with our implementation of the Implicit integrator with Adaptive time-Stepping of 15th-order \citep[IAS15,][]{IAS15} taking into account Saturn's gravitational moments (\cref{tab:phySat}). Subsequently, we integrate the satellite system for 50, $5\times10^3$, $5\times10^4$, $5\times10^5$, and $5\times10^6$ yr. We use the geometric orbital elements to calculate several libration angle combinations, among all satellites in \cref{tab:phyMoons} and Pallene.

Finally, we study the evolution of the diffuse ring through two distinct scenarios: (a) particles initially co-orbital to the satellite and (b) by considering the temporal evolution of particles launched from Pallene's surface. The study is performed considering the system's gravitational effects and also non-gravitational forces acting in the region, such as solar radiation force, plasma drag, and the electromagnetic force. Using an adapted version of the Mercury6 package which includes the effects of these forces and Saturn's gravitational moments, we integrated the system formed by Pallene, the six large moons, and a set of 5,000 test particles until all the particles were removed from the simulation.

\section{Pallene's Current Dynamical Context}
\label{sec:dynconx}

\subsection{Characterisation Through Frequency Map Analysis}
\label{ssec:FMA}

\begin{figure*}
    \centering
    \includegraphics[width=1.0\textwidth]{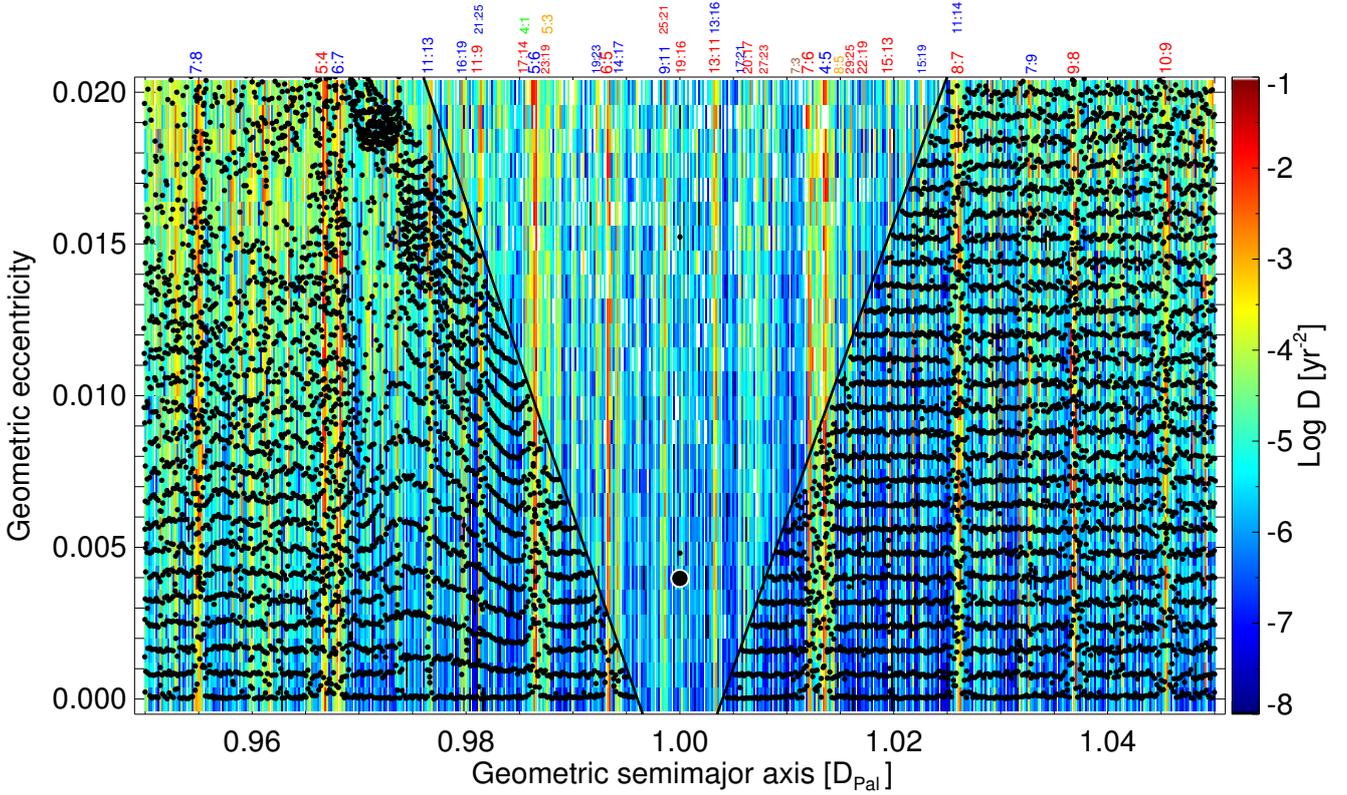}
    \caption{Diffusion map for a wide region of the geometric semi-major axis - eccentricity phase-space plane around Pallene. A colour scale indicates the stability of orbits, where bluer regions represent the more stable, and redder ones the more unstable. Locations where particles were ejected or collided with Pallene before the end of the simulation are coloured in white. The solid black lines stand for the constant pericentre and apocentre distances of Pallene, delimiting the collision region of the small moon. All the MMR ratios which were explored for libration in the simulations of \cref{ssec:resanalysis}, going from first to fourth order, are labelled at the top of the figure. Colours correspond to MMRs with Mimas (blue), Enceladus (red), Tethys (orange), Dione (brown), and Rhea (green). The final conditions of a longer simulation ($10^4$ yr), of the same particles used to create the diffusion map, are over-plotted on the map (black dots) to highlight the predictive power of the frequency analysis technique for the characterisation of the dynamical stability of wide regions of phase space.}
    \label{fig:diffmap}
\end{figure*}

To gain a better understanding of the dynamical behaviour and future stability of Pallene, as well as of the micrometric dust particles in its vicinity, we carried out a frequency map analysis \citep[FMA,][]{Laskar90,Laskar92,Robutel01} of a broad region of the geometric $a$--$e$ phase space plane, surrounding Pallene. We performed a short-term numerical integration (of $\sim18$~yr, or approximately $5\,700$ Pallene orbital periods and $2\,000$ orbital periods of the most external particle in the map). We used this time span since at least $2\,000$ revolutions of each particle are required to confidently recover the main orbital frequencies. 

We included $13\,025$ test particles distributed in a homogeneous grid covering the $a$--$e$ geometric plane, with the following conditions: $a$ is sampled from 0.95 to 1.05~$D_{\mathrm{Pal}}$ (where $D_{\mathrm{Pal}}$ is the normalised average geometric semi-major axis of Pallene) in steps of size $\Delta a=5\times10^{-5}$. In $e$ we sampled from 0 to 0.02 in steps of $\Delta e=2\times10^{-3}$. The remaining orbital elements are all set to zero for simplicity, namely, inclination $I$, longitude of pericentre $\varpi$, longitude of the ascending node $\Omega$, and mean anomaly $M$. We recall that test particles are subject to the gravitational perturbations of an oblate Saturn, Pallene, and the six gravitationally dominant moons in our region of interest.

A frequency analysis for each test particle in the grid was performed, using the algorithm of \citet{Sidli96}, over the dynamical variable:
\begin{equation}
\xi(t)=a(t)\exp(i\lambda(t)),
\end{equation}
where $a(t)$ and $\lambda(t)$ are the semi-major axis and mean longitude of each particle, respectively. The variable $\xi(t)$ expresses a combination closely related to a formal combination of the action and angle variables ($J_i$,$\eta_i$), of each orbit, expressed as $\xi'_i(t)=J_i\exp{\eta_i}$. Though it is clear that $\xi(t)$ and $\xi'(t)$ are not equal, they are still related as $\xi(t) = f(\xi'_1, \xi'_2, ..., \xi'_n)$, being $f$ is a function close to unity \citep{Laskar93}. 

When we perform a frequency analysis of $\xi(t)$, we obtain a decomposition of the form 
\begin{equation}
\xi(t)=\alpha_0\exp(i\beta_0)+\sum_{k=1}^{N}\alpha_k\exp(i\beta_k).
\label{eq:zt}
\end{equation}

For a Keplerian orbit, the decomposition of $\xi(t)$ would have only one term, i.e. $\alpha_0=a$ and $\beta_0=n$, where $\beta_0$ is what we call the ``mean frequency'', while $a$ and $n$ are the semi-major axis and mean motion of the particle, respectively. For non-Keplerian orbits, the decomposition given in \cref{eq:zt} contains many periodic terms. Nonetheless, frequency analysis ensures that if a particle remains in a stable orbit, the conditions expressed by the approximations $\alpha_0\approx a$ and $\beta_0\approx n$ will prevail; also, for stable orbits $\alpha_0\gg\alpha_k$. These conditions do not hold for particles following unstable orbits, for which $\beta_0$ will change dramatically from one time interval to the next, since the evolution of chaotic orbits does not remain on the surface of KAM tori.

To compute the change of the main frequencies, we perform a frequency analysis of $\xi(t)$ in two adjacent time intervals of length $T$, equal to half the total integration time. We call $\beta_{01}$ and $\beta_{02}$ the main frequencies obtained in each interval, respectively. Finally, we define a diffusion parameter, $D$, which provides a measure of the stability of the orbits. Following \citet{Correia05,Munoz17} we have
\begin{equation} 
D=\frac{\left|\beta_{01}-\beta_{02}\right|}{T}.
\end{equation} 
It can be seen that small values of $D$ will be obtained for stable trajectories, while larger values of $D$ are the result of unstable orbital evolution.

\subsection{Diffusion Map of Pallene's Neighbourhood} \label{ssec:pallneigh}

A diffusion map for the region around Pallene, shown in \cref{fig:diffmap}, was obtained after applying the above procedure to all the grid particles covering the geometric $a$--$e$ plane. A coloured rectangle is plotted for each particle according to its initial location in the plane, where colour is scaled according to the value of the logarithm of $D$. Redder colours indicate more unstable orbits, while bluer colours represent the more stable trajectories. Particles that are lost from the simulation before it finished, mainly due to collisions with Pallene, are coloured white. Solid black lines delimit Pallene's collision region, i.e. the region in which, at their apocentric or pericentric excursions, particles will cross Pallene's orbit, thus having a higher probability of colliding with the small moon.

The diffusion map provides a quick method to globally characterise the dynamical state of a vast region of phase-space, at a low computational cost, i.e. using only short-term numerical simulations. Unstable regions are immediately highlighted by the colour contrast. We can quickly identify MMRs, as well as their relative strength. The semi-major axis parameter space from 0.98 to 1.02~$D_{\mathrm{Pal}}$ encompasses completely both Pallene's orbit and the co-orbital dusty ring. In this region the strongest MMRs are due to first-order commensurabilities with either Mimas or Enceladus, however, higher-order MMRs with Dione, Tethys, and Rhea can also be observed. The location of all the existing commensurabilities with the six major moons (up to order 4 and degree 30) are indicated at the top of \cref{fig:diffmap}; outside this interval we only indicate the location of first-order MMRs with Mimas and Enceladus. The stronger resonances are characterised by thin vertical structures of homogeneous yellow to orange colour, such as the 4:5, 5:6, 6:7, and 7:8 MMRs with Mimas (blue labels), as well as the 5:4, 6:5, 7:6, 8:7, 9:8, and 10:9 with Enceladus (red labels). Second-, third-, and fourth-order MMR bands are thinner than first-order resonances. Furthermore, MMR chords are less stable than the broader, non-resonant, blue bands, regardless of eccentricity. Aside from possible exceptions at MMRs, lower eccentricity orbits are far more stable in general throughout the map. 

From \cref{fig:diffmap}, it is apparent that Pallene, whose location is indicated by the large black circle, is not currently trapped inside any strong MMR, despite the very close proximity of three resonances: the 9:11 with Mimas and the 19:16 and 25:21 with Enceladus.

Moreover, two interesting regions stand out from the map, corresponding to the clustering of several MMRs with various moons. The first of such regions, $b_1$, is located at $\sim0.986$~$D_{\mathrm{Pal}}$, where the 5:6 MMR with Mimas, the 17:14 and 23:19 MMRs with Enceladus, the 5:3 MMR with Tethys (orange label), and the 4:1 MMR with Rhea (green label) lie in close proximity to each other. The second region, $b_2$, is located around $\sim1.014$~$D_{\mathrm{Pal}}$; in this region two first-order resonances, the 4:5 with Mimas and the 7:6 with Enceladus, are in close proximity to the 8:5 MMR with Tethys (orange label), and the 7:3 MMR with Dione. It is apparent that the interaction of several low-order resonances results in especially unstable regions at these locations. A similar case occurs at $\sim0.966$~$D_{\mathrm{Pal}}$, where the two first-order resonances, 6:7 with Mimas and 5:4 with Enceladus, produce a particularly wide unstable region.

To reassess the predictive power of the frequency analysis technique, we integrated up to $10\,000$~yr the same set of $13\,025$ particles of the grid covering the geometric $a$--$e$ phase-space plane. The final conditions of this simulation were over-plotted on the diffusion map of \cref{fig:diffmap} with black dots.

To the left of the collision region of Pallene, the largest perturbations in eccentricity are observed for particles located in the bands of MMRs, as expected. The most unstable region, however, is the one located in the top left corner of the map, roughly above 0.015 in eccentricity and to the left of the 11:13 MMR with Mimas; here the direct influence of Mimas is stronger and particles are removed faster. To the right of the collision region, all the particles remain nearly unperturbed, except for the $b_2$ band where several resonances converge, as well as at the locations of other first-order MMRs with Mimas and Enceladus.

Notably, inside the collision region of Pallene, only three particles survive after 10~kyr, one of them is a co-orbital with the small moon; a second lies at the location of the 19:16 MMR with Enceladus, and the last one lies inside the band of the 11:9 MMR with Enceladus, which overlaps with the 21:25 MMR with Mimas. 

Both the map and the long-term simulation of the particles serve as an indication of the future evolution of large dust particles, with radii larger than $\sim30$~$\mu$m, i.e. those unaffected by non-gravitational forces known to act in this region. Towards the internal zone of the Pallene collision region, even this kind of large particles would be removed (though at times greater than 10~kyr) due to perturbations from Mimas. Exterior to the Pallene collision region, large particles could in principle survive for very long times. This indicates that the Pallene ring would find greater stability towards semi-major axes larger than that of the small moon, increasing the eccentricity of its conforming particles as they find MMR regions with Enceladus. On the other hand, ring-forming particles within the Pallene collision region could survive mainly as co-orbitals; however, with only one co-orbital and two apparently resonant particles surviving in this region in the 10~kyr simulation, we cannot provide quantifiable predictions for the behaviour of the ring, based exclusively on the diffusion map. For a more in-depth analysis of the possible origin of the ring and its future evolution, we performed a large set of detailed simulations, presented in \crefrange{ssec:ringdynmod}{ejectedmaterial} of this paper.

\section{Dynamical evolution of Pallene in different timescales} 
\label{sec:dymevol}

\subsection{Tidal Evolution}
\label{ssec:tides}

To gain an appropriate perspective on the timescales of Pallene's dynamical evolution, we first look at Pallene's tidal evolution in between Mimas and Enceladus. Although more complex analyses of tidal evolution in the Saturn system have recently been done \citep[e.g.][]{Fuller16, Lainey20}, here we employ a simpler model to gain a general understanding of the context in which Pallene may have evolved. Using Equation 4.213 from \citet{MurrayDermott}, we can calculate previous semi-major axes 
\begin{equation}\label{tidal_equation}
    a=a_0\left[1-\left(\frac{k_2}{Q}\frac{39M_mR_S^5}{2a_0^{13/2}}\sqrt{\frac{G}{M_S}}t\right)\right]^{2/13},
\end{equation}
assuming that the tidal dissipation mechanism is linear and that $Q$ is frequency-independent. 

For our tidal evolution calculations, we take our value for Saturn's Love number, $k_2=0.390$, from \citet{Lainey17}. We estimate a quality factor $Q=2000$ also based on \citet{Lainey17} and similar to what is used in \citet{Cuk16}, which was based on the earlier work of \citet{Lainey12}, though there is less agreement on this value and it is meant to apply only near the semi-major axes roughly around Mimas and Enceladus. Previous estimates of $Q$ an order of magnitude higher were due to the assumption that Mimas was primordial \citep{MurrayDermott,Meyer08}. However, recent studies that argue Saturn's rings and the mid-sized moons are probably young, use a $Q$ value in the range we have assumed \citep{Cuk16, Fuller16, Lainey17, Neveu19, Hesselbrock19}. 
Other values for this calculation are given in \cref{tab:phySat,tab:phyMoons}. 

Using these values, we measured the change in semi-major axis with respect to today's semi-major axis value $\frac{\Delta a}{a}$ over the past five million years for Mimas, Pallene, Enceladus, Tethys, and Dione. Out of these measurements, Mimas has $\frac{\Delta a}{a} = 0.0017$, which is the largest among these moons. Because this change in semi-major axis due to tidal evolution is small, we expect our long-term simulations of 5~Myr without the inclusion of tidal evolution to be accurate enough.


From the semi-major axis calculations, if Pallene is old enough, it may have recently escaped the 4:5 resonance with Mimas (40~Myr ago with $Q=2000$). Prior to escape, Pallene could have migrated with Mimas for a substantial period of time. 
For this reason, it becomes difficult to project Pallene's previous tidal evolution with much certainty. If Pallene was not captured in any resonance with Mimas for a significant period of time, which is unlikely because their orbits are converging, then further in the past Pallene's orbit may have crossed that of Enceladus (400~Myr ago with $Q=2000$), suggesting that Pallene could be a fragment from Enceladus, similar to the way \citet{Showalter19} propose that Hippocamp could have fragmented off of Proteus, possibly from a cometary impact. 

Hippocamp is close to the orbit that is synchronous with Neptune's rotation, which, together with the fact that it is the least massive of Neptune's moons, implies that the rest of Neptune's moons are diverging from Hippocamp. In contrast, Pallene's orbit is converging with Mimas's orbit. For this reason, Pallene is expected to have been captured into resonance with Mimas at each resonance crossing, but it is difficult to determine the duration of the capture in each resonance.

Proteus and Hippocamp have mean radii of 203.8~km and 17.4~km \citep{Showalter19}, while Enceladus and Pallene have mean radii of 252~km and 2.23~km \citep{Roatsch09, Thomas13}. 
Using these mean radii and masses of $1.08 \times 10^{20}$~kg for Enceladus \citep{Jacobson06} and $4.4 \times 10^{19}$~kg for Proteus (multiplying the volume from \citet{Stooke94} by an assumed density of 1.3~g/cm$^3$), the escape velocity $v_\mathrm{esc} = \sqrt{2GM_m/R_m}$ from the surface of Enceladus is 240~m/s, while for Proteus it is 170~m/s. Pallene has a smaller size ratio to Enceladus than Hippocamp has to Proteus, but perhaps Pallene is evidence of the proposed impactor in the south polar terrain of Enceladus \citep{Roberts17}. 

Not too long in the past, however, is the Mimas-Enceladus 3:2 resonance crossing (115~Myr ago with $Q=2000$). \citet{Meyer08} studied a triplet of Mimas-Enceladus 3:2 resonances and found that Mimas's eccentricity can be explained either by passage through the 3:2 $e$-Mimas resonance or the 6:4 $ee'$-mixed resonance (but not the 3:2 $e$-Enceladus resonance), and found dynamical escape to be possible for both of these resonances.
\citet{Cuk16} proposed that Tethys, Dione, and Rhea all formed in one event about 100~Myr ago, and suggests that Mimas and Enceladus could have formed during the same epoch or could be even younger. 
\citet{Neveu19}, however, have suggested that Mimas could be significantly younger than Enceladus. 
This last scenario allows for the possibility of Pallene migrating away from Enceladus after an impact before the formation of Mimas. 

Thus, given a constant $Q$ tidal model, it looks like Pallene has crossed some resonances, which, especially if it had been trapped in any of them for some period of time, could have affected its eccentricity and inclination. 
However, the new tidal models indicate the evolution of the satellites could be more complex than previously thought \citep{Fuller16, Lainey20}.
Still, small moons such as Pallene are likely sensitive probes of this tidal evolution \citep[see, for example,][]{Moutamid17} and so should be considered in those contexts.

\subsection{Resonance Analysis}
\label{ssec:resanalysis}

In view of the rich dynamical structure of the phase-space close to Pallene, where many resonances are in close proximity to each other, we seek to determine whether any particular resonance between Pallene and one or more of the major Saturnian moons drives the evolution of Pallene, or could be a possible mechanism to confine the particles of the dusty ring. Hence, we ran five sets of numerical N-body simulations with different integration times, i.e. 50 (or approximately $15\,766$ Pallene's orbits), $5\times10^3$, $5\times10^4$, $5\times10^5$, and $5\times10^6$~yr. The output interval in each integration is always a multiple of Pallene's orbital period, $P\approx1.2$~d, so that in each output file there are a total of $15\,220$ data points. For each integration, several libration angles from the direct and indirect arguments of the disturbing function were explored, up to fourth-order \citep{MurrayDermott}. Due to the uncertainties in Pallene's density and therefore its mass, three different densities were considered as described in \cref{sec:methods}, which means that in total 15 realisations were performed, three per each integration time; we designate these as density-sets per integration time.

We referred to the resonant arguments of the disturbing function for two reasons: (1) the number of possible arguments is constrained and (2) in the case that one of these arguments librates, then the corresponding argument would facilitate its use in future secular theory calculations of this system. The libration angle among an outer (with primed orbital elements) and an inner satellite (un-primed elements), is expressed as 
\begin{equation}
    \phi = j \lambda' + (k-j) \lambda + \gamma(\varpi', \varpi, \Omega', \Omega),
    \label{eq:libang}
\end{equation}
where $k$ is the order, $j$ the degree, and $\gamma$ is a linear combination of $\varpi',\, \varpi,\, \Omega'$, and $\Omega$. The examined libration angles range in order $k$ from 1 to 4, while the degree $j$ corresponds to possible resonances within 0.98 and 1.02 $D_\mathrm{Pal}$. The linear combination $\gamma(\varpi', \varpi, \Omega', \Omega)$ in \cref{eq:libang} is determined from the direct and indirect arguments of the disturbing function described in \citet{MurrayDermott}, which have the form $\gamma = k_1 \varpi' + k_2 \varpi + k_3 \Omega' + k_4 \Omega$, where $k_1+k_2+k_3+k_4 = k$. 

In the rest of this section, we denote the libration angles of a given moon with Pallene by their capitalised initials, e.g., $\phi_\mathrm{PM}$ for the Pallene-Mimas libration angle, except for Tethys which will be denoted by a ``t'' to distinguish it from Titan. For the semi-major axis interval considered above, the possible resonant combinations are summarised in \cref{tab:res_comb}. The majority of explored direct arguments involve either Pallene and Mimas, or Pallene and Enceladus. In contrast, the combination between Pallene and Titan lacks possible resonant combinations in this semi-major axis interval. For completeness, additional zeroth-order resonances were also evaluated for degrees $j=$~0 to 15.
\begin{table}
    \centering
    \caption{Order $k$ and degree $j$ explored for libration angles with Pallene}
    \label{tab:res_comb}
    \begin{tabular}{ccc}
        \hline 
        \hline
        Moon & $k$ & $j$ \\ 
        \hline
        \multirow{4}{*}{Mimas} & 1 & 5, 6 \\ 
        & 2 & 10 -- 12 \\
        & 3 & 15 -- 19 \\
        & 4 & 20 -- 25 \\
        \hline
        \multirow{4}{*}{Enceladus} & 1 & 6, 7 \\ 
        & 2 & 11 -- 15 \\
        & 3 & 17 -- 22 \\
        & 4 & 22 -- 30 \\
        \hline
        \multirow{3}{*}{Tethys} & 2 & 5 \\
        & 3 & 8 \\
        & 4 & 10 \\
        \hline
        Dione & 4 & 7 \\ 
        \hline
        Rhea & 3 & 4 \\
        \hline
    \end{tabular} 
\end{table}

\begin{figure}
    \centering
    \subfloat[$\psi_\mathrm{DP} = \lambda' + 3\lambda - 2\varpi' - 2\varpi$ \label{fig:ind_argA}]{\includegraphics[width=\linewidth,trim=0 5 0 0,clip=True]{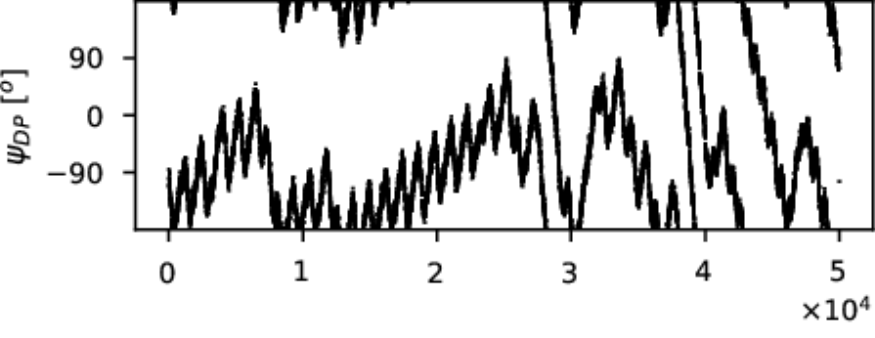}} \\ [-1pt]
    \subfloat[$\psi_\mathrm{TP} = 3\lambda' + \lambda - 2\varpi' - 2\Omega$ \label{fig:ind_argB}]{\includegraphics[width=\linewidth]{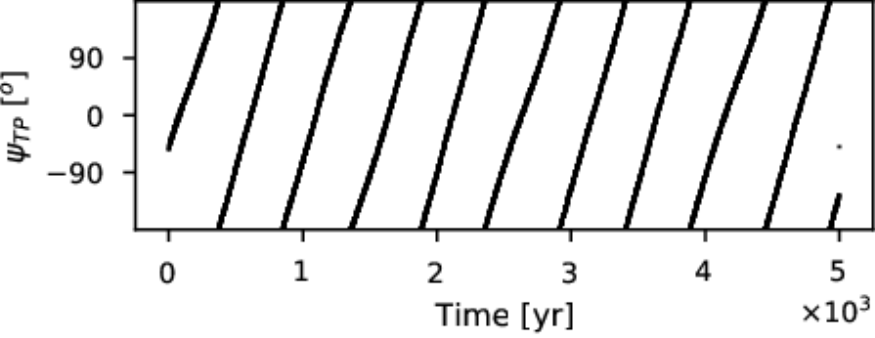}}
    \caption{
        Unique indirect arguments in our search with either librating properties or long period circulation. The remaining 448 arguments displayed short period circulation. 
    }
    \label{fig:ind_arg}
\end{figure}

We inspected 75 indirect arguments per moon pair, i.e., 450 in total, denominated as $\psi$ to distinguish them from the direct arguments, $\phi$. Most of the indirect arguments explore all the angular range in every timescale. Only two fourth-order indirect arguments show interesting behaviour: the Dione-Pallene argument $\psi_\mathrm{DP} = \lambda' + 3\lambda - 2\varpi' - 2\varpi$ displays temporal libration (\cref{fig:ind_argA}) for about 30~kyr; while the Titan-Pallene argument $\psi_\mathrm{TP} = 3\lambda' + \lambda - 2\varpi' - 2\Omega$ (\cref{fig:ind_argB}) presents a long circulation period of 494~yr.

In contrast, the direct arguments displayed a broader variety of phenomena depending on the timescale of the integration: circulation, alternating intervals of circulation, libration, or overall circulation with `steps' of near constant value. In \crefrange{sec:short_real}{sec:long_reals}, we only present angles that show resonant-like features and that coincide within a given density-set; we display the evolution of $\rho_\mathrm{Pal} = 0.25$~g/cm$^3$ integrations only. Nevertheless, when the resonant-like libration angles are compared within density-sets, we find that for integrations longer than $5\times10^3$~yr the angles evolve similarly within $5 \times 10^4$~yr but differ after this threshold. Consequently, the effect of Pallene's mass in its dynamical evolution is small and only noticeable after $10^4$~yr or $\sim 10^6$ Pallene orbits.

We divide our analysis into short (50~yr) and long-term ($t \geq 5\times10^3$~yr), demonstrating that Pallene has different resonant behaviour with one or more Saturnian satellites depending on the timescale, some emerging just in either short- or long-term simulations.

\subsection{Short-term evolution of direct arguments}
\label{sec:short_real}

\begin{figure}
    \centering
    \subfloat[$\phi_\mathrm{EP}=19\lambda' - 16\lambda - \varpi - 2\Omega$  \label{fig:EP50a} ]{\includegraphics[width=\linewidth,trim=0 25 0 0,clip=True]{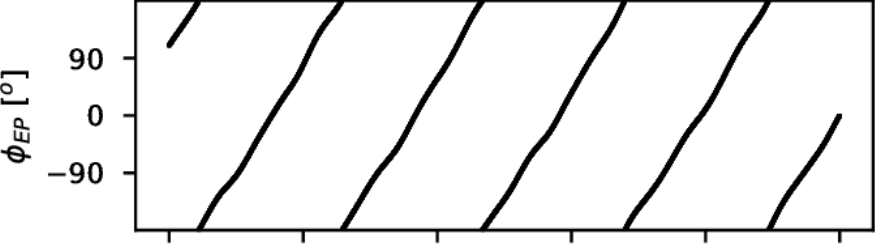} } \\ [-1pt]
    \subfloat[$\phi_\mathrm{EP}=19\lambda' - 16\lambda - \varpi' - 2\Omega'$ \label{fig:EP50b} ]{\includegraphics[width=\linewidth]{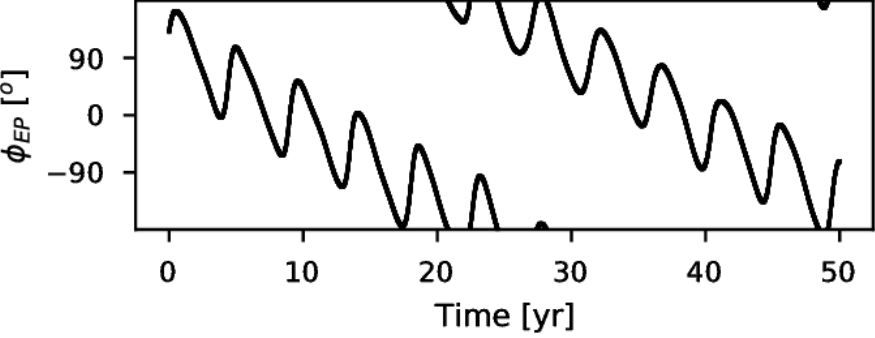} }
    \caption{
        Two different 19:16 libration angles between Enceladus and Pallene over 50~yr. The top panel corresponds to the resonant angle suggested by \citet{Spitale06} and the bottom panel corresponds to our finding. Although both libration angles circulate, the argument in \cref{fig:EP50b} has a circulation period 3 times longer than the one in \cref{fig:EP50a}. 
    }
    \label{fig:1916EP50}
\end{figure}

The intention of the 50~yr simulations was to re-examine the suggested third-order resonance between Pallene and Enceladus \citep{Spitale06}. We probed for libration all ten possible direct arguments with $19\lambda' - 16\lambda$, finding one additional combination with interesting behaviour in this interval. 
\Cref{fig:1916EP50} shows a comparison between the resonant angle suggested by \citet{Spitale06} (\cref{fig:EP50a}) and our finding (\cref{fig:EP50b}). The angle $\phi_\mathrm{EP}=19\lambda' - 16\lambda - \varpi - 2\Omega$, circulates with a period of $10.6$~yr. Similarly, \citet{Munoz17} found this angle to circulate but with a period 1.8 times shorter. The angle $\phi_\mathrm{EP}=19\lambda' - 16\lambda - \varpi' - 2\Omega'$ differs from that suggested in \citeauthor{Spitale06}, in that the longitudes of ascending node and pericentre belong to the outer satellite instead of the inner one. The evolution of this argument exhibits a softer negative slope that circulates with a period of $\sim 30$~yr.

In contrast to other small moons in the region, clearly trapped in first-order MMRs with Mimas, such as Aegaeon in the 7:6 \citep{Hedman10,Madeira18}, Methone in the 14:15 \citep{Spitale06,Hedman09,Callegari21}, and Anthe in the 10:11 \citep{Cooper08,Callegari20}, our short-term (and long-term) simulations indicate that Pallene's evolution is not characterised uniquely by any MMR, either with Mimas or Enceladus. Although some of the 19:16 libration angles between Pallene and Enceladus present features associated with a near-resonance, they all clearly circulate in longer timescales. It is likely that Pallene is just outside the parameter space that characterises the 19:16 MMR with Enceladus. Similarly, several of the libration angles shown outside of the Mimas-Anthe 10:11 MMR \citep[Fig. 8][]{Callegari20} resemble the evolution of some of the direct arguments we studied in this work. The latter suggest that an analysis of the ``individual dynamic power spectra'' \citep[IPS in][]{Callegari20} of the 19:16 MMR between Pallene and Enceladus could disclose the nature of the current resonant state of Pallene (\cref{fig:1916EP50}), however, we consider such analysis beyond the scope of the current work.

\subsubsection{Simultaneous zeroth-order direct argument among all moons}
\label{sec:0rdall}

\begin{figure}
    \centering
    \includegraphics[width=\linewidth,trim=0 25 0 0,clip=True]{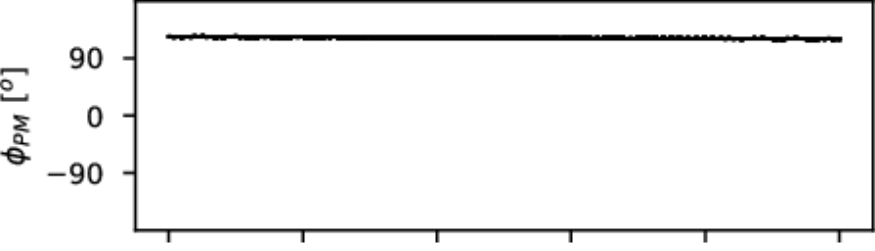} \\
    \includegraphics[width=\linewidth,trim=0 25 0 0,clip=True]{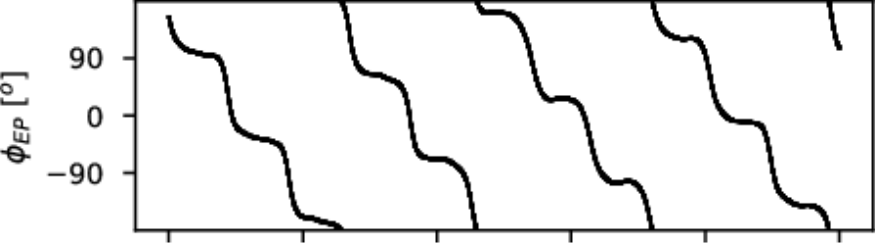} \\
    \includegraphics[width=\linewidth,trim=0 25 0 0,clip=True]{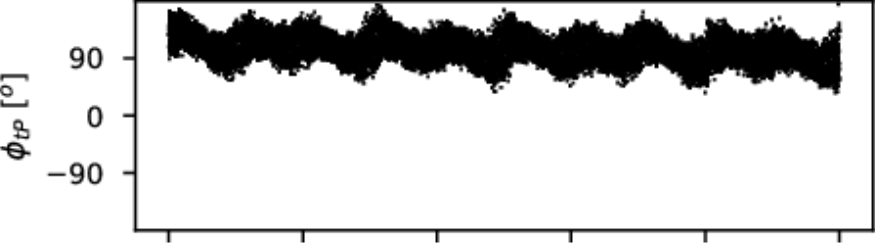} \\
    \includegraphics[width=\linewidth,trim=0 25 0 0,clip=True]{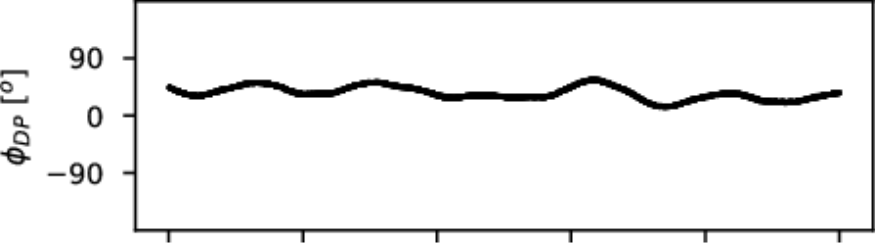} \\
    \includegraphics[width=\linewidth,trim=0 25 0 0,clip=True]{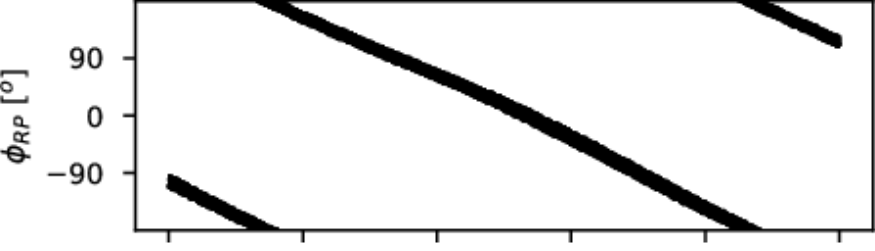} \\
    \includegraphics[width=\linewidth]{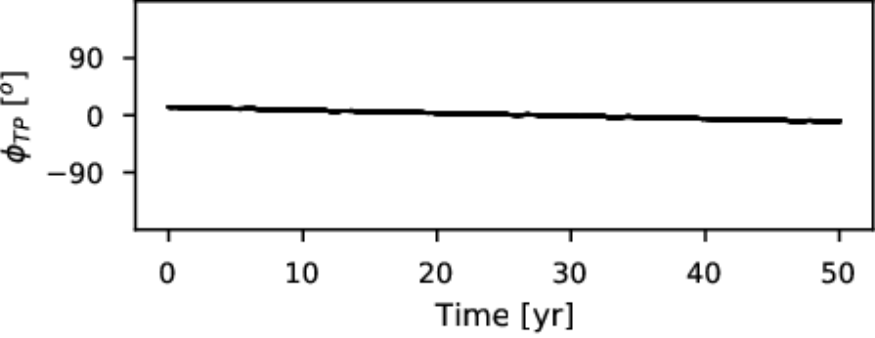}
    \caption{
        50~yr evolution of the libration angle $\Phi = \varpi' - \varpi + \Omega' - \Omega$ between each moon in \cref{tab:phyMoons} and Pallene. The known libration angle between Pallene and Mimas \citep[][top panel]{Callegari10}, also presents resonant behaviour between Pallene and three other moons: Tethys, Dione, and Titan (3rd, 4th, and 6th panels from top to bottom). In contrast, $\Phi_\mathrm{EP}$ exhibits circulation with semi-constant `steps', whereas $\Phi_\mathrm{RP}$ (5th panel) circulates.}
    \label{fig:all_nodiff}
\end{figure}

While examining the zeroth-order direct arguments of the 50~yr simulations, a simultaneous resonant libration angle was detected between Pallene and four other moons: Mimas, Tethys, Dione, and Titan. Here `simultaneous' means that more than one pair of satellites (Pallene and another large Saturnian moon) displays apparent resonant properties for the same libration angle expression. In this case, this simultaneity emerged for $\Phi \equiv \varpi' - \varpi + \Omega' - \Omega$ as presented in \cref{fig:all_nodiff}. In this time interval, $\Phi$ appears to be constant with small oscillations, except for the pairs Enceladus-Pallene and Rhea-Pallene, which circulate with a period of 12 and 36~yr, respectively. Nonetheless, Enceladus displays a semi-resonant behaviour due to the step-like oscillation of $\Phi_\mathrm{EP}$. Each ``step'' has a semi-constant value that changes in each full circulation. For example, in the first step (from 1 to 4~yr) the nearly-constant value is $90^\circ$ while on the fourth step (from 14 to 18~yr) the corresponding value is $60^\circ$, therefore, there are $\sim 4$~yr intervals where this angle librates followed by a shift of $\sim 130^\circ$ during $\sim 1.5$~yr to another semi-constant step.

\citet{Callegari10} suggested this quasi-resonant relationship between Mimas and Pallene ($\Phi_\mathrm{PM}$) and demonstrated that it has a long circulation period \citep[$\sim 5000$ yr, later confirmed by][]{Munoz17}). In order to explore possible circulation of $\Phi$ for Tethys, Dione, and Titan, we looked for circulation of this direct argument in our $5\times10^3$ yr integrations and, if circulation existed, determined the corresponding period using Fourier frequency analysis. \Cref{tab:all_nodiff} lists the circulation periods of $\Phi$ for each moon pair, including our estimate for $\Phi_\mathrm{PM} = 4708$~yr. The measured circulation periods for Tethys-Pallene (tP), Dione-Pallene (DP) and Titan-Pallene (TP), are 872~yr, 844~yr, and 794~yr, respectively. Even though these angles are not resonant, their long circulation relative to Pallene's orbital period might significantly affect the dynamics of Pallene in the short-term.
\begin{table}
    \centering
    \caption{Period of libration angle $\phi = \varpi' - \varpi + \Omega' - \Omega$ for the moon pairs in \cref{fig:all_nodiff}. }
    \label{tab:all_nodiff}
    \begin{tabular}{ccc}
        \hline \hline
        Moon pair & $P_\mathrm{circ}$ & No. orbits \\
        & [yr] & [$10^3$]\\
        \hline 
        PM & 4708 &  1495.5  \\
        EP & 12 &  3.9  \\
        tP & 872 &  277.0  \\
        DP & 844 &  268.2 \\
        RP & 36 &  11.3  \\
        TP & 794 &  252.4  \\
        \hline
    \end{tabular}
\end{table}

The possible existence of a quasi-resonance with the same combination of angles that excludes the mean longitudes suggests an alignment of the lines of nodes and apses of Pallene, Mimas, Tethys, and Dione, most likely with Titan. In other terms, a combination of the eccentricity and inclination vectors of these satellites may be aligned to some extent to Titan's. This is not entirely unexpected, since secular resonances could lead to apsidal alignments; in the Saturnian system an example of this has long been known to occur between Rhea and Titan \citep[see][and references therein]{Greenberg75}. The well known example of Tethys-Mimas 2:4 MMR ($\phi_\mathrm{tM} = 4\lambda' - 2\lambda - \Omega' - \Omega$), for which the variation in inclination drives the resonance \citep{Greenberg73,Allan96}, is another important example of node alignment. Moreover, alignment of the nodes has been discussed in several works involving the dynamics of compact extrasolar systems \citep[e.g.,][]{Kaib11,boue14b,Granados18}; the later works refer to this alignment as the interaction of an outer massive planet/companion with an inner compact system (of planets) which affects the inner system as if it were a rigid body. In the case of the Saturnian moon system, a study of the compactness of the orbits interior to Titan could reveal whether this phenomenon also occurs in this system. Nonetheless, a detailed study of this scenario is currently beyond the scope of this paper focused on Pallene dynamics; thus we consider this idea for future work. 

\subsection{Long-term evolution of direct terms}
\label{sec:long_reals}

We performed four long-term simulations, lasting $5\times10^3$, $5\times10^4$, $5\times10^5$, and $5\times10^6$~yr. In these simulations, most of the explored arguments circulate. Although a handful of angles display resonant characteristics during definite time intervals, there is not a single case in which the libration angle has a constant value for the total length of the simulations. 

In \crefrange{ssec:long1st}{ssec:long0th}, we present libration angles of interest separated by order, at least one per order, from first to fourth-order finishing with zeroth-order. The second-order arguments in \cref{ssec:long2nd} failed to produce similar behaviour among the density-set in all timescales. However, we include the results of two Tethys-Pallene arguments (each with distinct densities for Pallene) displaying temporal libration to exemplify the long-term effect of Pallene's mass in determining its resonant state.

\subsubsection{First-order arguments}
\label{ssec:long1st}

\begin{figure}
    \centering
    \includegraphics[width=\linewidth]{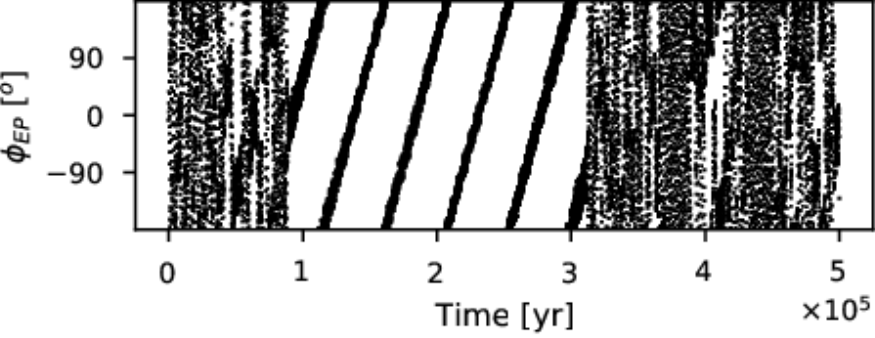} 
    \caption{
        First-order argument $\phi_\mathrm{EP}=7\lambda' - 6\lambda + \varpi - 2\Omega$ between Enceladus and Pallene over $5\times10^5$~yr. A change in the circulation frequency is observed between 100 to 300~kyr.
    }
    \label{fig:angs1st}
\end{figure}

Only one first-order argument presenting unusual features was recovered from our simulations (\cref{fig:angs1st}). Although it circulates at all times in the $5\times10^5$~yr integration, this 7:6 argument between Enceladus and Pallene shows a change in circulation frequency that slows down and holds for more than $2\times10^5$~yr, a considerable interval in terms of Pallene's orbital period. The exact resonance is located at 1.012$D_{\mathrm{Pal}}$ and is one of the strongest resonances in the region considered in this work (see map of \cref{fig:diffmap}). However, due to its semi-major axis being far from the 7:6 MMR location, it is unlikely that Pallene would be trapped or suffer strong perturbations from Enceladus through this resonance.

\subsubsection{Second-order arguments}
\label{ssec:long2nd}

\begin{figure}
\centering
    \subfloat[$\phi_\mathrm{tP}=5\lambda' - 3\lambda + \Omega' - 3\Omega$  \label{fig:53tPa} ]{\includegraphics[width=\linewidth,trim=0 25 0 0,clip=True]{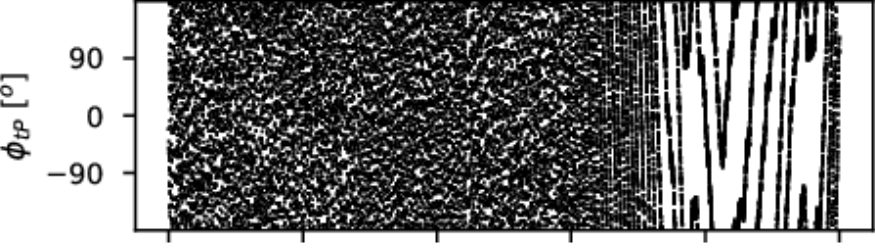} }\\[-1pt]     
    \subfloat[$\phi_\mathrm{tP}=5\lambda' - 3\lambda + \varpi' - \varpi + \Omega' - \Omega$  \label{fig:53tPb} ]{\includegraphics[width=\linewidth]{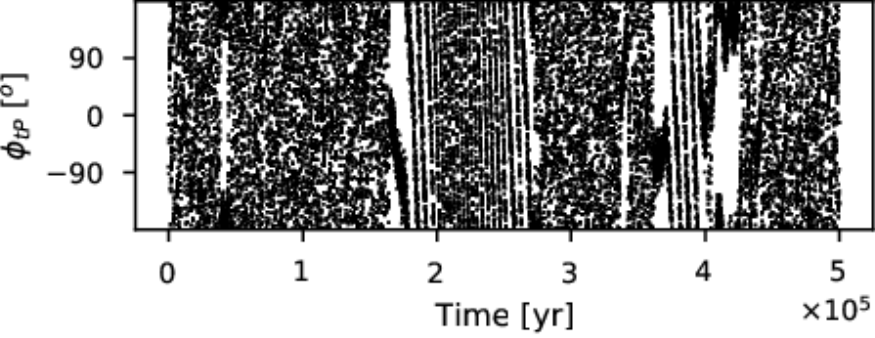} }     
    \caption{
        Second-order 5:3 MMR between Tethys and Pallene. Both arguments present temporal libration at different times that last for thousands of years.
    }
    \label{fig:angs2nd}
\end{figure}

\cref{fig:angs2nd} presents the evolution of two second-order libration angles with the same degree, 5:3, over $5\times10^5$~yr. The argument involving the longitudes of ascending nodes of Tethys and Pallene (\ref{fig:53tPa}), corresponding to the simulation with $\rho_\mathrm{Pal} = 0.19$~g/cm$^3$, exhibits two librating intervals, one between 350 and 400~kyr and another extending from 450 to 500~kyr, with a slow circulation period enclosed by both intervals. On the other hand, the second argument (\ref{fig:53tPb}) is an outcome of the $\rho_\mathrm{Pal} = 0.34$~g/cm$^3$ simulation and involves both the nodal and apsidal longitudes. This argument briefly librates at different intervals of the simulation, the most notable of which covers the 400 to 450~kyr interval. 

\subsubsection{Third-order arguments}
\label{ssec:long3rd}

\begin{figure}
    \centering
        \subfloat[$\phi_\mathrm{EP}=22\lambda' - 19\lambda - \varpi' - 2\varpi$  \label{fig:2219EP} ]{\includegraphics[width=\linewidth,trim=0 25 0 0,clip=True]{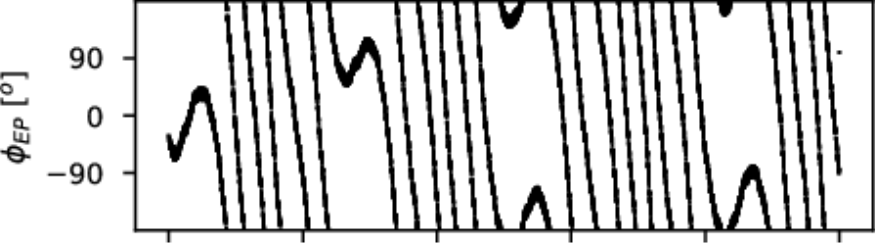} }\\[-1pt]
        \subfloat[$\phi_\mathrm{tP}=8\lambda' - 5\lambda - \varpi' - 2\varpi$  \label{fig:85tP} ]{\includegraphics[width=\linewidth,trim=0 5 0 0,clip=True]{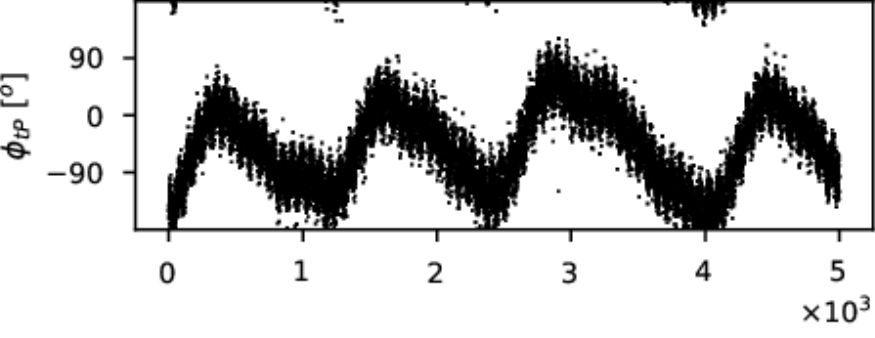} }\\[-5pt]
        \subfloat[$\phi_\mathrm{EP}=19\lambda' - 16\lambda - 3\varpi$ \label{fig:1916EPlong} ]{\includegraphics[width=\linewidth,trim=0 25 0 0,clip=True]{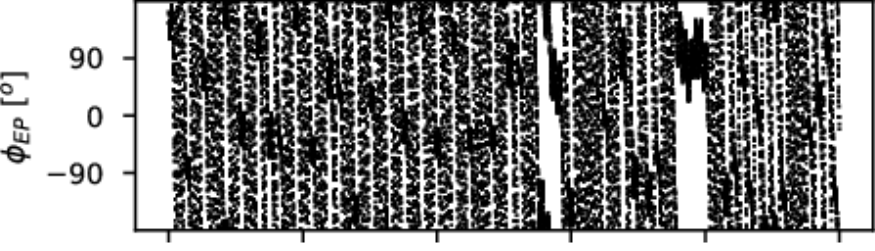} }\\[-1pt] 
        \subfloat[$\phi_\mathrm{tP}=8\lambda' - 5\lambda - \varpi' - 2\varpi$  \label{fig:85tPlong} ]{\includegraphics[width=\linewidth]{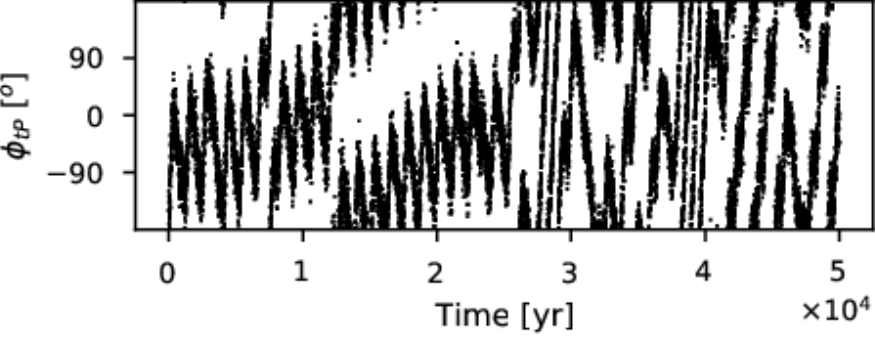} }
    \caption{
       Third-order direct arguments between Enceladus and Pallene and between Tethys and Pallene on different timescales. From all the arguments with librating properties, the argument $\phi_\mathrm{tP}=8\lambda' - 5\lambda - \varpi' - 2\varpi$ (\cref{fig:85tP,fig:85tPlong}) librates for about 10~kyr, the longest time amongst our findings in \cref{sec:dymevol}.
    }
    \label{fig:angs3rd}
\end{figure}
In total, three different third-order arguments were found (\cref{fig:angs3rd}), associated with resonances with Enceladus and with Tethys. On the $5\times10^3$~yr timescale, the direct argument $\phi_\mathrm{EP}=22\lambda' - 19\lambda - \varpi' - 2\varpi$ between Enceladus and Pallene (\cref{fig:2219EP}) circulates for most of the integration yet displays intervals of libration which last about 400~yr, and, similar to the argument in \cref{fig:angs1st}, shifts the constant value at which it librates, e.g., in the first 400~yr it librates close to 0$^\circ$ and then shifts to librate close to 90$^\circ$ from the 1200 to 1600~yr interval. The argument associated with the 8:5 MMR between Tethys and Pallene, $\phi_\mathrm{tP}=8\lambda' - 5\lambda - \varpi' - 2\varpi$, exhibits a clear ample libration for the duration of the integration. \Cref{fig:85tP} is the clearest example of libration found in our exhaustive exploration of resonant arguments between Pallene and a major Saturnian moon. 

On the $5\times10^4$~yr realisations (\cref{fig:1916EPlong,fig:85tPlong}), we recover a Tethys-Pallene 8:5 argument and find an additional 19:16 direct argument between Enceladus and Pallene. The latter argument (\cref{fig:1916EPlong}), $\phi_\mathrm{EP}=19\lambda' - 16\lambda - 3\varpi$, presents a distinct libration interval between $3.8$ and $4\times10^4$~yr around 90$^\circ$. Finally, the Tethys-Pallene 8:5 argument is displayed in \cref{fig:85tPlong}. We observe that the soft slope visible in the shorter timescale (\cref{fig:85tP}) is maintained in this scale, which then steepens after $\sim1.2\times10^4$~yr until the argument initiates an erratic behaviour, followed by alternating circulation and libration intervals. Similar behaviour occurs in the $5\times10^5$ realisation, but not in our longest integrations ($5\times10^6$~yr) where the 8:5 argument no longer exhibits signs of libration, just circulation. 

\subsubsection{Fourth-order arguments}
\label{ssec:long4th}

\begin{figure}
    \centering
        \subfloat[$\phi_\mathrm{DP}=7\lambda' - 3\lambda - 4\Omega'$ \label{fig:73DP}]{\includegraphics[width=\linewidth,trim=0 5 0 0,clip=True]{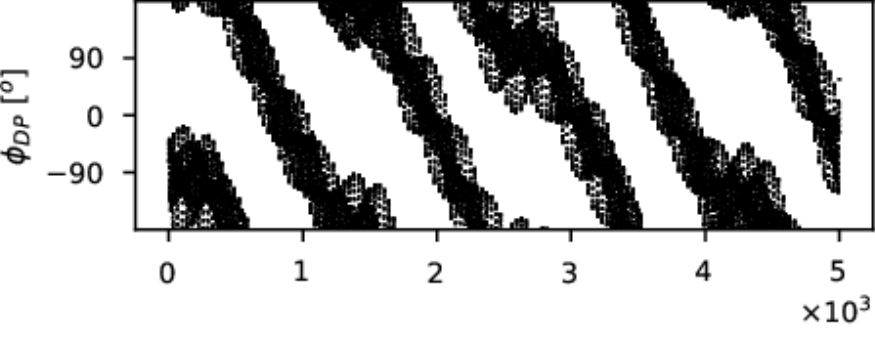} } \\[-5pt]
        \subfloat[$\phi_\mathrm{EP}=22\lambda' - 18\lambda -\Omega' - 3\Omega$ \label{fig:2218EP}]{\includegraphics[width=\linewidth,trim=0 25 0 0,clip=True]{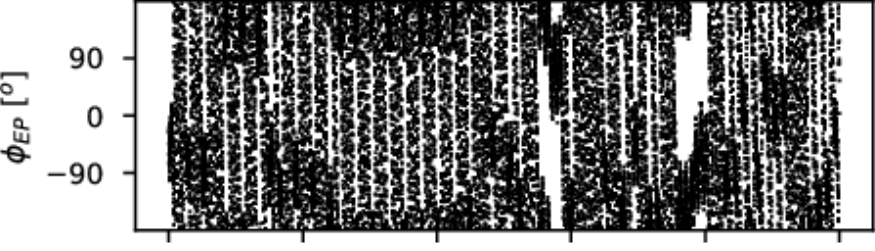} } \\[-1pt]
        \subfloat[$\phi_\mathrm{EP}=25\lambda' - 21\lambda - \varpi' - \varpi - 2\Omega'$ \label{fig:2521EP}]{\includegraphics[width=\linewidth]{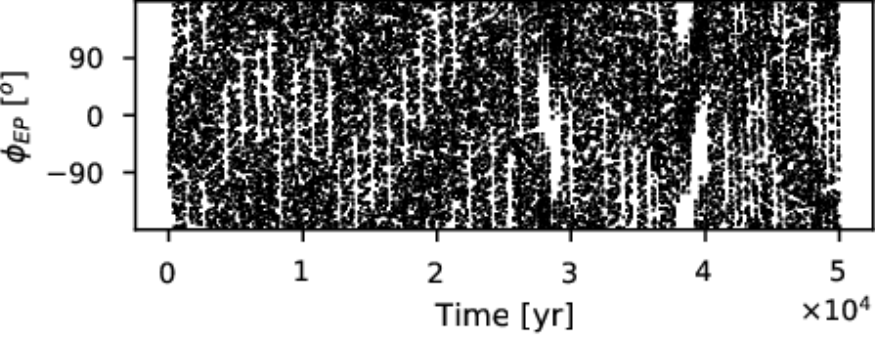} } \\[-1pt]
    \caption{
        Evolution of fourth-order direct arguments (of Pallene) with Dione and with Enceladus on different timescales. 
    }
    \label{fig:angs4th}
\end{figure}

We identified three direct arguments of fourth-order with temporal libration with Dione and with Enceladus. \Cref{fig:73DP} illustrates the argument $\phi_\mathrm{DP}=7\lambda' - 3\lambda - 4\Omega'$ between Dione and Pallene; this inclination-type resonance involves only the longitude of the ascending node of Dione; it was recovered in the timescale of $5\times10^3$~yr only. Despite the general circulation of this argument, some libration intervals with large amplitude about 180$^\circ$ are observed.

The remaining arguments are between Enceladus and Pallene (\cref{fig:2218EP,fig:2521EP}) over $5\times10^4$~yr which have the particularity that they coincide in the location of their temporal libration and have similar width in \cref{fig:diffmap}. The argument $\phi_\mathrm{EP}=22\lambda' - 18\lambda -\Omega' - 3\Omega$ corresponds to the Enceladus-Pallene 11:9 MMR location in \cref{fig:diffmap}, while the argument in \cref{fig:2521EP}, i.e., $\phi_\mathrm{EP}=25\lambda' - 21\lambda - \varpi' - \varpi - 2\Omega'$, is situated at 0.9983~$D_\mathrm{Pal}$ almost overlapping the Mimas-Pallene 9:11 MMR.

\subsubsection{Zeroth-Order arguments}
\label{ssec:long0th}

\begin{figure}
    \centering
        \subfloat[$\phi_\mathrm{DP} = 3 \lambda' - 3\lambda + \varpi' + \varpi -2\Omega'$ \label{fig:33DP} ]{\includegraphics[width=\linewidth,trim=0 25 0 0,clip=True]{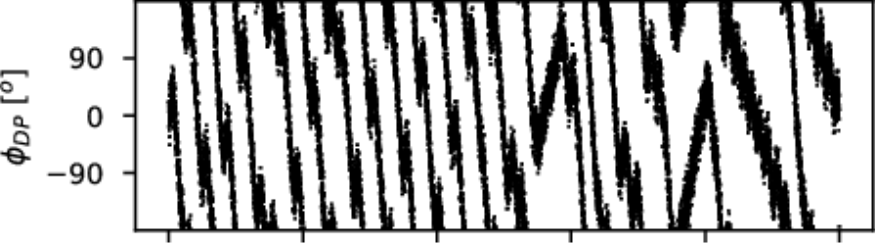} }\\[-1pt]
        \subfloat[$\phi_\mathrm{RP} = 13\lambda' - 13\lambda + \varpi' - \varpi - \Omega' + \Omega$ \label{fig:1313RP} ]{\includegraphics[width=\linewidth,trim=0 25 0 0,clip=True]{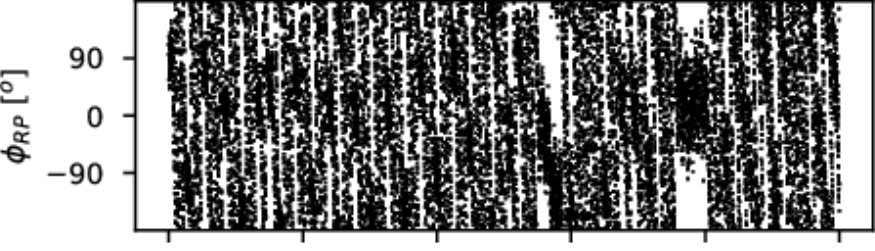} }\\[-1pt]
        \subfloat[$\phi_\mathrm{TP} = 5\lambda' - 5\lambda + \varpi' + \varpi - \Omega' - \Omega $ \label{fig:55TPa} ]{\includegraphics[width=\linewidth,trim=0 25 0 0,clip=True]{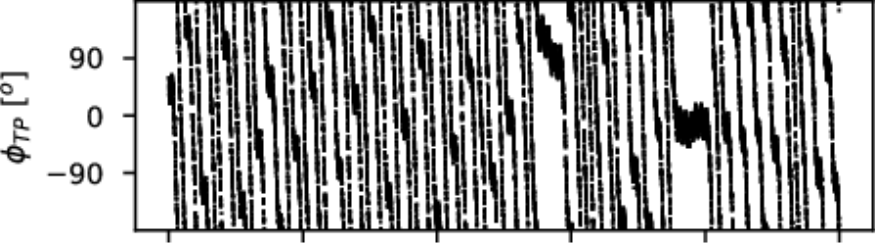} }\\[-1pt]
        \subfloat[$\phi_\mathrm{TP} = 5\lambda' - 5\lambda + 2\varpi - 2\Omega' $ \label{fig:55TPb}]{\includegraphics[width=\linewidth,trim=0 5 0 0,clip=True]{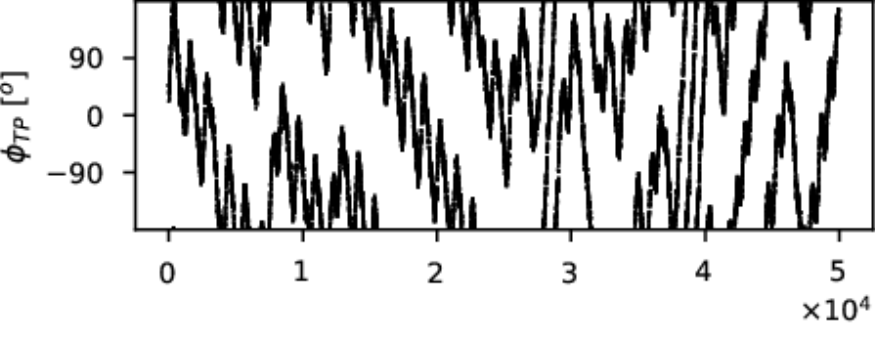} }\\[-5pt]
        \subfloat[$\phi_\mathrm{DP} = \varpi' - \varpi $ \label{fig:DPzero} ]{\includegraphics[width=\linewidth,trim=0 25 0 0,clip=True]{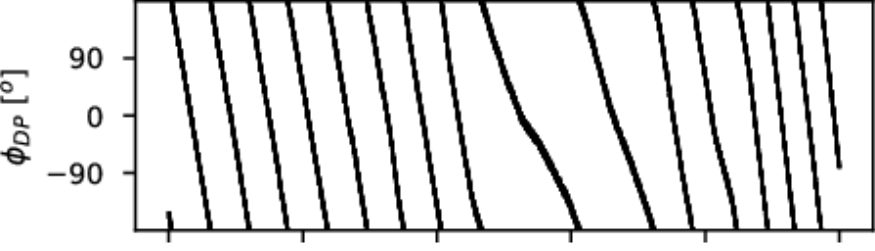} }\\[-1pt] 
        \subfloat[$\phi_\mathrm{RP} = 2\varpi - \Omega' - \Omega$ \label{fig:RPzero} ]{\includegraphics[width=\linewidth]{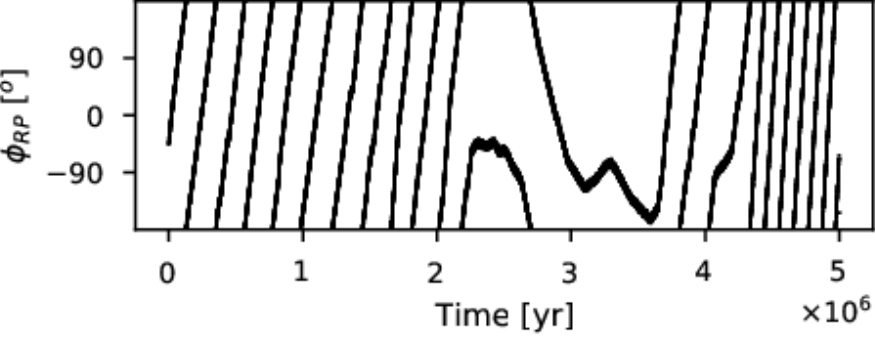} }
    
    \caption{
        Several zeroth-order arguments over timescales of $10^4$ and $10^6$~yr. In these timescales, none of the zeroth-order arguments repeat as in \cref{sec:0rdall}. Furthermore, two Titan-Pallene direct arguments of degree 5 were recovered (\cref{fig:55TPa,fig:55TPb}); the latter displays a clearer temporal libration about 0$^\circ$ at $3.75\times10^4$~yr.
    }
    \label{fig:angsZERO}
\end{figure}

We recovered several zeroth-order arguments with various degrees of Pallene with Dione, Rhea, and Titan from the $5\times10^4$~yr and $5\times10^6$~yr integrations. The clearest libration occurs in the argument $\phi_\mathrm{TP} = 5\lambda' - 5\lambda + \varpi' + \varpi - \Omega' - \Omega $ between Titan and Pallene (\cref{fig:55TPa}) which coincides with the libration intervals of $\phi_\mathrm{RP} = 13\lambda' - 13\lambda + \varpi' - \varpi - \Omega' + \Omega$ (\cref{fig:1313RP}) and the reversal of circulation of $\phi_\mathrm{DP} = 3\lambda' - 3\lambda + \varpi' + \varpi - 2\Omega'$ (\cref{fig:33DP}). A different argument involving Titan and Pallene also with degree 5 is shown in \cref{fig:55TPb}. It displays a slow circulation with intervals of faster circulation coincident with the libration period of the argument in \cref{fig:55TPa}.   

Finally, the bottom two panels in \cref{fig:angsZERO} only involve the apsidal and nodal longitudes. The longer and most evident circulation found in our simulations occurs between the longitudes of pericentres of Dione and Pallene (\cref{fig:DPzero}), while a reversal in the circulation of the argument $\phi_\mathrm{RP} = 2\varpi - \Omega' - \Omega$ between Rhea and Pallene (\cref{fig:RPzero}) takes place around 2 and 3~Myr, producing a temporary libration in this interval.

\subsection{What all these arguments mean}

In our exhaustive search for resonant behaviour, we did not find any clear libration for either first- or second-order resonances among any of the Pallene pairings with the six large moons considered in this work. This means that the proposed 19:16 MMR between Enceladus and Pallene does not exist.

The quasi-resonant zeroth-order argument suggested by \citet{Callegari10} between Pallene and Mimas is also present with other moons. Taking into account the values of $e$ and $I$, we consider that the most important contribution of this combination to the disturbing function would be the one arising from the Mimas-Pallene pair, followed by the Titan-Pallene pair. The small discrepancy in the circulation period found in this paper with respect to the value found in \citet{Munoz17} may be due to the updated values of both $GM_m$ and Saturn's zonal harmonics.

The clearest librations, observed for arguments of third- and fourth-order resonances, would, however, have only a slight contribution to the disturbing function, given the small eccentricities and inclinations of both Pallene and the other moons. For the same reason, we do not expect that any resonance of major order, or with the same order but of a larger degree, would result in any significant contribution to the evolution of Pallene.

Based on our analysis, we can conclude that Pallene is not currently trapped in any two-body MMR of any order or degree. This does not exclude the possibility of the existence of a more complex, three-body resonance, involving Pallene and some of the other moons, not exclusive to Mimas and Enceladus. Although a preliminary analysis of this possibility does not show any clear signs for the existence of such a configuration, an in-depth analysis of three-body resonances is left for future work.

We find no significant variations in the overall results of simulations shorter than $\sim2\times10^4$~yr as a function of density (this includes all the simulations referring to the evolution of Pallene's ring). For longer simulations, the accumulation of numerical errors, resulting from differences in the $GM_m$ values of order $10^{-7}$ - $10^{-8}$, and the weak chaotic nature of the N-body problem, lead to a loss of coherence among different simulations; nonetheless, statistically, all the longer-term simulations are equivalent to each other up to our longest integration time of $5\times10^6$~yr. Despite the shift in angular phases, the main orbital elements, $(a,e,I)$, remain confined and evolve regularly up to 5~Myr.

\section{Origin and Dynamical Evolution of the Pallene Ring}
\label{sec:ring}

Pallene shares its orbit with a complete dusty ringlet \citep{Hedman09, Hedman10} seen by \emph{Cassini} images in high phase angle, while a concentration of large particles ($\gtrsim 100~\mu m$) was detected in other phase angle images \citep{Hedman09}. These data indicate that the ring is composed of micrometre-sized particles and denser bodies. \citet{Hedman09} found that the ring has a radial full-width of $\sim$2500~km and a vertical profile with a full-width at half-maximum (FWHM) of $\sim$50~km, that is, the ring is vertically thin. More recently, \citet{Spahn19} measured the FWHM of the Gaussian vertical profile as $\sim$270~km while obtaining the same radial full-width as \citet{Hedman09}. \citet{Spahn19} also found that the radial mean position of the ring is shifted radially outwards by $\sim$1100~km. 

\subsection{Pallene's Mass Production by Impacts} \label{productions}

\defcitealias{Madeira18}{M18}
\defcitealias{Madeira20}{M20}
In theory, satellites of a few kilometres in radius are efficient sources of debris for rings and arcs due to their reasonably large cross-section and low escape velocity \citep{Poppe16}. However, \citet[][hereafter M18]{Madeira18} and \citet[][hereafter M20]{Madeira20} found that Saturn's three smallest moons (Aegaeon, Anthe, and Methone) do not replenish the material lost by their associated arcs due to non-gravitational forces. It raises the question of whether Pallene can maintain its diffuse ring in a steady state, as proposed by \citet{Hedman09}. In this section, we compute the amount of debris ejected from Pallene and analyse the fate of the ejecta in Section~\ref{ejectedmaterial}.

The production of material by Pallene is the result of energetic collisions between the surface of the satellite and fluxes and interplanetary dust projectiles (IDPs) \citep{Grun85,Divine93}. Typically, IDPs are supplied by families of comets \citep[Jupiter-family, Halley-type, and Oort-Cloud comets,][]{Dikarev05,Nesvorny10,Poppe11} and by the Edgeworth-Kuiper Belt \citep[EKB,][]{Landgraf02}. Data obtained by the Student Dust Counter (SDC) on board the New Horizons spacecraft indicate that the Saturn neighbourhood is dominated by EKB dust \citep{Piquette19,Poppe19} corresponding to the population that reaches the orbits of Saturn's satellites.

In addition to the impacts with IDPs, Pallene may produce material due to impacts with the E~ring particles (ERPs). The icy-dust emission from Enceladus's volcanism is the principal source of the E~ring \citep{Spahn06,Kempf10}, producing a dense debris that impacts the surface of satellites immersed in the E~ring \citep{Spahn06}. The mass production rate by Pallene (or any other satellite) is given by \citep{Krivov03}:
\begin{equation}
 M^+=\pi R_m^2(F_{\rm IDP}Y_{\rm IDP}+F_{\rm ERP}Y_{\rm ERP}) 
\end{equation}
where $R_m$ is the satellite radius, $F_{\rm IDP}$ and $F_{\rm ERP}$ are the mass flux of impactors due to IDPs and ERPs, respectively, and $Y_{\rm IDP}$ and $Y_{\rm ERP}$ are the ejecta yields associated to each projectile-type.

The ejecta yield is the ratio between the mass produced during the impact and the impactor's mass. This quantity is calculated using the empirical prescription obtained by \citet{Koschny01} for pure-ice satellites:
\begin{equation}
 Y=\frac{6.69\times10^{-8}}{2^{1.23}~{\rm kg/m^3}}\left(\frac{1}{927~{\rm kg/m^3}}\right)^{-1}\left(\frac{m_{\rm imp}}{\rm kg}\right)^{0.23}~\left(\frac{v_{\rm imp}}{\rm m/s}\right)^{2.46}   \label{yield} 
\end{equation}
where $m_{\rm imp}$ and $v_{\rm imp}$ are the mass and velocity of the impactor.

Pallene, Aegaeon, Anthe, and Methone are likely porous satellites \citep{Hedman20}, due to their bulk densities, $\rho_m$, being lower than the density of ice ($\rho_{\rm ice}$=927~kg/m$^3$). Since an impact on a porous body is expected to generate more material than an impact on a non-porous surface, we artificially modified \Cref{yield} by introducing a porosity ratio of ${\rm \alpha_p=\rho_m/\rho_{ice}}$: 
\begin{equation}
 Y_p=\frac{(6.69\times10^{-8})^{\alpha_p}}{2^{1.23}~{\rm kg/m^3}}\left(\frac{\alpha_p}{927~{\rm kg/m^3}}\right)^{-1}\left(\frac{m_{\rm imp}}{\rm kg}\right)^{0.23}~\left(\frac{v_{\rm imp}}{\rm m/s}\right)^{2.46}.   \label{yieldp}  
\end{equation}

We must point out that \Cref{yieldp} is theoretical, and there is no experimental evidence that it actually rules the yield for a porous body. In this work, we will use \Cref{yieldp} only as an artifice to demonstrate the uncertainties related to the collision yield. The parameters assumed for the two projectile populations are presented below.

\subsubsection{Interplanetary Dust Projectiles} \label{idpsec}

In Saturn's vicinity, the (unfocused) IDP mass flux is estimated to be $F_{\rm IDP}^{(\infty)}=10^{-16}$~kgm$^{-2}$s$^{-1}$ \citep{Altobelli18,piquette2019thesis}. We assume the IDPs' velocity near Saturn as the median speed of EKB grains, $v_{\rm imp}^{(\infty)}=3.1$~km/s \citep{Poppe16} and the mass of the impactors as $m_{\rm imp}=10^{-8}$~kg. When IDPs enter Saturn's Hill sphere, the planet's gravitational force is responsible for enhancing the flux and velocity of the projectiles \citep{Krivov03}. Respectively, the mass flux and velocity of IDPs at an orbital radius $r$ are \citep{Colombo66,Krivov03}:
\begin{multline}
    \frac{F_{\rm imp}}{F_{\rm imp}^{(\infty)}}=\frac{1}{2}\left(\frac{v_{\rm imp}}{v_{\rm imp}^{(\infty)}}\right)^2+\frac{1}{2}\frac{v_{\rm imp}}{v_{\rm imp}^{(\infty)}}\left[\left(\frac{v_{\rm imp}}{v_{\rm imp}^{(\infty)}}\right)^2 \right. \\ 
    \left. -\left(\frac{R_\mathrm{S}}{r}\right)^2\left(1+\frac{2GM_S}{R_\mathrm{S}(v_{\rm imp}^{(\infty)})^2}\right)\right]^{1/2} ,
\end{multline} 
and
\begin{equation}
    \frac{v_{\rm imp}}{v_{\rm imp}^{(\infty)}}=\sqrt{1+\frac{2GM_S}{r\left(v_{\rm imp}^{(\infty)}\right)^2}} .
\end{equation}

\subsubsection{E~Ring Impactors} \label{erpsec}

We assume the E~ring is composed of sub-micrometric ejecta from Enceladus onto highly eccentric orbits \citep{Nicholson96,Kempf08a,Postberg08,Ye14a}. The average mass of impactors is assumed to be $m_{\rm imp}=2.3\times 10^{-15}$~kg \citep[$0.65~\mu$m,][]{Spahn06} and the impact velocity is given by \citep{Hamilton94,Spahn06}:
\begin{equation}
 v_{\rm imp}=\frac{1}{2}\sqrt{\frac{GM_S}{r}}  
\end{equation}

The flux of impactors on the equator plane is assumed to be $F_{\rm ERP}=m_{\rm imp}v_{\rm imp}N_{\rm ERP}$, where $N_{\rm ERP}$ is the particle number density in the E~ring, extracted from the Cosmic Dust Analyser data \citep{Kempf08a}:
\begin{equation}
    N_{\rm ERP}(r)=N_0\exp\left(-\frac{z_0(r)^2}{2\sigma(r)^2}\right)\left\{\begin{array}{ll} \left(\frac{r}{3.98~R_\mathrm{S}}\right)^{50} & \textrm{for}~r\leq 3.98~R_\mathrm{S} \\ 
    \left(\frac{r}{3.98~R_\mathrm{S}}\right)^{-20} & \textrm{for}~r> 3.98~R_\mathrm{S}, \end{array}\right. 
\end{equation}
with
\begin{equation}
    \sigma(r)=1826~{\rm km}+(r-3.98~R_\mathrm{S})\left\{\begin{array}{ll} -\frac{467~{\rm km}}{0.82~R_\mathrm{S}} & \textrm{for}~r\leq 3.98~R_\mathrm{S} \\ 
    \frac{510~{\rm km}}{0.77~R_\mathrm{S}} & \textrm{for}~r> 3.98~R_\mathrm{S}, \end{array}\right. 
\end{equation}
and,
\begin{equation}
    z_0(r)=\left\{\begin{array}{ll} -1220\left(\frac{r-3.98~R_\mathrm{S}}{0.82~R_\mathrm{S}}\right)~{\rm km} & \textrm{for}~r\leq 3.98~R_\mathrm{S} \\ 
    0 & \textrm{for}~r> 3.98~R_\mathrm{S}, \end{array}\right.    ,
\end{equation}
where $N_0$ is the maximum particle number density -- near Enceladus' radius -- set as $N_0=1~$m$^{-3}$ \citep{Ye14b}.

\subsubsection{Mass Production Rate of Aegaeon, Anthe, Methone, and Pallene}

Following the prescription described in Sections~\ref{idpsec} and \ref{erpsec} and using \cref{yield}, we estimate the mass production rate of Pallene as
\begin{equation}
 M^+\sim 7.4\times 10^{-4}~{\rm kg/s} .
\end{equation}

In order to determine whether Pallene can maintain the ring, we need to estimate the mass of the structure and compare it with the lifetime of the ejected material, which is obtained by N-body numerical simulations in Section~\ref{ejectedmaterial}. If the time $\mathcal{T}$ for Pallene to produce the amount of mass observed in the ring is shorter than the particles' lifetime, then the satellite is an efficient source for the ring and the structure will be in a steady state. On the other hand, if $\mathcal{T}$ is longer than the lifetime of the particles, the ring will disappear unless another source keeps it in a steady-state.

The time for the satellite to produce the observed mass of the ring is \citepalias{Madeira20}
\begin{equation}
\mathcal{T}=M_\mathrm{Ring}/M^+ ,  
\end{equation}
if $M_\mathrm{Ring}$ is the mass of a ring (or arc), as given by \citep{Sfair12}:
\begin{equation}
M_\mathrm{Ring}=A\left(\frac{4}{3}\pi \rho_{\rm ice}\right)\int_{0.1~\mu m}^{100~\mu m} C\pi s^{3-q}ds ,   
\end{equation}
where $s$ is the physical radius of the particles, $C$ is a constant, and $q$ is the slope of the size distribution of the particles. The surface area is $A=r\Delta\theta \Delta r/2$ \citepalias{Madeira20}, where $\Delta\theta$ is the angular width of the ring/arc in radians and $\Delta r$ is the radial width. The constant $C$ can be obtained from the observed optical depth $\tau$ \citep{Sfair12}
\begin{equation}
\tau=\int_{0.1~\mu m}^{100~\mu m} C\pi s^{2-q}ds .   
\end{equation}

The distribution of particles in Pallene's ringlet is not constrained by observational data. However, the data regarding the size distribution of the E~ring provides us with a range of possible slopes $q$ for the ringlet, with values ranging from 1.9 to 5 \citep{Horanyi08,Kempf08a,Ye14a,Srama20}. For instance, \citet{Horanyi08} estimated from numerical simulations that the grain density in the E~ring follows a power law distribution with $q=2.5$, while \citet{Kempf08a} obtained slopes between 4 and 5 for $s>0.9~\mu$m from Cassini data. The slopes reported by \citet{Ye14a} vary between 3 and 4 for $s>10~\mu$m. 
To cover all possible values of $q$, we assume slopes between 1 and 6.

\begin{figure}
\includegraphics[width=\columnwidth]{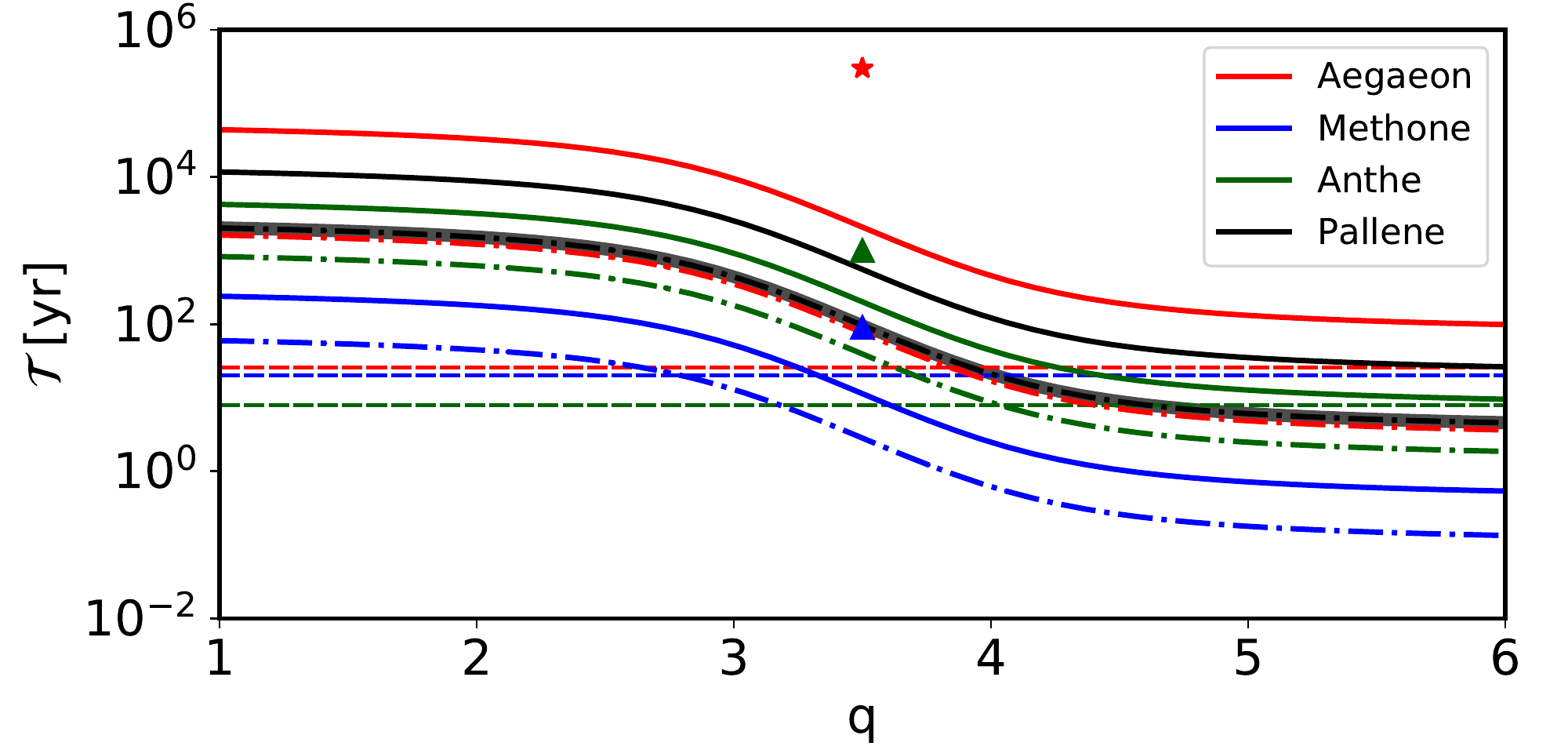}
\caption{
    Estimated time $\mathcal{T}$ for Aegaeon, Methone, Anthe, and Pallene to produce the mass of their associated arc/ring as a function of the slope $q$ of the particle radius distribution. The solid and dashed lines correspond to the time calculated following the prescription given in \cref{productions}. The solid (dash-dotted) black line corresponds to Pallene's system assuming a non-porous (porous) satellite and the grey area gives the error in the calculation of $\mathcal{T}$ due to the uncertainties in Pallene's bulk density. The coloured red, blue, and green lines correspond to the arcs of Aegaeon, Methone, and Anthe, respectively. The arc lifetime is given by different coloured dashed lines. The red star gives $\mathcal{T}$ obtained for Aegaeon by \citetalias{Madeira18} and the triangles the times obtained for Methone (blue) and Anthe (green) by \citetalias{Madeira20}. 
\label{figmplus}}
\end{figure}

\Cref{figmplus} shows the time $\mathcal{T}$ for Pallene to produce the ringlet mass (solid black line) for slopes between 1 and 6, assuming a non-porous satellite (\cref{yield}). The figure also shows the time for the moons Aegaeon, Methone, and Anthe to produce the material of their associated arcs (solid coloured lines). Meanwhile, the dash-dotted lines provide the estimated production time $\mathcal{T}$ assuming that the satellites are porous. For Aegaeon, Anthe, and Methone, we assume a bulk density of 500~kg/m$^3$, while for Pallene this value is 250~kg/m$^3$. The filled region surrounding the dashed black line gives the $\mathcal{T}$ calculated using the minimum and maximum bulk densities estimated for Pallene ($\rho_\mathrm{Pal}$=190-340~kg/m$^3$). The mass production rate depends only on the cross-section of the satellite, so if we assume a non-porous Pallene, the uncertainties regarding its bulk density do not affect the mass production, since the physical radius of the satellite is constrained by observational data \citep{Hedman09}.   

\citetalias{Madeira18} and \citetalias{Madeira20} estimated $\mathcal{T}$ following a simple prescription assuming production due to IDP impacts of cometary origin (with lower focused fluxes and velocities than the EKB grains), and assumed a single slope, $q=3.5$. The prescription here presented goes a step further in relation to their model because it incorporates recent data and the production due to ERP impacts. The time $\mathcal{T}$ obtained in \citetalias{Madeira18} for the arc of Aegaeon is shown by the red star in \cref{figmplus} and the times obtained in \citetalias{Madeira20} for the arcs of Methone and Anthe are the triangles with matching colours. The dashed lines correspond to the lifetime of ${\rm 10~\mu}$m-sized particles, obtained by \citetalias{Madeira18} and \citetalias{Madeira20}.

Our times are shorter than those estimated in previous works. \citetalias{Madeira18} obtained that Aegaeon's arc will most likely disappear if it is composed exclusively of micrometre-sized grains. Here, we also obtained that a non-porous Aegaeon cannot replenish the arc material when we disregard other sources in the arcs,\footnote{We do not compute production due to ERPs because Aegaeon is immersed in the G~ring.} since $\mathcal{T}$ is at least an order of magnitude higher than the lifespan of the particles. However, if we mimic the effect of porosity on the yield, the satellite can maintain the arc for $q\gtrsim4$. Unlike \citetalias{Madeira20}, Methone can replenish the arc material for $q>3.3$ regardless of its porosity. Although the lifetime of the particles in Anthe's arc is shorter than our $\mathcal{T}$ for the non-porous case, the radial width of the arc is unknown \footnote{We assume the same radial width as Methone's arc due to the proximity of the systems and the similar evolution of the particles under the effects of the 14:15 and 10:11 corotation resonance.} and we cannot be sure if the satellite can produce by itself the amount of material necessary to keep the arc in a steady-state or not. Assuming a porous limit, the Anthe arc seems to be in a steady-state for $q\gtrsim4$.
\begin{table}
\centering
\caption{Radial width ($\Delta r$), angular width ($\Delta\theta$), and optical depth ($\tau$) assumed for the systems of Aegaeon, Methone, Anthe, and Pallene \citep{Hedman09,Hedman10,Hedman20,Sun17,Spahn19}. The table shows the fractions of yield $Y$, flux $F$, and mass rate $M^+$ between the IDP and ERP, and the total mass rate production in kg/s.} \label{tab:mplus}
\begin{tabular}{lcccc} \hline \hline
          & Aegaeon & Methone & Anthe & Pallene \\ \hline
$\Delta r$ [km] & 250 & 1000 & 1000 & 2500 \\
$\Delta \theta$ [$^\circ$] & 60 & 10 & 20 & 360 \\
$\tau$ & $10^{-5}$ & $10^{-6}$ & $10^{-6}$ & $10^{-6}$ \\ \hline
$Y_{\rm IDP}/Y_{\rm ERP}$  & -- & ${\rm  447}$ & ${\rm  448}$ & ${\rm  449}$ \\
$F_{\rm IDP}/F_{\rm ERP}$  & -- & ${\rm  10}$ & ${\rm  4}$ &  ${\rm  10^{-1}}$ \\
$M^+_{\rm IDP}/M^+_{\rm ERP}$  & -- & ${\rm  4\times 10^3}$ & ${\rm  2\times 10^3}$ & ${\rm  50}$ \\
$M^+$[kg/s] & ${\rm  2.6\times 10^{-5}}$ & ${\rm  3.7\times 10^{-4}}$ & ${\rm  4.2\times 10^{-5}}$ & ${\rm  7.4\times 10^{-4}}$ \\ \hline 
\end{tabular}
\end{table}

\Cref{tab:mplus} summarises the initial ring (arc) parameters and the estimated fraction of yield, flux, and mass production between the IDP and ERP populations. We also include the total mass production for Aegaeon, Methone, Anthe, and Pallene for the non-porous case. Ejecta production due to IDP impacts is the most efficient for all systems. For the arcs of Aegaeon, Methone, and Anthe, production due to ERPs can be disregarded because the $M^+$ due to IDP impacts is more than 1000 times higher than for ERPs. The production due to ERPs corresponds to 2\% of the total amount produced by Pallene.

\subsection{Dynamical Model}
\label{ssec:ringdynmod}

We study the evolution and fate of Pallene's ringlet by analysing the temporal evolution of two distinct sets of particles: i) particles initially co-orbital to the satellite (\cref{coorbitalmaterial}) and ii) particles ejected from Pallene's surface (\cref{ejectedmaterial}). The first set corresponds to a scenario in which the ringlet, and perhaps Pallene, would have formed by a disruption of an ancient satellite; while the second, mimics the evolution of the material produced by impacts into the satellite (\cref{productions}).

The numerical simulations were performed using Mercury6 \citep{Chambers99} with the Bulirsch-Stoer algorithm. We used 5,000 particles with micrometric sizes ranging from 0.1~$\mu$m to 100~$\mu$m, and integrated the system until either all particles collide with Mimas, Pallene, or Enceladus or migrate outwards beyond the orbit of Enceladus. We adopted the collision detection treatment between particles and satellites as implemented in Mercury6 \citep[for details, see][]{Chambers99,Liu16}.

Micrometre-sized particles are affected by non-gravitational forces that decrease their lifetimes. Thus it is necessary to include these effects in the system. In our simulations, the particles are under the effect of a total force,
\begin{equation}
\vec{\rm F}=\vec{\rm F}_{\rm SR}+\vec{\rm F}_{\rm PD}+\vec{\rm F}_{\rm EM} + \vec{\rm F}_{\rm G} , \label{eq:totalF}
\end{equation}
where $\vec{\rm F}_{\rm SR}$ is the solar radiation force, $\vec{\rm F}_{\rm PD}$ is the plasma drag force, $\vec{\rm F}_{\rm EM}$ is the electromagnetic force, and $\vec{\rm F}_{\rm G}$ corresponds to the sum of the gravitational forces of the system: Saturn (including its gravitational coefficients), Mimas, Enceladus, Tethys, Dione, Rhea, Titan, and Pallene.

\subsubsection{Non-Gravitational Forces}

The solar radiation force ($\vec{\rm F}_{\rm SR}$) includes two components \citep{Burns79,Mignard84}: the radiation pressure (RP) caused by collisions of solar radiation on the dust grain, 
\begin{equation}
\vec{\rm F}_{\rm RP}=\frac{\Phi \pi s^2}{c}Q_{pr}\frac{\vec{r}_{sp}}{r_{sp}} ,
\label{rpforce}
\end{equation}
and the Poynting-Robertson drag (PR), caused by the re-emission of the solar radiation absorbed by the particles,
\begin{equation}
\vec{\rm F}_{\rm PR}=-\frac{\Phi \pi s^2}{c}Q_{pr}\left\{\frac{\vec{V}_{P} + \vec{V}}{c}+\left[\left( \frac{\vec{V}_{P}}{c} + \frac{\vec{V}}{c} \right)\cdot\frac{\vec{r}_{sp}}{r_{sp}} \right] \frac{\vec{r}_{sp}}{r_{sp}} \right\} ,
\label{prforce}
\end{equation}
where $c$ is the speed of light, $\Phi$ is the solar flux \citep{Burns79}, and $\vec{V}$ is the velocity vector of the particle relative to the planet. The solar radiation pressure efficiency $Q_{pr}$ (in \cref{rpforce,prforce}) depends on the radius of the particle and is computed from Mie theory \citep{Irvine65,Mishchenko99,Mishchenko02} assuming spherical ice grains. The particle is in a circumplanetary orbit $\vec{r}$ ($r = |\vec{r}|$), and the planet in a circular heliocentric orbit. The heliocentric position of Saturn $\vec{r}_{sp}$ ($r_{sp} = |\vec{r}_{sp}|$) and the magnitude of the planet's velocity $\vec{V}_{P}$ are considered constants. We also assume that Saturn shields particles from solar radiation when the planet eclipses the Sun from the particle's perspective, i.e., the solar radiation force is neglected when the particle is in the planet's shadow, which happens when
$\vec{r}\cdot\vec{r}_{sp}<0$ and $(r^2-R_\mathrm{S}^2)r_{sp}-|\vec{r}\cdot\vec{r}_{sp}|^2<0$ \citep{Liu16}.

The principal source of plasma for Saturn's magnetosphere in the E~ring region is the ionisation of neutrals provided by the Enceladus plume. The E~ring region is dominated by water group ions, i.e., O$^{+}$, OH$^{+}$, H$_2$O$^{+}$, and H$_3$O$^{+}$, the O$^{+}$ ion being the most abundant \citep{Cassidy10,Tseng10,Tseng11,Sittler15}. Direct collision of the plasma with the ring particles is responsible for a drag force ($\vec{\rm F}_{\rm PD}$) \citep{Morfill79,Morfill93,Horanyi08}, given by
\begin{equation}
\vec{\rm F}_\mathrm{PD}=\pi s^2m_iN_ia^2(n-\Omega_\mathrm{S})^2\hat{u}_t ,
\label{plasmaforce}
\end{equation}
where $n$ is the mean motion of the particle, $m_i$ and $N_i$ are the mass and number density of the plasma ions, respectively, and $\hat{u}_t$ is the unit vector in the tangential direction to the osculating orbit of the particle. 

\textit{Cassini} measurements have shown seasonal variations in ion densities ranging from $N_i\sim 40~{\rm cm}^{-3}$ to $N_i\sim 120~{\rm cm}^{-3}$ in Pallene's vicinity \citep{Elrod14,Persoon15,Persoon20}. For simplicity, we assume the plasma in the Pallene region is only composed of O$^{+}$ ions (molecular mass of 16~a.m.u.) with constant number density $N_i=65.9~{\rm cm}^{-3}$ \citep{Persoon15}. Moreover, we neglect the indirect Coulomb interaction between charged ring particles and the plasma material, since this effect is at least two orders of magnitude weaker than the direct collisions \citep{Northrop82,Grun84,Sun15}.

The ring particles are also influenced by Saturn's magnetosphere due to the charging of the particles by the ambient plasma and electrons photoemission (solar UV). Therefore, the electromagnetic force ($\vec F_{\rm EM}$) \citep{Northrop82,Burns85}, is included in our simulations as 
\begin{equation}
\vec{\rm F}_\mathrm{EM}=\frac{4\pi\epsilon_0sV}{c}\left\{\left[\vec{V}-\Omega_\mathrm{S}(\hat{u}_n\times\vec{r})\right]\times\vec{B}\right\} ,
\end{equation}
where $\epsilon_0=8.8542\times 10^{-12}$~F/m is the vacuum permittivity \citep{Chapman40}, $V$ is the electric potential, $\vec{B}$ is the magnetic field vector, and $\hat{u}_n$ is the unit vector perpendicular to the planet's equatorial plane. We adopt an equilibrium potential of $V=-3$~V for the Pallene region, as determined by \citet{Hsu11} in their investigation of the dynamics of the Saturnian stream particles.

We assumed the Saturnian magnetic field as a composition of an aligned dipole and a quadrupole \citep{Chapman40,Hamilton93}:
\begin{equation}
\vec{B}=g_{1.0}R_\mathrm{S}^3\vec{\nabla}\left(\frac{\cos{\zeta}}{r^2}\right)+\frac{g_{2.0}}{2}R_\mathrm{S}^4\vec{\nabla}\left(\frac{3\cos^2{\zeta}-1}{r^3}\right)
\end{equation}
where $g_{1.0}=0.21$~G is the Saturnian dipole momentum and $g_{2.0}=0.02$~G, the quadrupole momentum \citep{Hamilton93,Belenkaya06}; $\zeta$ is the angle between $\hat{u}_n$ and $\vec{r}$.

\subsubsection{Orbital Elements Of One Representative Particle}
The non-gravitational forces are responsible for variations in the shape and orientation of the orbits, affecting the temporal evolution of the particles. The mean temporal variations of the osculating orbital elements of a particle with mass $m$ are \citep{Mignard84,Hamilton93,Madeira20}
\begin{equation}
\dot{a}=-\frac{2na^2\alpha_{\rm r}}{c}\frac{5+\cos^2{I}}{6}+\frac{2|\vec{F}_\mathrm{PD}|}{mn}\sqrt{1-e^2} , \label{adot}
\end{equation}
\begin{multline}
\dot{e}=\alpha_{\rm r}\sqrt{1-e^2}(\cos{\Omega}\sin{\omega}+\sin{\Omega}\cos{\omega}\cos{I})\\ -\frac{3}{2}\frac{e|\vec{F}_\mathrm{PD}|}{mna}\sqrt{1-e^2}-\frac{qg_{1.0}R_\mathrm{S}^3\Omega_\mathrm{S}}{4mcna^3}e\sqrt{1-e^2}\sin^2{I}\sin{2\omega} , \label{edot}
\end{multline}
\begin{multline}
\dot{I}=\frac{\alpha_{\rm r}e}{\sqrt{1-e^2}}\sin{\Omega}\cos{\omega}\sin{I}+\frac{3}{2}\frac{|\vec{F}_\mathrm{PD}|}{mna}\sqrt{1-e^2}\sin{I}\\
+\frac{qg_{1.0}R_\mathrm{S}^3\Omega_\mathrm{S}}{8mcna^3}\frac{e^2}{\sqrt{1-e^2}}\sin{2I}\sin{2\omega}, \label{Idot} 
\end{multline}
\begin{multline}
\dot{\Omega}=-\dot{\Omega}_{\rm obl}+\frac{\alpha_{\rm r}e}{\sqrt{1-e^2}}\sin{\Omega}\sin{\omega}-(2-e)\frac{|\vec{F}_\mathrm{PD}|}{mna}\cos{I}\sqrt{1-e^2}\\
+\frac{qg_{1.0}R_\mathrm{S}^3\Omega_\mathrm{S}}{mcna^3}\frac{1}{\sqrt{1-e^2}}\left[\cos{I}-\frac{1}{(1-e^2)}\left(\frac{n}{\Omega_\mathrm{S}}\right)\right],     
\end{multline} 
and
\begin{multline}
\dot{\varpi}=\dot{\varpi}_{\rm obl}+\frac{\alpha_{\rm r}\sqrt{1-e^2}}{e}(\cos{\Omega}\cos{\omega}-\sin{\Omega}\sin{\omega}\cos{I})\\
+(2-e)\frac{|\vec{F}_\mathrm{PD}|}{mna}\sqrt{1-e^2}+\frac{qg_{1.0}R_\mathrm{S}^3\Omega_\mathrm{S}}{mcna^3}\frac{2\cos{I}}{(1-e^2)^{3/2}}\left(\frac{n}{\Omega_\mathrm{S}}\right),   
\end{multline} 
where
\begin{equation}
\alpha_{\rm r}=\frac{3\Phi\pi s^2}{2mcna}Q_{pr}. 
\end{equation}
$\dot{\Omega}_\mathrm{obl}$ and $\dot{\varpi}_\mathrm{obl}$ are the temporal variation of longitude of ascending node and argument of pericentre, respectively, due to the non-sphericity of Saturn \citep[see][]{Renner06}. 

\begin{figure}
\centering
\includegraphics[width=\columnwidth]{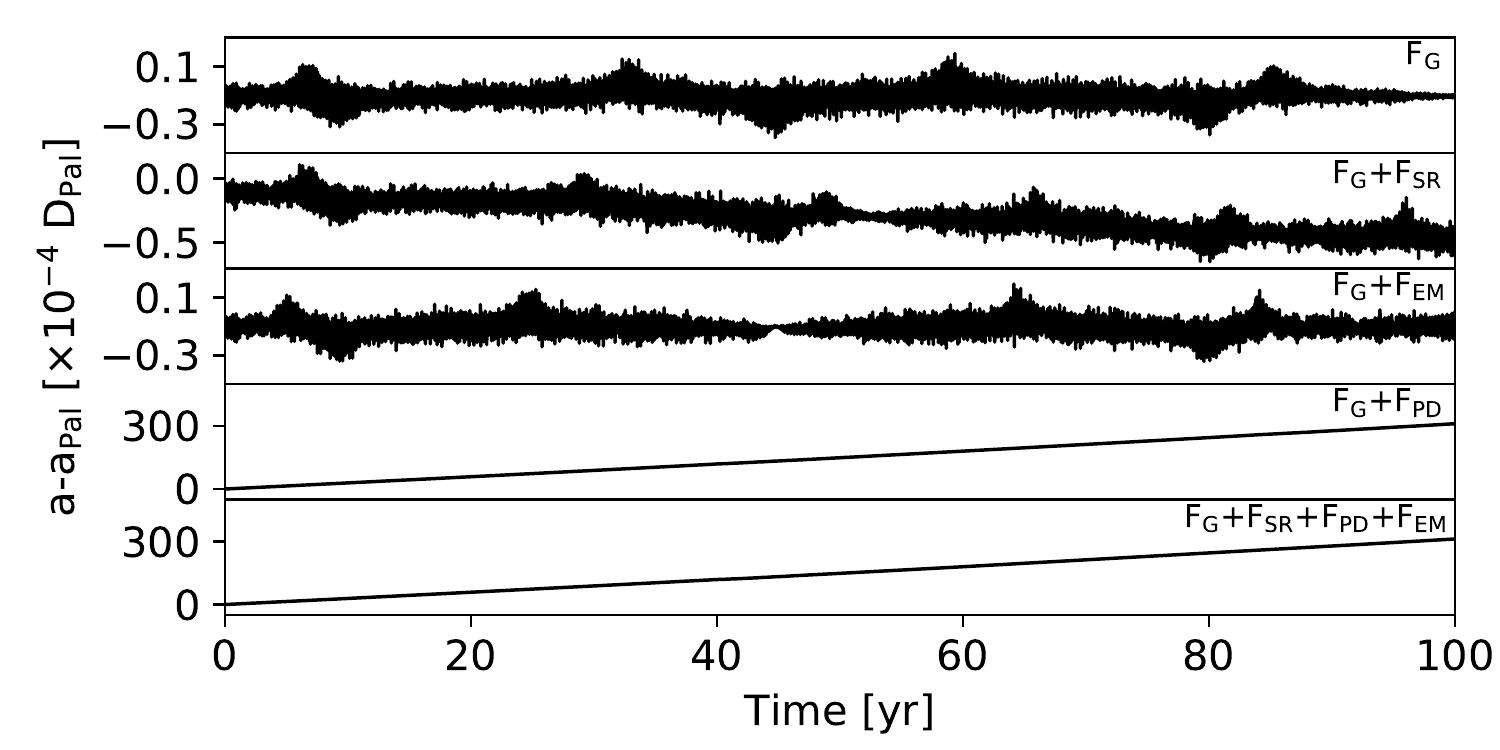}\\
\includegraphics[width=\columnwidth]{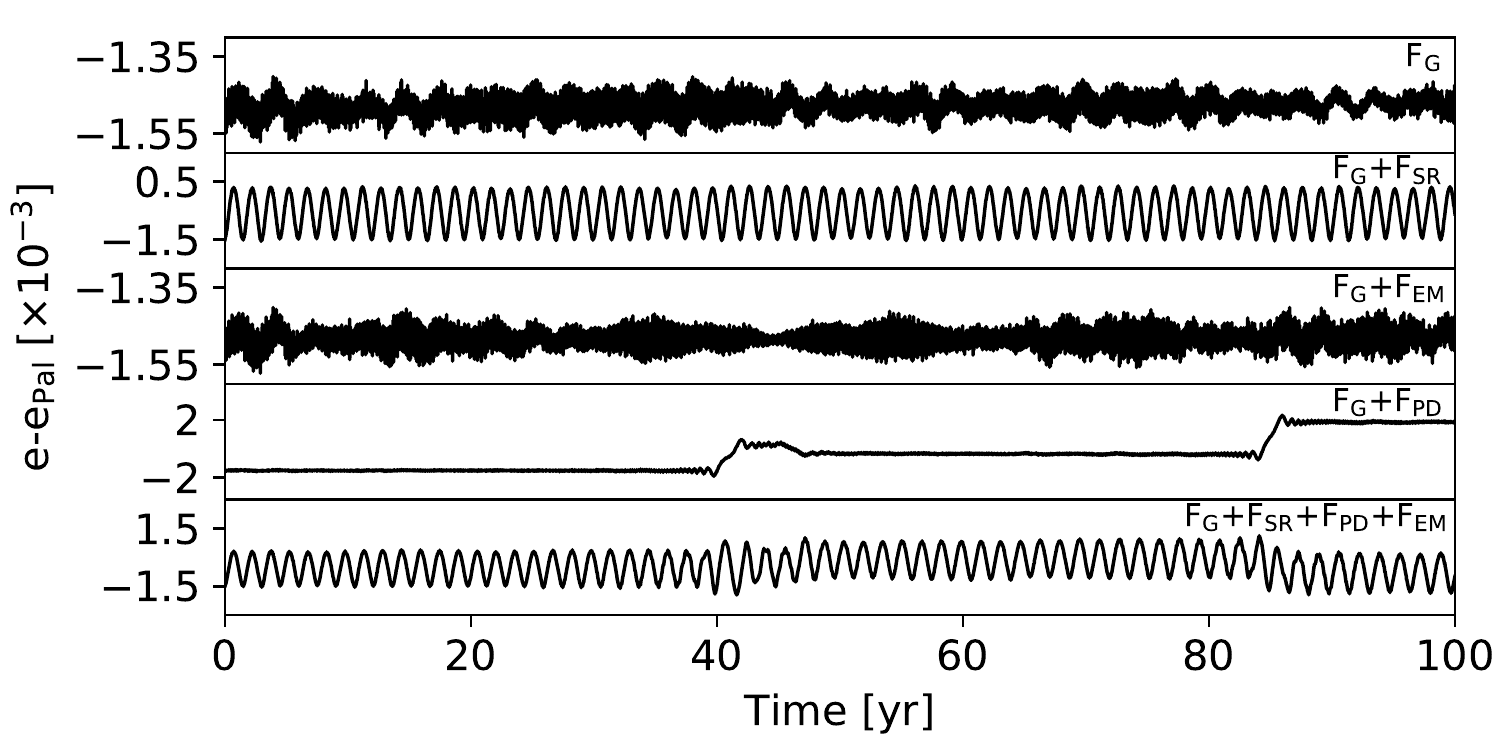}\\
\includegraphics[width=\columnwidth]{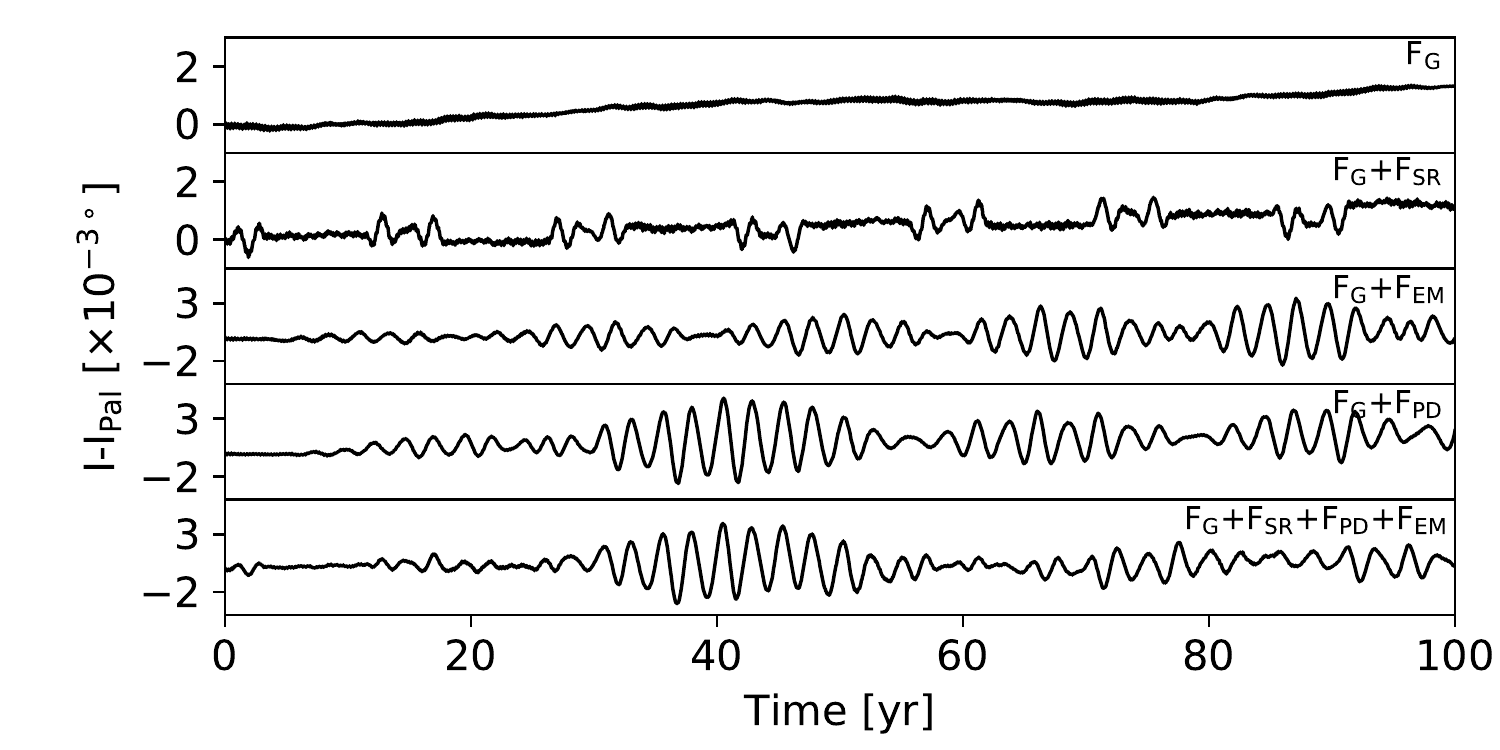}\\
\includegraphics[width=\columnwidth]{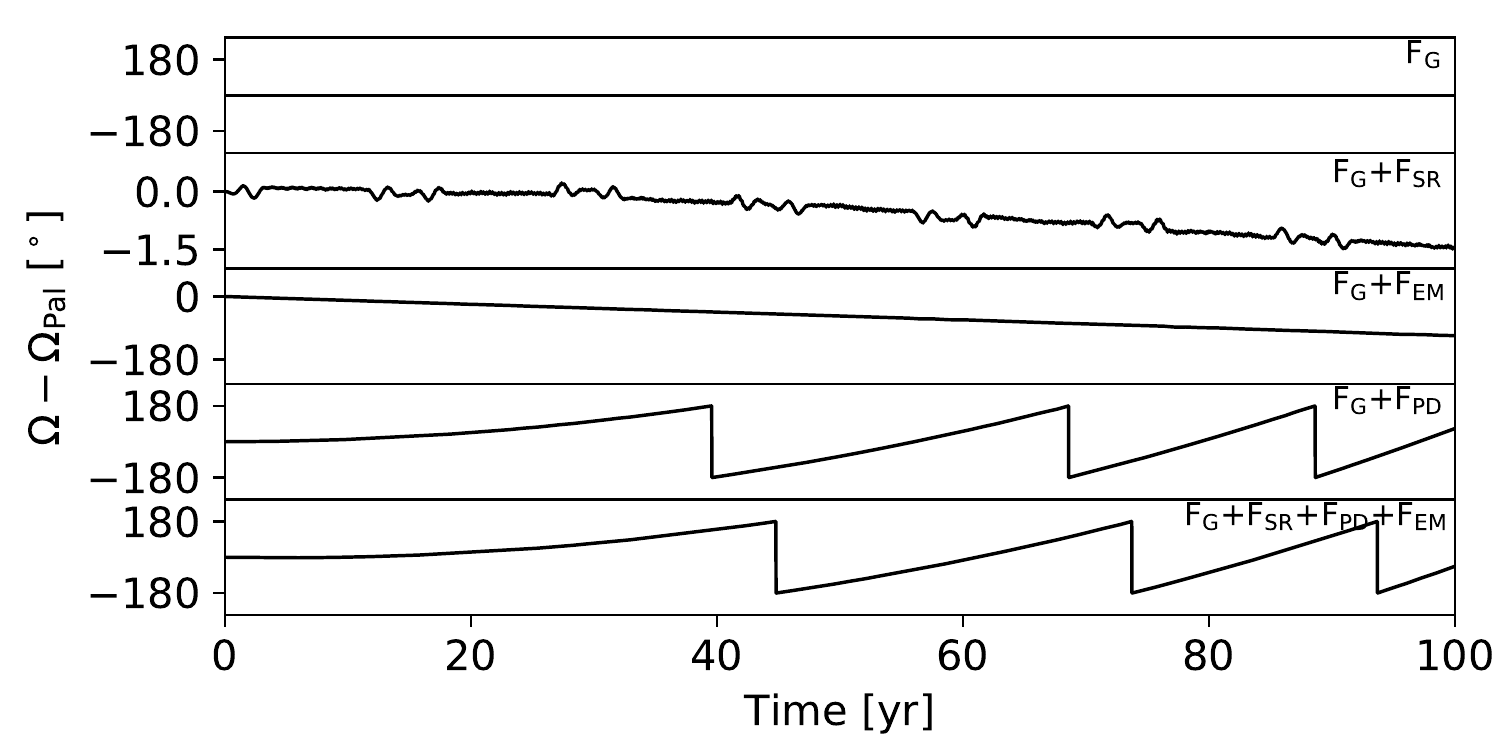}\\
\includegraphics[width=\columnwidth]{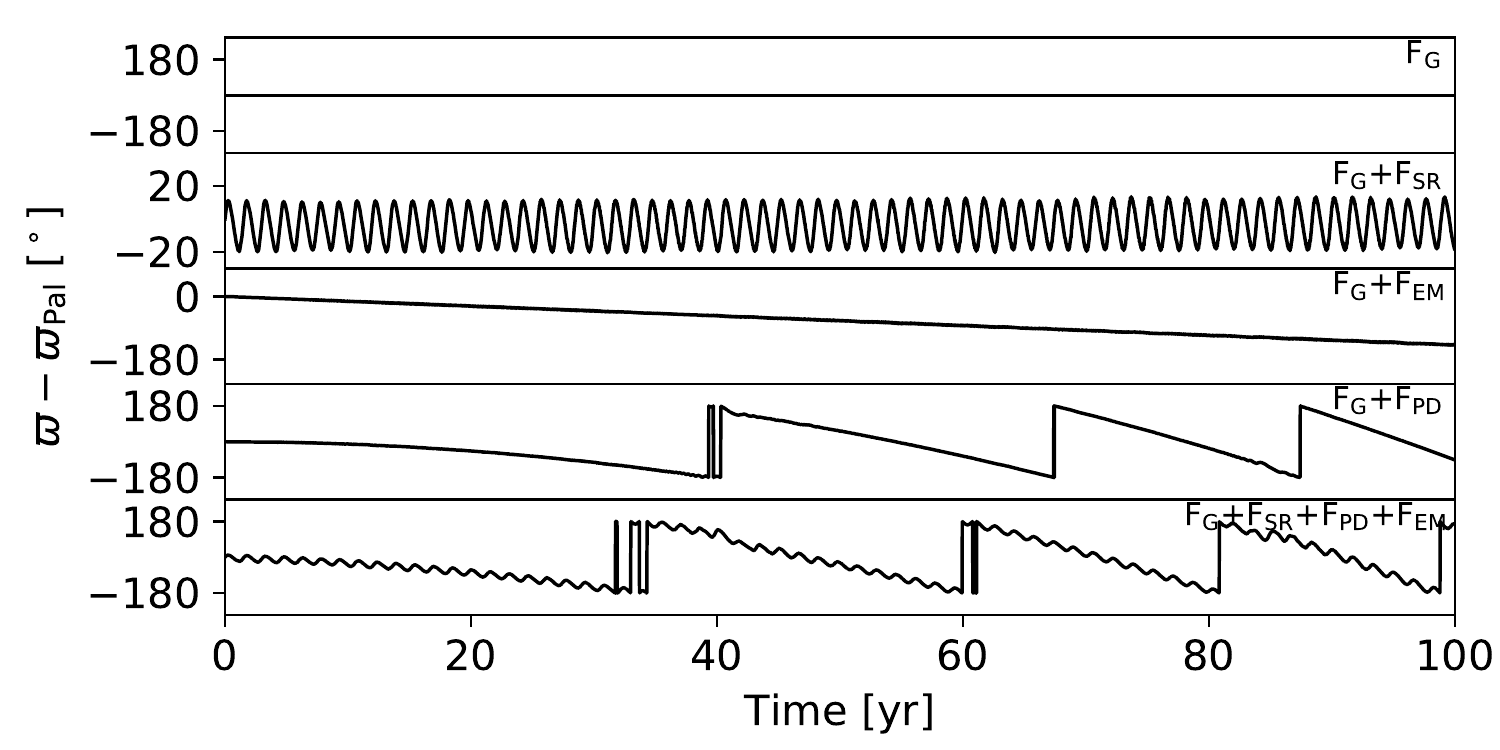}
\caption{From top to bottom: Geometric semi-major axis, eccentricity, inclination, longitude of ascending node, and argument of pericentre of a ${\rm 10~\mu}$m-sized particle co-orbital to Pallene with displacement in the mean anomaly of 180$^\circ$ in relation to the satellite. The top row of each panel shows the orbital elements when only  gravitational effect is included. The following rows display the evolution of the particle when different non-gravitational forces are included (i.e., solar radiation force, electromagnetic force, and plasma drag). Finally, the bottom row of each panel shows the effect of all forces.}
\label{efa}
\end{figure}

\Cref{efa} illustrates the variation of geometric orbital elements ($a$, $e$, $I$, $\Omega$ and $\varpi$) of one representative ${\rm 10~\mu}$m particle due to each non-gravitational force and the total force (\cref{eq:totalF}). The particle is initially co-orbital to Pallene with $\lambda=\lambda_{\mathrm{Pal}}+180^{\circ}$, where $\lambda$ and $\lambda_{\mathrm{Pal}}$ are the mean longitude of the particle and Pallene, respectively. As one can see in the top panel of \cref{efa} (\cref{adot}), the semi-major axis is affected secularly by two distinct drag effects: the Poynting-Robertson component that produces an inward migration, and the plasma drag, which increases the semi-major axis of the particle. We find that the plasma drag is at least one order of magnitude stronger than the Poynting-Robertson component for all particle sizes. While the electromagnetic force only induces short-term variations in the semi-major axis, the net outcome is that grains migrate outward when all the effects are included.

In the eccentricities, we have that the electromagnetic and solar radiation forces produce oscillations with constant period and amplitude for the same particle size \citep{Hamilton93,Madeira18,Gaslac20}. As we can see in \cref{edot}, the intensity of these effects depends on the radius of the particles, with $\dot{e}\propto s^{-3}$ for the electromagnetic force and $\dot{e}\propto s^{-1}$ for solar radiation. Thus, the effect of the electromagnetic force dominates over the solar radiation for smaller particles, while for larger sizes the electromagnetic force can be disregarded in relation to the solar radiation. 

Plasma drag, on the other hand, produces only short-term variations in the eccentricities \citepalias{Madeira20}. The jumps of this element, seen in \cref{efa}, result from the crossing of the particle with resonances with Enceladus, as will be shown in \cref{coorbitalmaterial}. For Pallene ringlet particles, the electromagnetic force dominates for ${\rm s\leq5~\mu}$m, while the solar radiation force is the most important effect on the eccentricity of ${\rm s>5~\mu}$m particles. We obtain that the non-perturbative forces produce only small variations in the inclination ($I\sim10^{-3}$~deg) for the time intervals considered by us in this section.

\begin{figure*}
    \centering
    \includegraphics[width=0.32\textwidth,trim=100 0 15 0,clip=True]{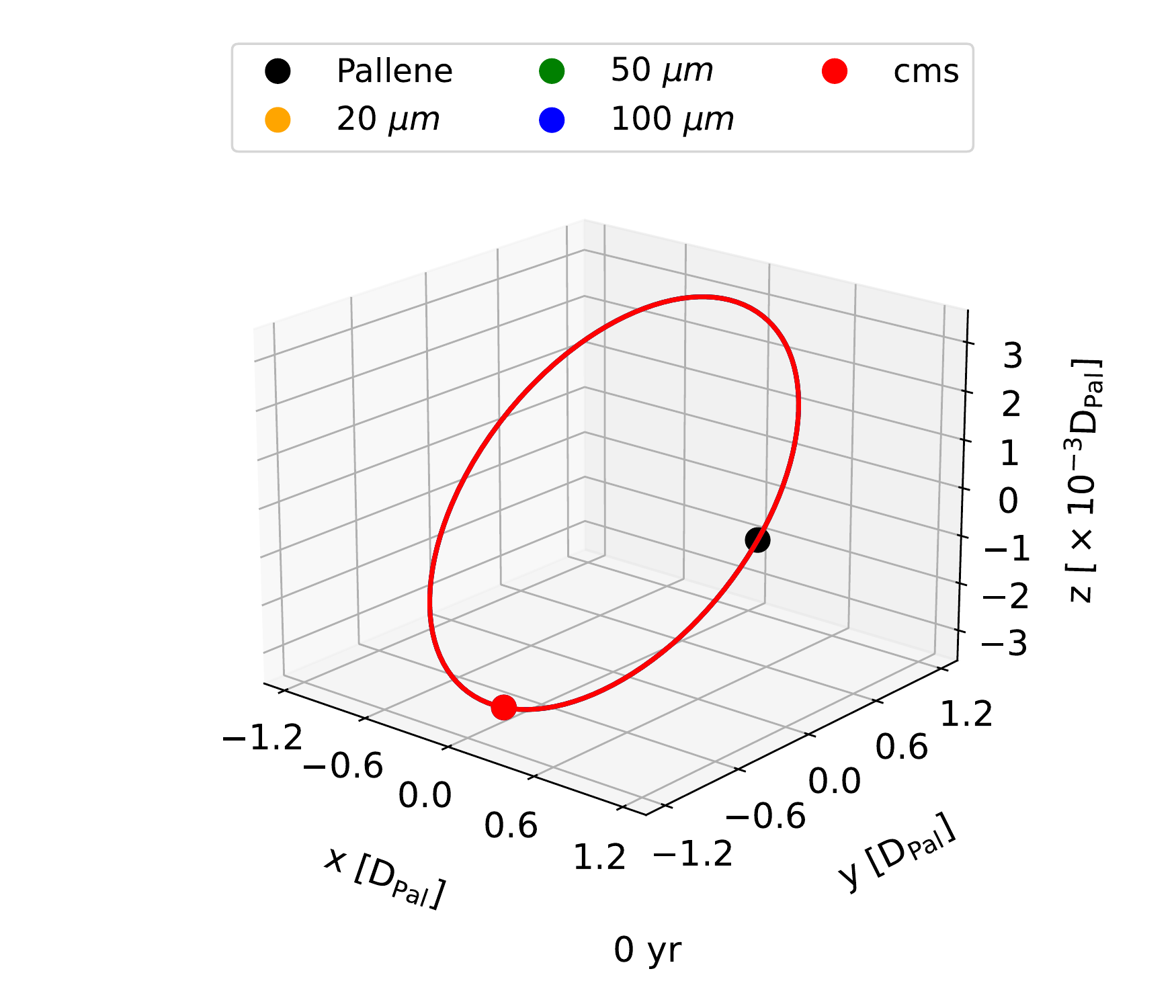}
    \includegraphics[width=0.32\textwidth,trim=100 0 15 0,clip=True]{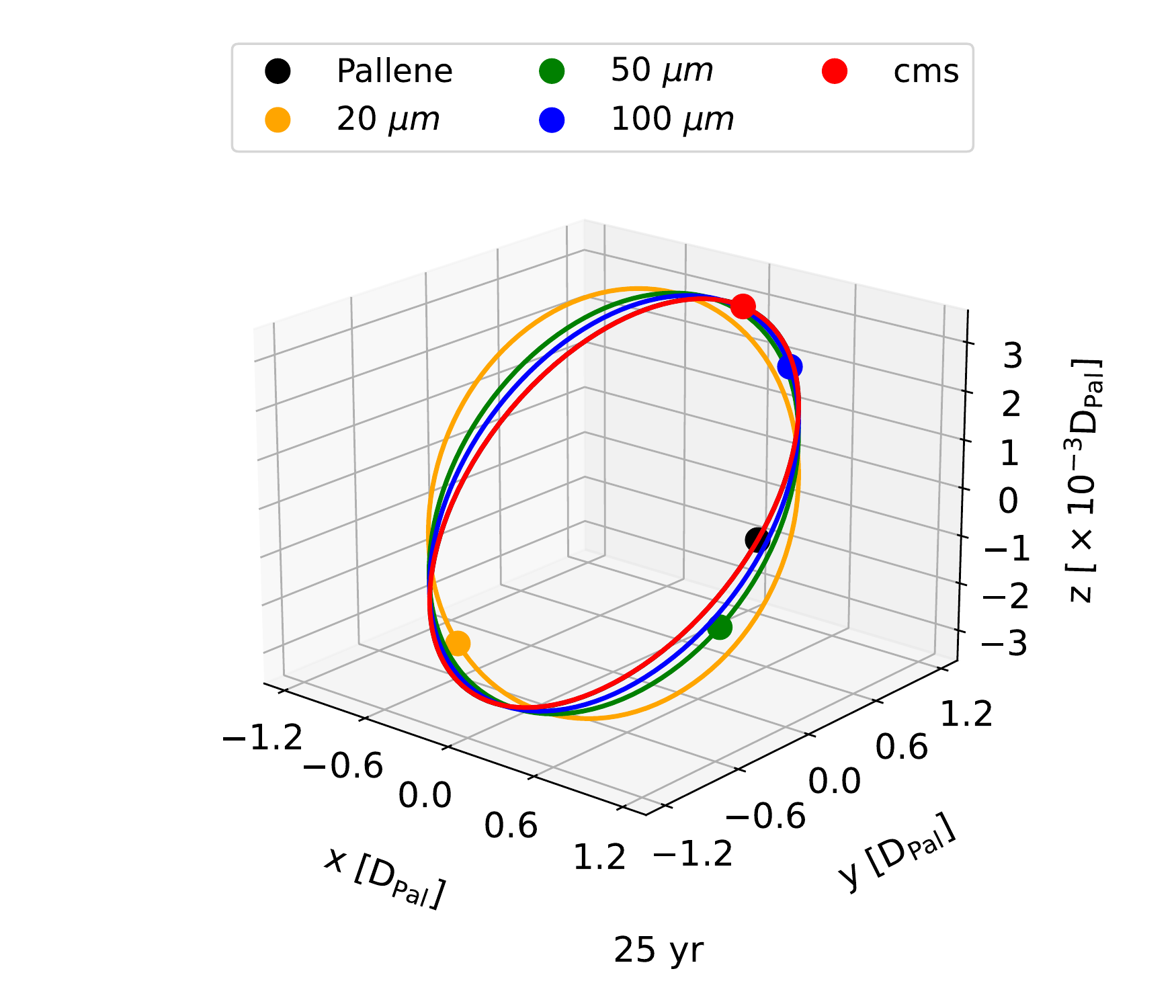}
    \includegraphics[width=0.32\textwidth,trim=100 0 15 0,clip=True]{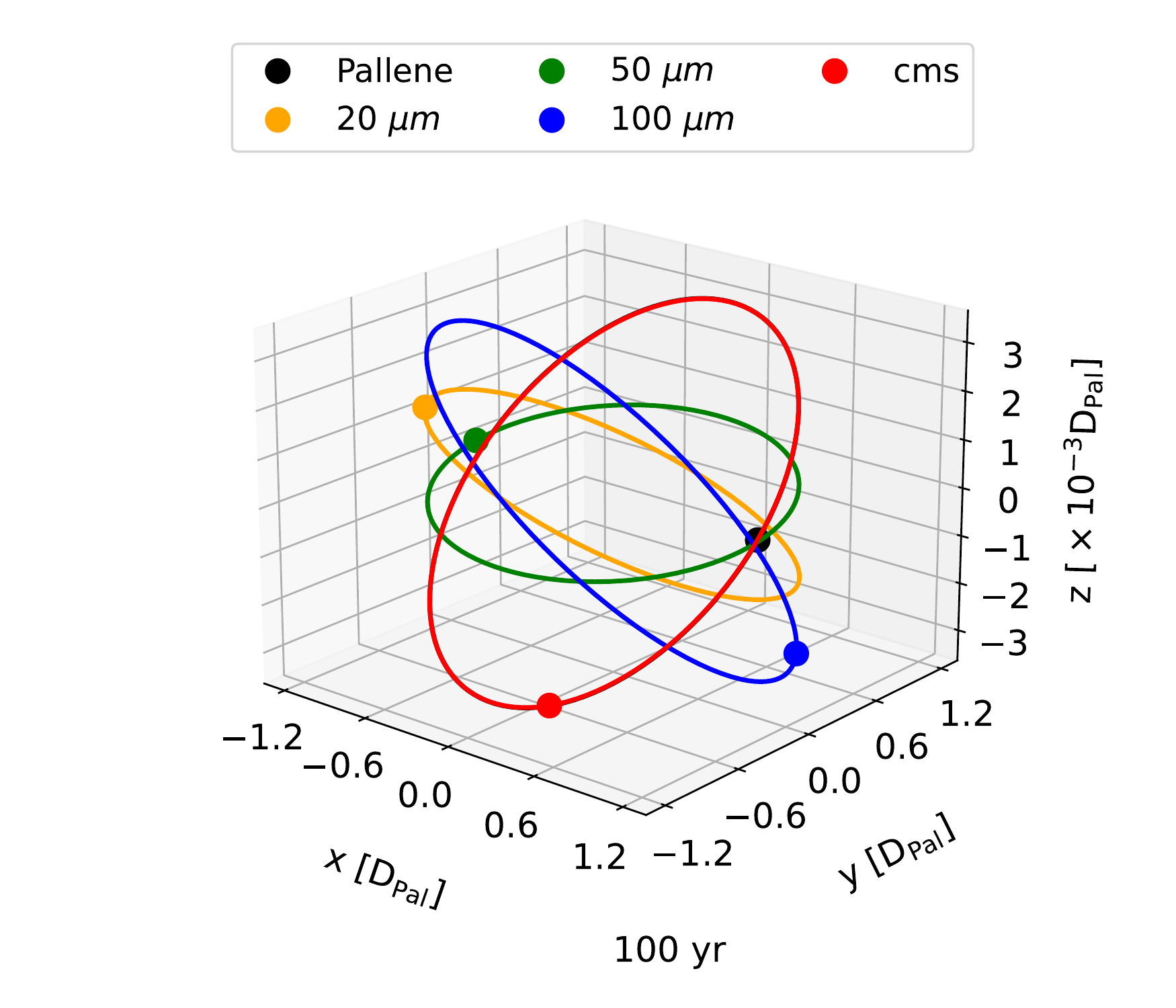}
    \caption{Snapshots of the osculating orbit (solid lines) and spatial position (dots) of Pallene (in black) and of a co-orbital particle with $\lambda=\lambda_P+90^{\circ}$. The colour indicates the body, as labelled. We assume the single particle has a radius of either ${\rm 20~\mu}$m, ${\rm 50~\mu}$m, or ${\rm 100~\mu}$m. Displayed in red, we include the case solely with gravitational forces (``cms''). The orbits are provided in the rotating frame in which Pallene is stationary at $x=1$~${\rm D_{Pal}}$. An animation of this figure is included in the electronic version; it requires Adobe Reader version $\geq$9 or similar.}
    \label{pericentre}
\end{figure*}

The longitude of ascending node and argument of pericentre are mainly affected by the plasma drag, which is responsible for the precession of the elements in relation to Pallene. \cref{pericentre} displays snapshots of the osculating orbit (solid lines) of a representative particle (coloured dots) and Pallene (black dot). We rotate the systems on each snapshot to keep Pallene in the fixed position $x=1$~${\rm D_{Pal}}$. We show particles with radius of ${\rm 20~\mu m}$, ${\rm 50~\mu m}$, ${\rm 100~\mu m}$, as well as with radius of centimetres, which corresponds to the case with only gravitational forces.

As we can see in \cref{pericentre}, without non-gravitational forces, the particle remains in the same orbit as Pallene and lacks vertical variation in relation to the satellite's orbital plane. When the non-gravitational forces are included, the orbit precesses, exhibiting vertical excursions in relation to Pallene's orbital plane. This phenomenon could be responsible for the observed vertical width of $\sim 10^2~$~km of the ring \citep{Hedman09,Spahn19} indicating that the ringlet may evolve into a torus, as observed in the gossamer rings of Jupiter \citep{Burns99}. The formation of the torus occurs when the precession of the pericentre acts long enough to completely randomise the orientation of the particles' orbits. These results will be discussed in detail in \cref{coorbitalmaterial}.

\begin{figure}
\centering
\includegraphics[width=\columnwidth]{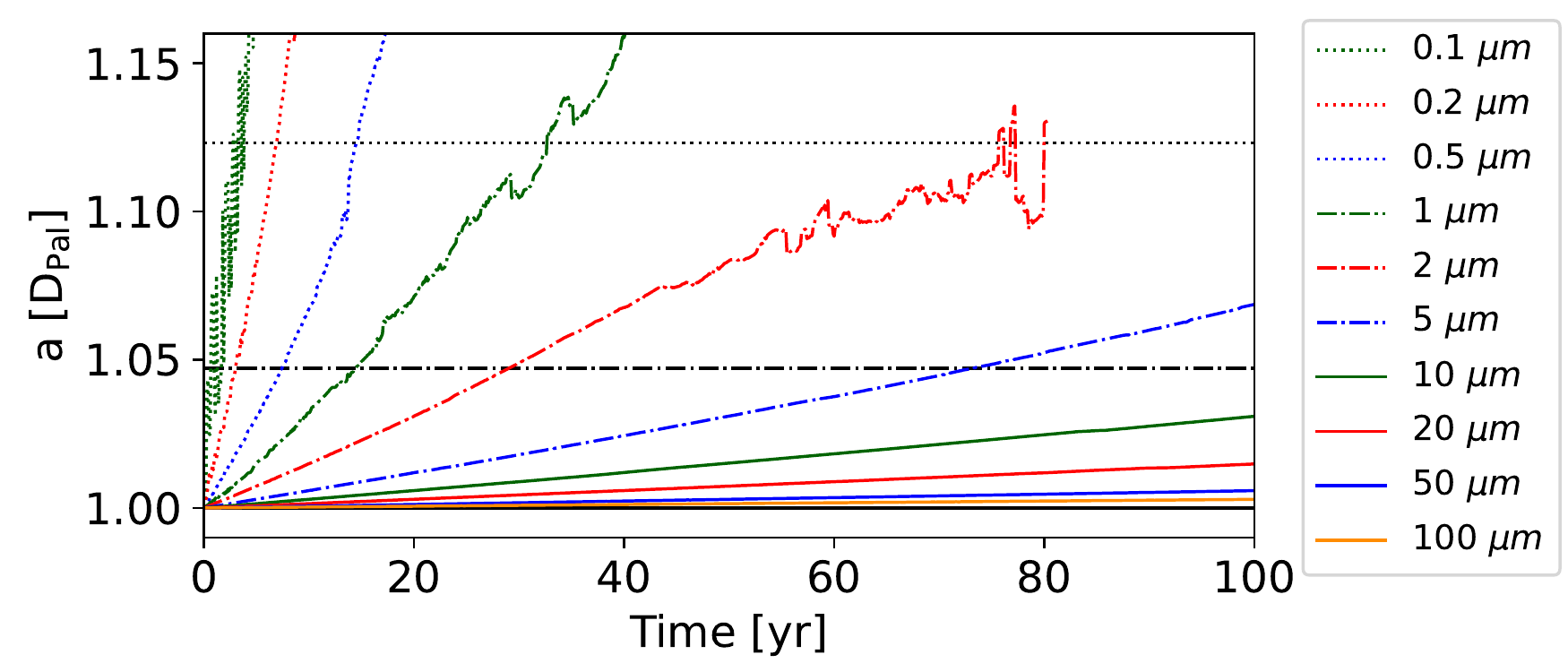}\\
\includegraphics[width=\columnwidth]{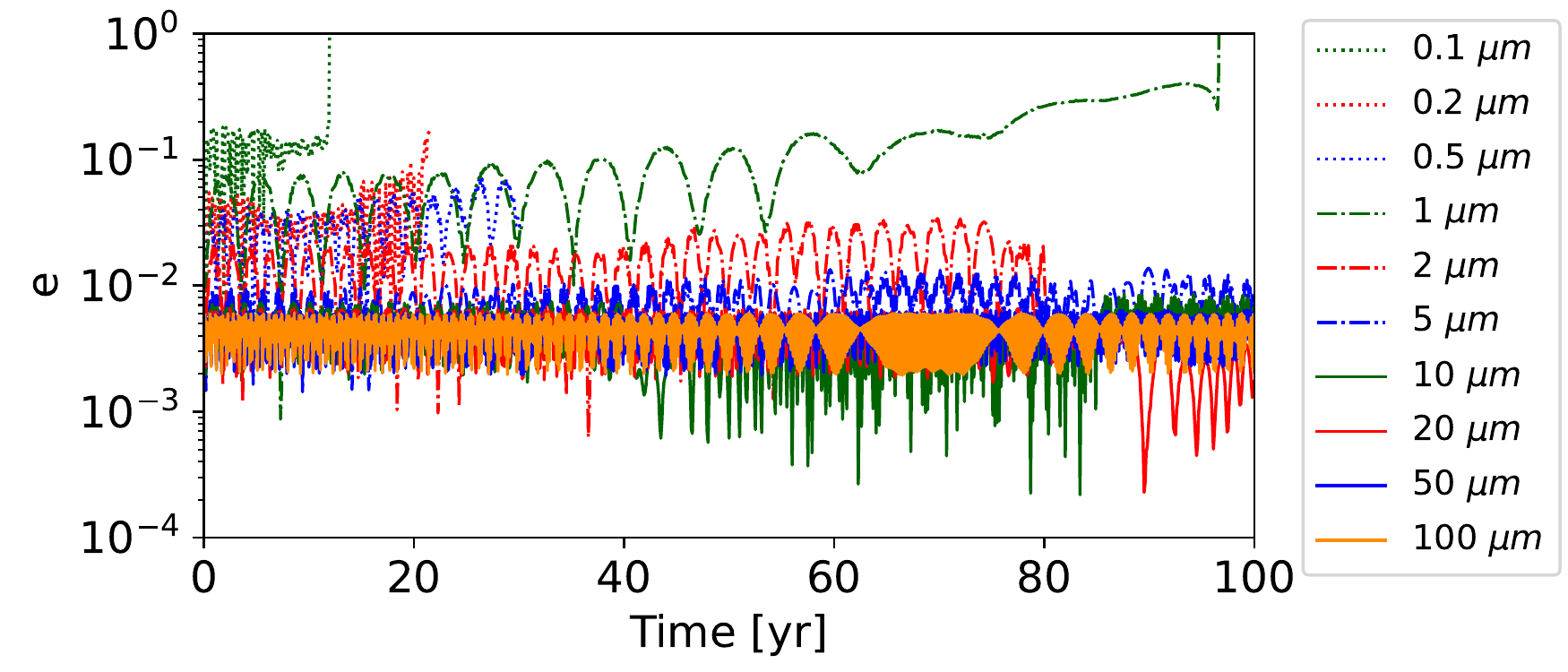}
\caption{Osculating semi-major axis and eccentricity of representative particles co-orbiting Pallene. The particles have a size of ${\rm 0.1~\mu}$m, ${\rm 0.2~\mu}$m, ${\rm 0.5~\mu}$m, ${\rm 1~\mu}$m, ${\rm 2~\mu}$m, ${\rm 5~\mu}$m, ${\rm 10~\mu}$m,  ${\rm 20~\mu}$m, ${\rm 50~\mu}$m, and ${\rm 100~\mu}$m (coloured lines). The horizontal dotted line indicates Enceladus's semi-major axis, while the horizontal dot-dashed line is the maximum semi-major axis of the particle to be considered as a ringlet particle. The particles are under the effects of the solar radiation force, plasma drag, and electromagnetic force.}
\label{pd}
\end{figure}

The osculating semi-major axis and eccentricity of a representative particle under the effects of the non-gravitational forces are presented in \cref{pd}. The lines correspond to numerical simulations where the physical radius of the single particle is modified (${\rm 0.1~\mu}$m, ${\rm 0.2~\mu}$m, ${\rm 0.5~\mu}$m, ${\rm 1~\mu}$m, ${\rm 2~\mu}$m, ${\rm 5~\mu}$m, ${\rm 10~\mu}$m, ${\rm 20~\mu}$m, ${\rm 50~\mu}$m, and ${\rm 100~\mu}$m). The solid and dotted horizontal lines indicate the orbits of Pallene and Enceladus, respectively. In this work, we consider a particle to be removed from the ringlet if it collides with a satellite or migrates outside the generous limit of $a_\mathrm{Pal} + 1100$~km (${\rm \sim1.05~D_\mathrm{Pal}}$). The latter can be seen in the figure by the horizontal dot-dashed line. 

Particles with ${\rm s\leq2~\mu}$m migrate beyond the orbit of Enceladus (horizontal dotted line) in less than 100~yr and reach $e>10^{-2}$. In the case shown in \cref{pd}, the particles of $0.1~\mu$m and $1~\mu$m are ejected from the Saturnian system ($e>1$) while the particles of $0.2~\mu$m and $0.5~\mu$m collide with a satellite outside the orbit of Enceladus. The ${\rm 2~\mu}$m-sized particle collides with Enceladus in about 80~yr. 

The effects of the non-gravitational forces are weaker for larger grains and particles with $s>{\rm 5~\mu}$m remain with eccentricities of the order of $10^{-3}$. These particles migrate outwards but still are considered ringlet particles according to our definition. These results roughly demonstrate that the permanence of the particles in the ring is strongly affected by non-gravitational forces and only particles with a radius of tens of micrometres or greater should have significantly long lifetimes in the ringlet (several hundreds of years). In the next sections, we perform full N-body simulations of the ring particles evolution.

\subsection{Particles co-orbital to Pallene} \label{coorbitalmaterial}

In this section, we analyse Pallene's ringlet as formed by a set of 5,000 particles co-orbital to the satellite. We assume particles with the same orbital elements as Pallene, except for the mean anomaly that was randomly selected from a uniform distribution between 0$^\circ$ and 360$^\circ$. The ring composed of co-orbital particles corresponds, e.g., to a scenario where the structure could be formed by the disruption of a proto-Pallene. In this scenario, the ring would also be composed of centimetre-sized or even larger particles. Nevertheless, we do not perform simulations for this size range since the effects of non-gravitational forces can be neglected. The orbital evolution of the centimetre-sized particles would correspond to the analysis in \cref{ssec:pallneigh} which demonstrated that most of the particles initially located inside the Pallene collision region would eventually collide with the satellite, reducing the survival rate of co-orbital particles.

As a general outcome, particles with $s\leq10~\mu$m present a dynamical evolution similar to those shown in \cref{efa}. The particles migrate towards Enceladus and show an increase in eccentricity. However, we obtain a more complex dynamical evolution for particles with $s\geq20~\mu$m caused by capture in resonances with Enceladus. Roughly speaking, a migrating particle is captured at a given resonance with a satellite if the migration timescale is shorter than the libration period of the resonance \citep{Batygin15}. In our case, this condition is achieved for the largest particles ($20~\mu$m, $50~\mu$m, and $100~\mu$m) which are captured, even for a short period of time, in the 7:6, 8:7, 9:8, and 10:9 $e$-type MMRs with Enceladus. 

\begin{figure*}
    \centering
    \includegraphics[height=2.85cm, trim=0 20 9 0, clip]{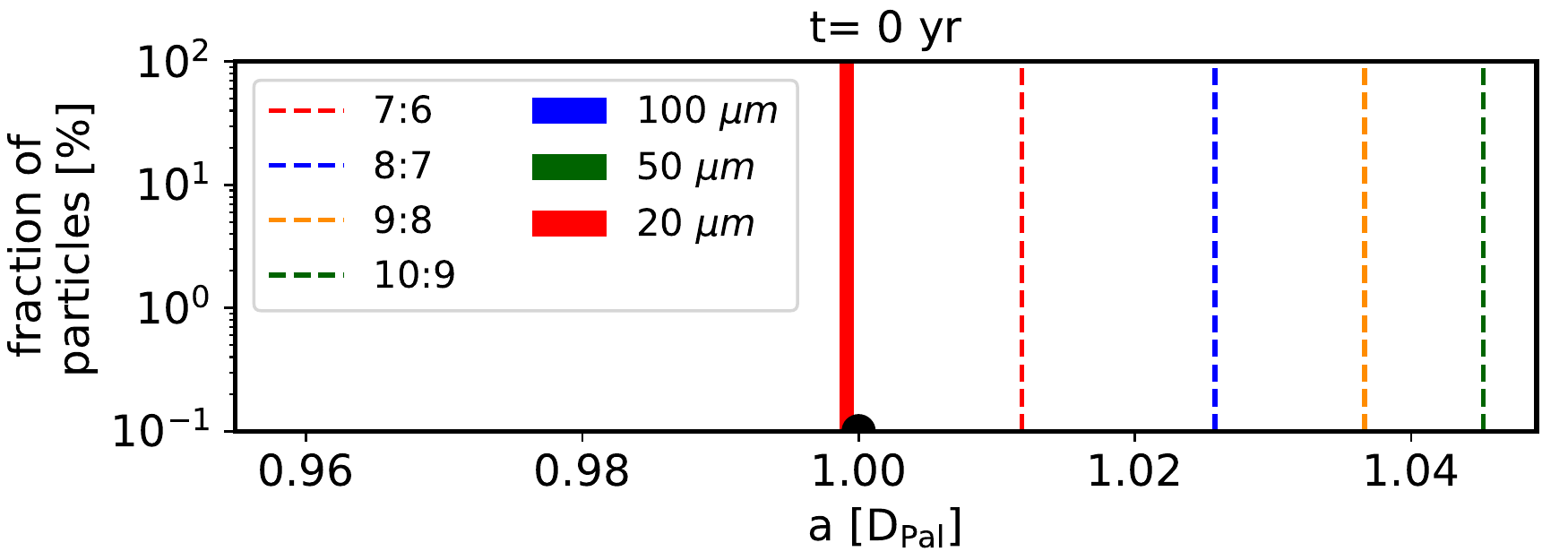} \hfill
    \includegraphics[height=2.85cm, trim=10 20 9 0, clip]{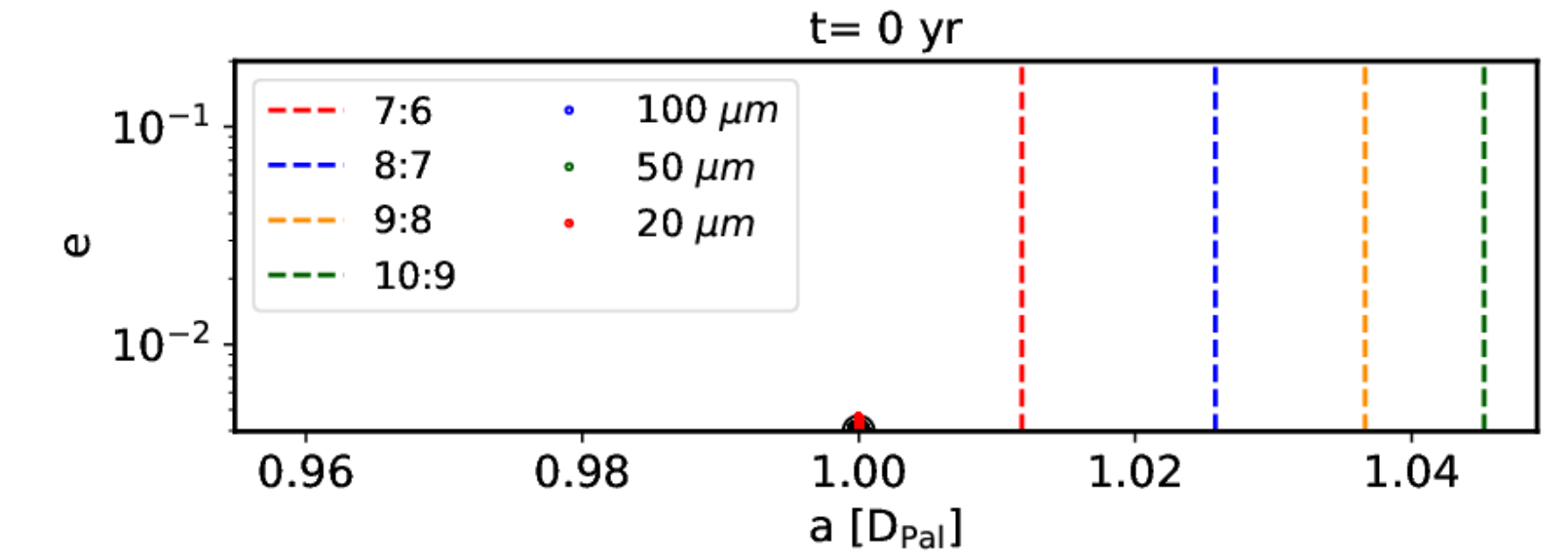} \\[2mm]
    \includegraphics[height=2.85cm, trim=0 20 9 0, clip]{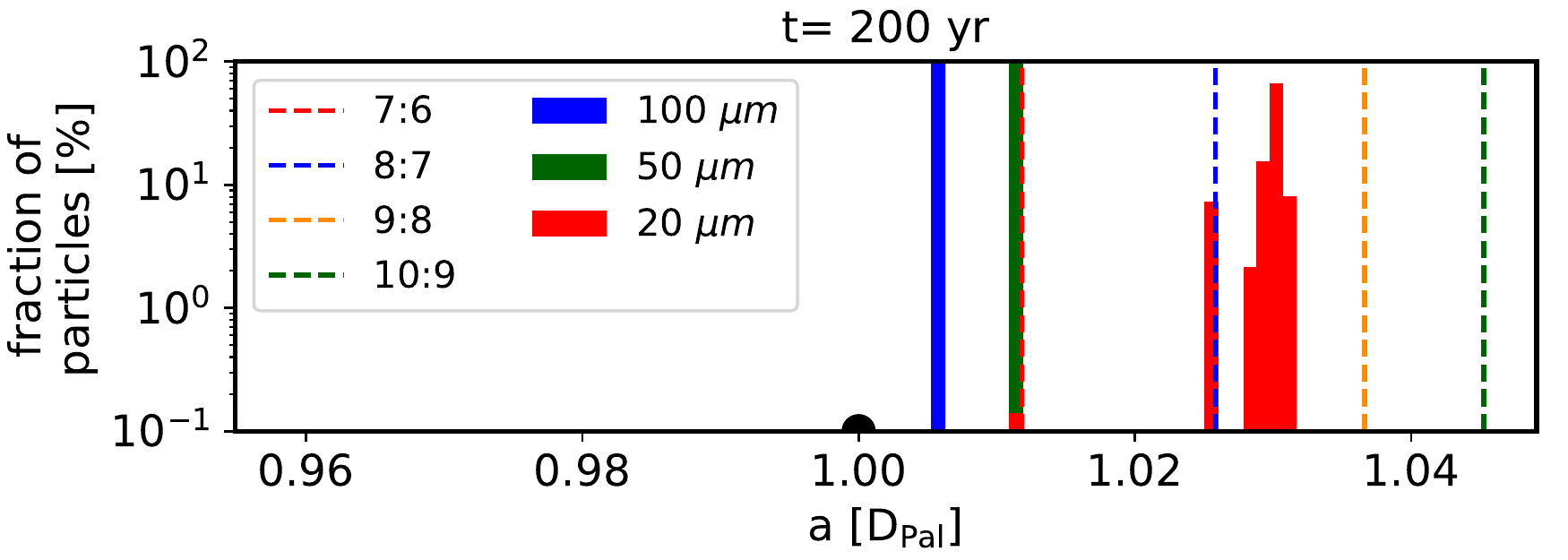} \hfill
    \includegraphics[height=2.85cm, trim=10 20 9 0, clip]{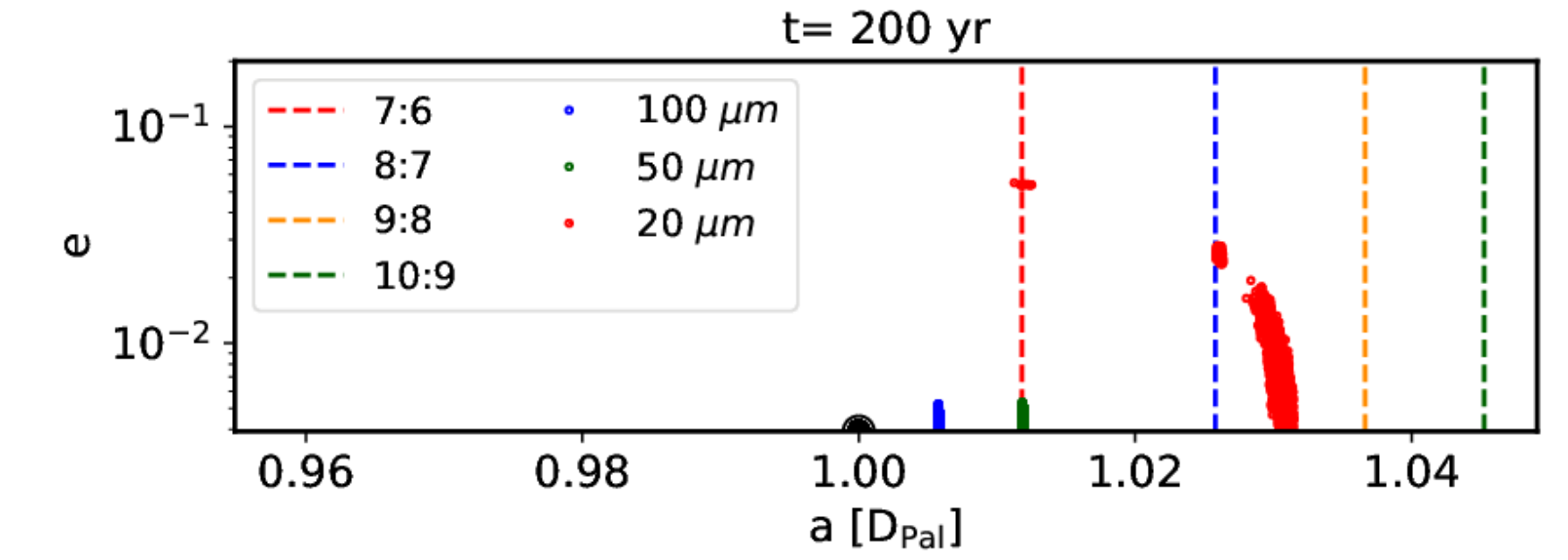} \\[2mm]
    \includegraphics[height=2.85cm, trim=0 20 9 0, clip]{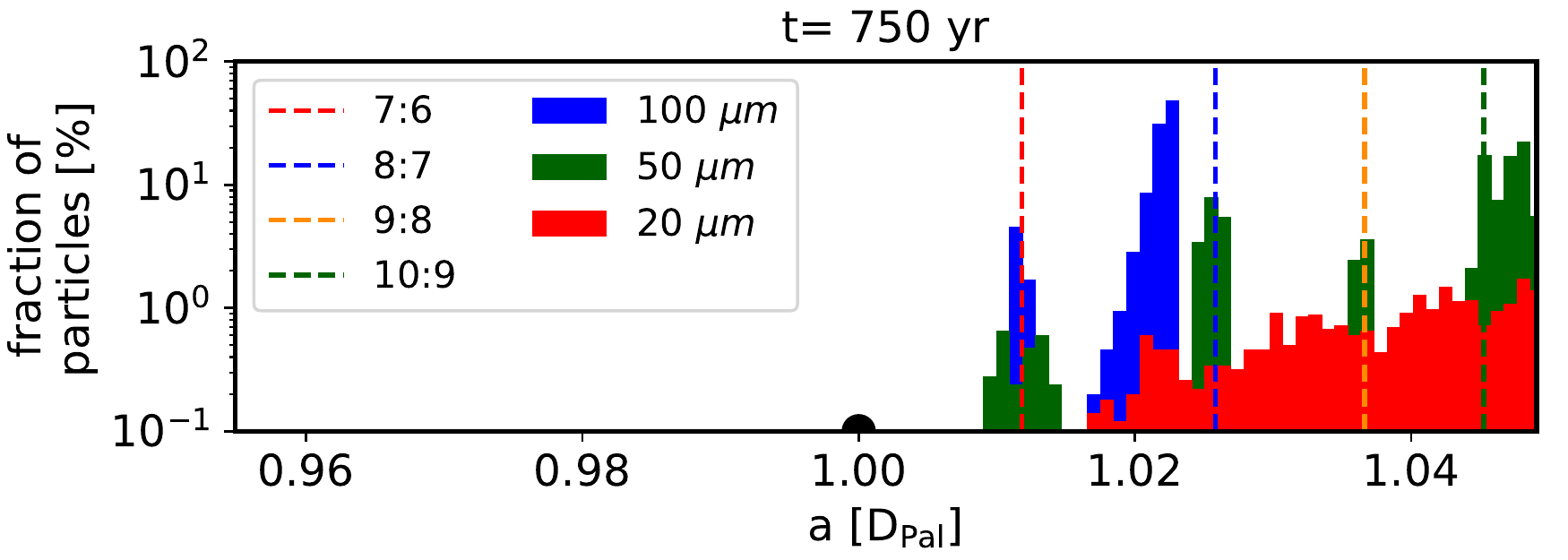} \hfill
    \includegraphics[height=2.85cm, trim=10 20 9 0, clip]{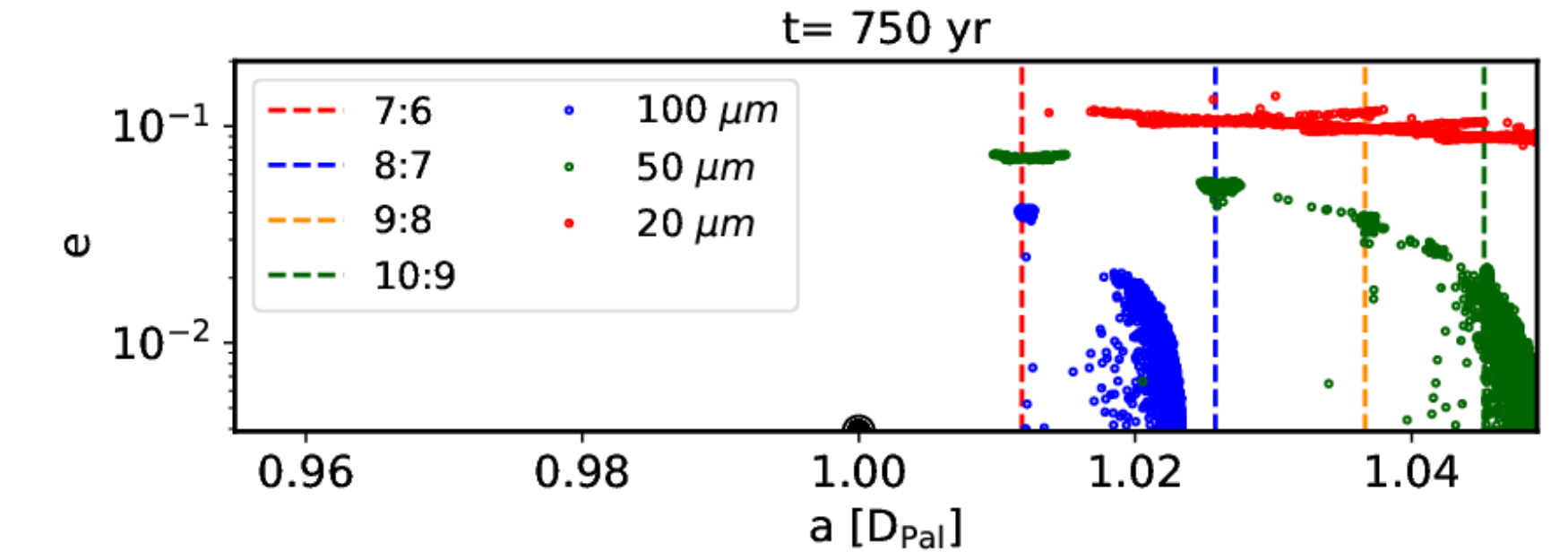} \\[2mm]
    \includegraphics[height=2.85cm, trim=0 20 9 0, clip]{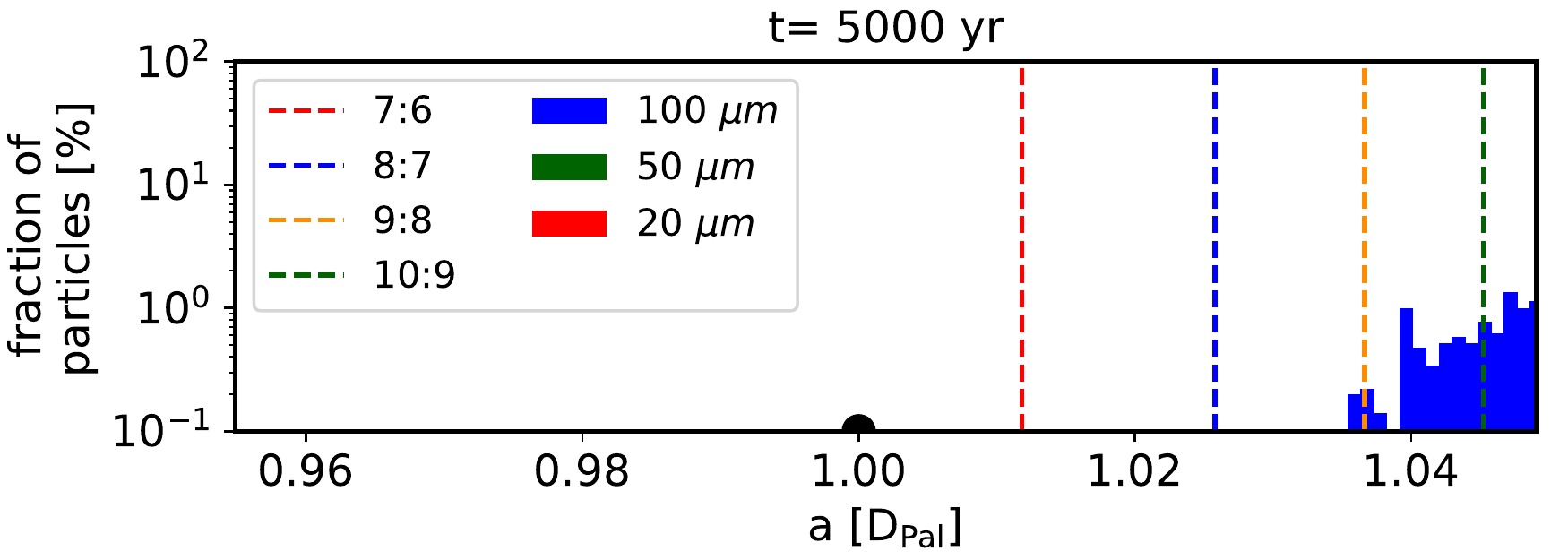} \hfill
    \includegraphics[height=2.85cm, trim=10 20 9 0, clip]{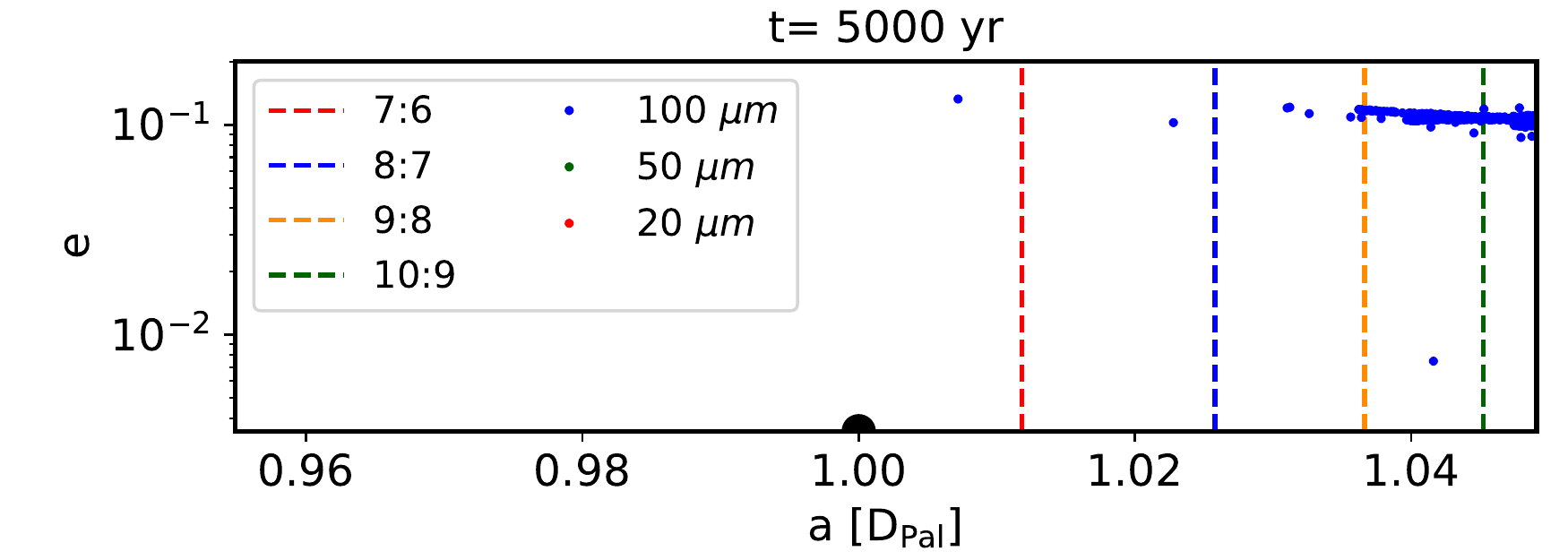} \\[2mm]
    \includegraphics[height=3.2cm, trim=0 0 9 0, clip]{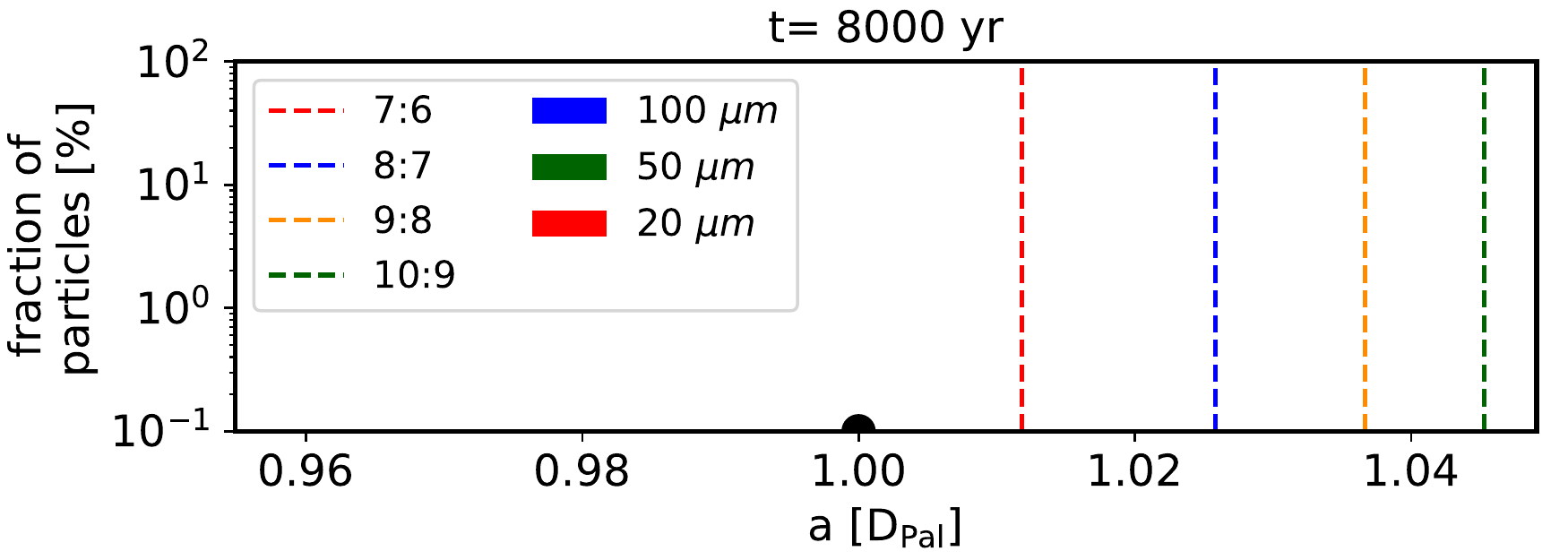} \hfill
    \includegraphics[height=3.2cm, trim=10 0 9 0, clip]{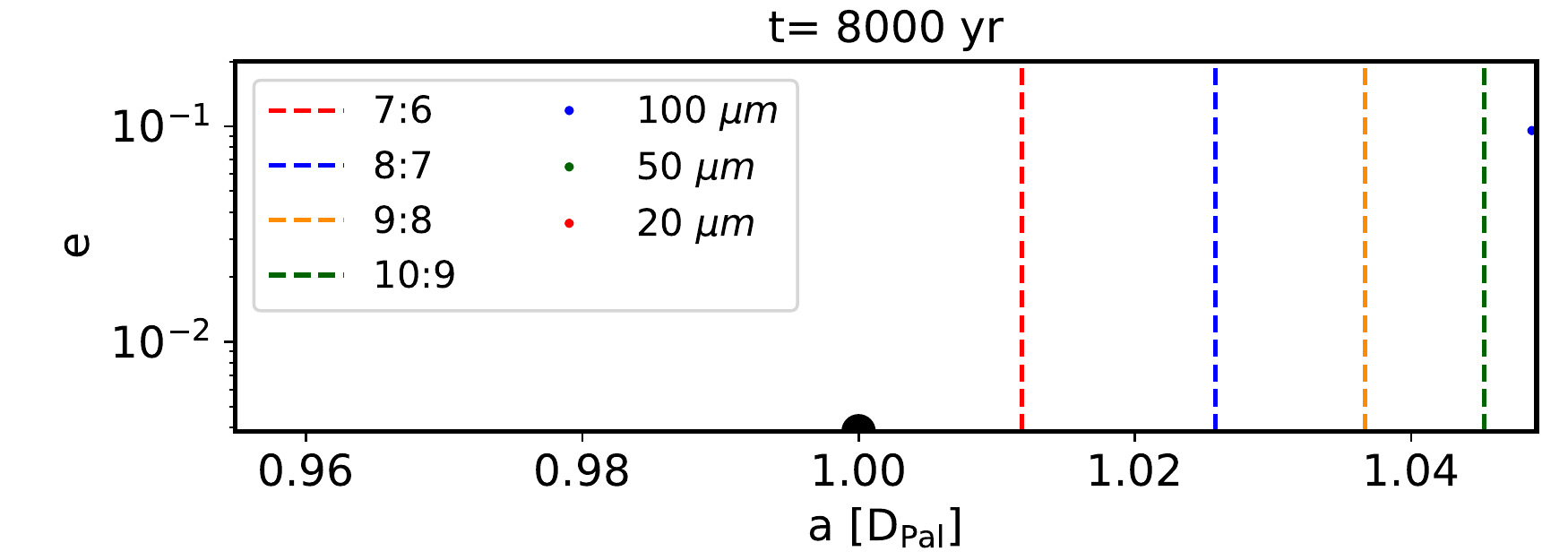} 
    \caption{
    Snapshots showing the percentage of particles as a function of the geometric semi-major axis (at left) and the geometric eccentricity vs. geometric semi-major axis (at right). From top to bottom, we show the data for 0, 200, 750, 5000, and 8000~yr. The ${\rm 20~\mu}$m, ${\rm 50~\mu}$m, and ${\rm 100~\mu}$m sized particles are shown in different colours, as indicated. Pallene is represented by a black filled-circle. The locations of MMRs with Enceladus are indicated by dashed vertical lines. Similarly to \cref{pericentre}, an animation of this figure is provided in the electronic version.}
\label{map_coor}
\end{figure*}
\Cref{map_coor} shows the evolution of the fraction of particles with $s>20~\mu$m (left column), as well as their geometric eccentricity (right column), as a function of the geometric semi-major axis. Initially, all particles have the same semi-major axis and eccentricity as Pallene (black dot). As the particles migrate outward, they cross resonances with Enceladus, increasing their eccentricities. After 200~yr, a fraction of $20~\mu$m-sized particles is trapped in the 7:6 and 8:7 MMRs, while most of the set is located between the 8:7 and 9:8 MMRs. Particles in the 7:6 MMR are confined for a longer period of time, reaching the highest eccentricity values ($\approx$0.05). The ${\rm 20~\mu}$m-sized particles that are not in MMRs at 200~yr had their eccentricity increased during the passage through the two innermost resonances, reaching values $\sim 0.01$. Particles with radius of $50~\mu$m and $100~\mu$m have not yet crossed any resonances and remain with the same initial eccentricity.

At 750~yr, the ${\rm 100~\mu}$m-sized particles have crossed the 7:6 MMR, and the ${\rm 50~\mu}$m-sized particles have crossed all four resonances. Most of the ${\rm 20~\mu }$m-sized particles migrated outside the limit of ${\rm \approx1.05~D_{Pal}}$, leaving only the particles confined in MMRs. A similar result is seen for 5,000~yr, when only ${\rm 100~\mu}$m-sized particles in MMRs remain in the ring, indicating that capture in resonances increases their longevity. Therefore, the vicinity of MMRs would correspond to brighter regions of the ring, as will be shown later. Finally, after 8000~yr, the ring is completely depleted of $\mu$m-sized particles.

\begin{figure}
\centering
\subfloat[\label{r_c-a}]{\includegraphics[width=\columnwidth]{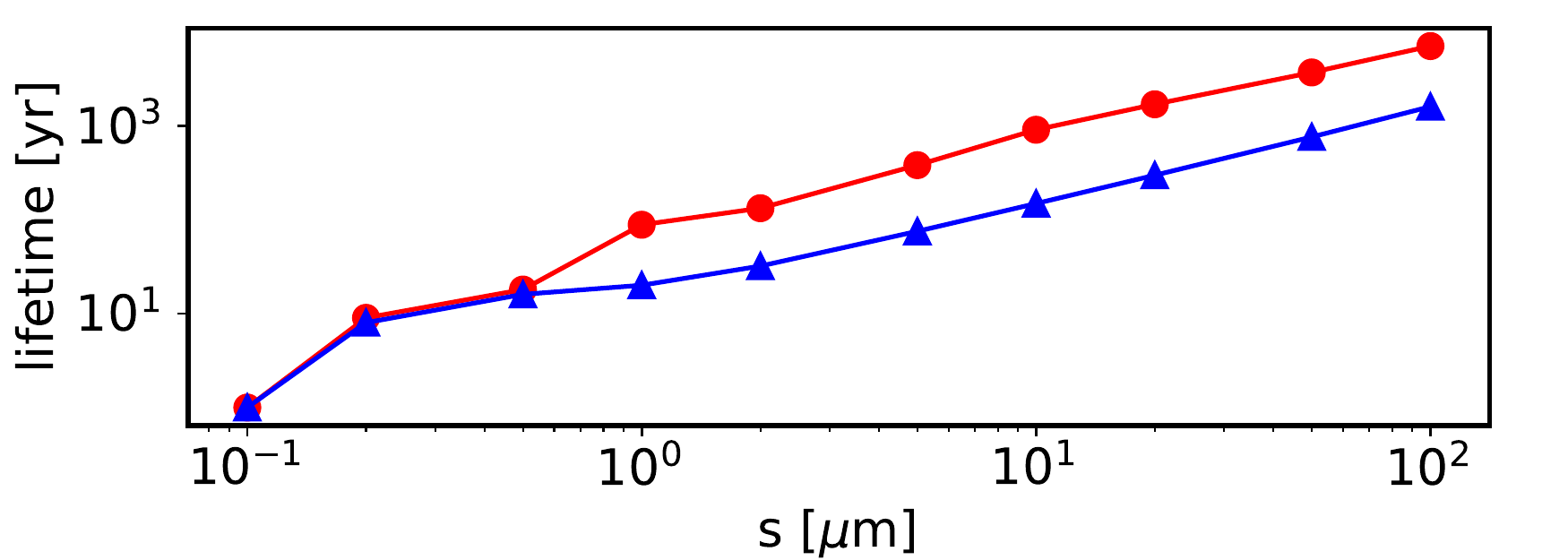}}
\quad
\subfloat[\label{r_c-b}]{\includegraphics[width=\columnwidth]{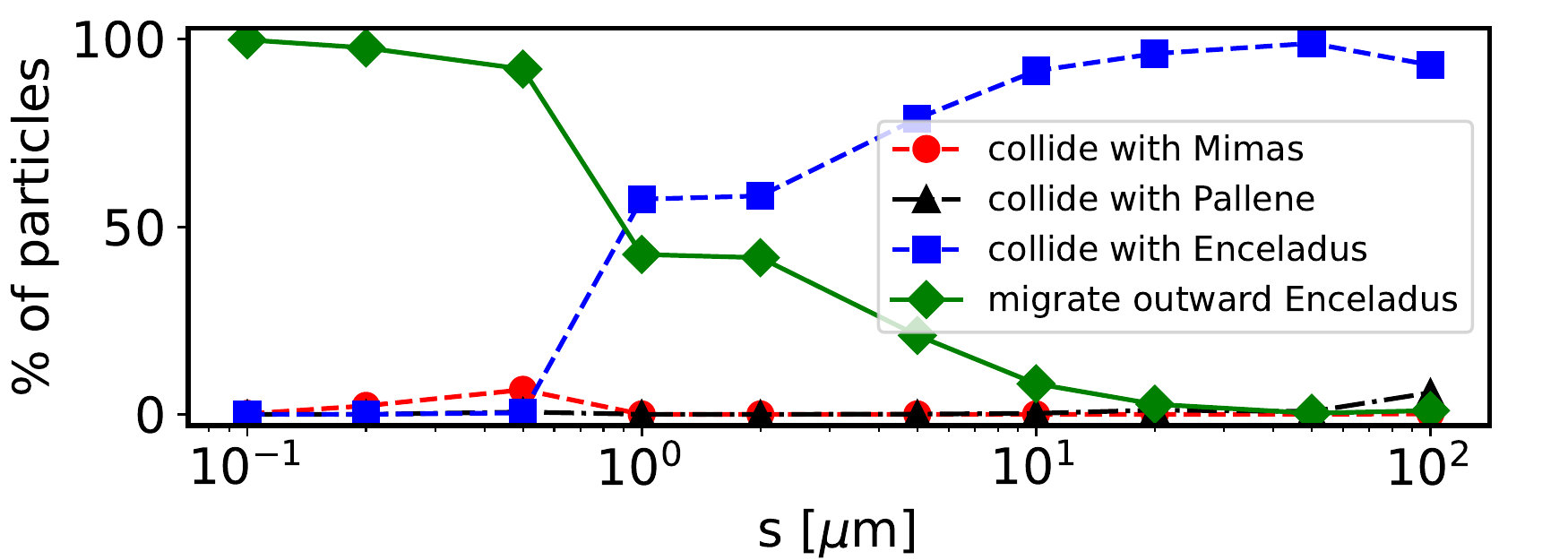}}
\quad
\subfloat[\label{r_c-c}]{\includegraphics[width=\columnwidth]{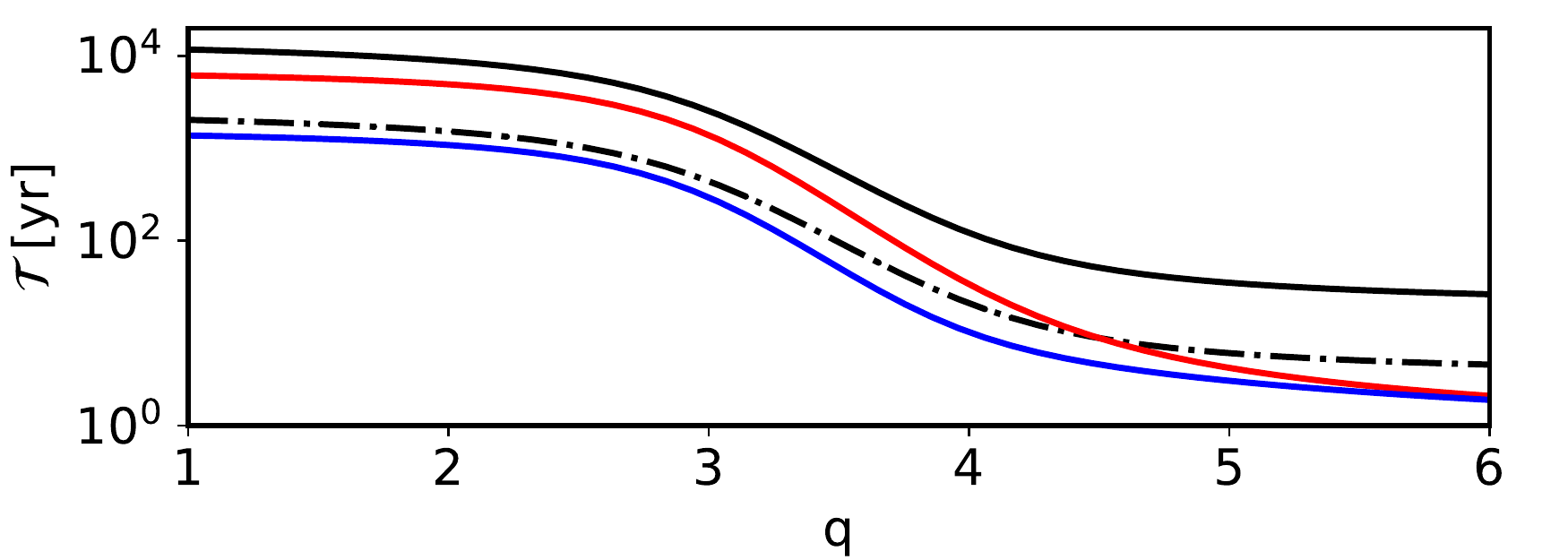}}
\caption{a) The half-life (in blue) and the lifetime (in red) of the ring as a function of the physical radius of the co-orbital particles. b) The fraction of the particles that collides with the satellites Mimas (in red), Pallene (in black), and Enceladus (in blue), and the fraction of particles that migrates out of the orbit of Enceladus (in green). c) The time $\mathcal{T}$ for the satellite to produce the mass of the ring, assuming a non-porous (black solid line) and a porous (black dot-dashed line) Pallene. The red and blue lines give the ring's lifetime and half-life, respectively, as a function of the slope $q$.}
\label{rate_coor}
\end{figure}

\Cref{r_c-a} shows two different timescales as a function of particle radius: in blue, the time required for 50\% of particles to collide with a satellite or migrate outside the limit of ${\rm \sim1.05~D_{Pal}}$ -- hereafter referred to as the ring's half-lifetime -- and in red the time required for all particles to be lost -- referred as the ring's lifetime. The ring is completely depleted of sub-micrometric particles in less than a decade, while particles of radius of $1-10~\mu$m have lifetimes of the order of ${\rm 10^2}$~yr. Particles that last longer are those with $s\geq20~\mu$m, with lifetimes of ${\rm \sim 10^3}$~yr -- same order of the time $\mathcal{T}$ for Pallene to produce the mass of the ring (see \cref{figmplus}).

Particle sinks are shown in \cref{r_c-b}. Due to the intense migration caused by the plasma drag, almost all the sub-micrometric particles migrate beyond the orbit of Enceladus and collide with an external satellite or are ejected from the system. By increasing the radius of the particles, the slower rate of migration increases the period that the particles interact gravitationally with Enceladus in the vicinity of the satellite. Consequently, the number of collisions with Enceladus increases, as seen in \cref{r_c-b}. Also due to migration, the number of particles that collide with Pallene is less than 5\% for all sizes; this rules out Pallene as an efficient secondary source of material, produced by subsequent impact with these particles. 

\Cref{r_c-c} shows in black lines the same curves shown in \cref{figmplus}: the solid line is the time for Pallene to produce the ring mass in the non-porous case, while the dot-dashed line is the same for the porous case. The red and blue lines indicate the ring's lifetime and half-lifetime, respectively, obtained by a time-weighted average:
\begin{equation}
\bar{T}=\frac{\sum_s m_s\left(\frac{s~{\rm (\mu m)}}{\rm 100~\mu m}\right)^{-q}T_s}{\sum_s m_s\left(\frac{s~{\rm (\mu m)}}{\rm 100~\mu m}\right)^{-q}} \end{equation}
where $m_s$ is the mass of a particle with radius $s$ and $T_s$ is the (half)-lifetime of the particles.

Focusing on the red curve in \cref{r_c-c}, we verify that the ring would not be in a steady-state, assuming ejection by Pallene as the only source of material. However, given the uncertainties in the yield calculation and the proximity of the values between the black and red solid curves, towards the lower values of $q$, we can conclude that Pallene might be able to maintain its ring if the particle distribution is given by ${\rm q\lesssim 3}$. Lower slope values mean that the ring has higher concentrations of larger particles, which seems to be the case of the ringlet of Pallene -- given that larger particles can be captured in MMRs with Enceladus, while smaller ones have lifetimes of only a few years. If the particle distribution in the ring is given by slopes ${\rm q\gtrsim4}$, Pallene by itself certainly cannot maintain the ring, since the lifetime is lower than $\mathcal{T}$ even for the porous limit.

\begin{figure*}
\centering
\includegraphics[height=2.8cm,trim=0 0 74 0,clip=True]{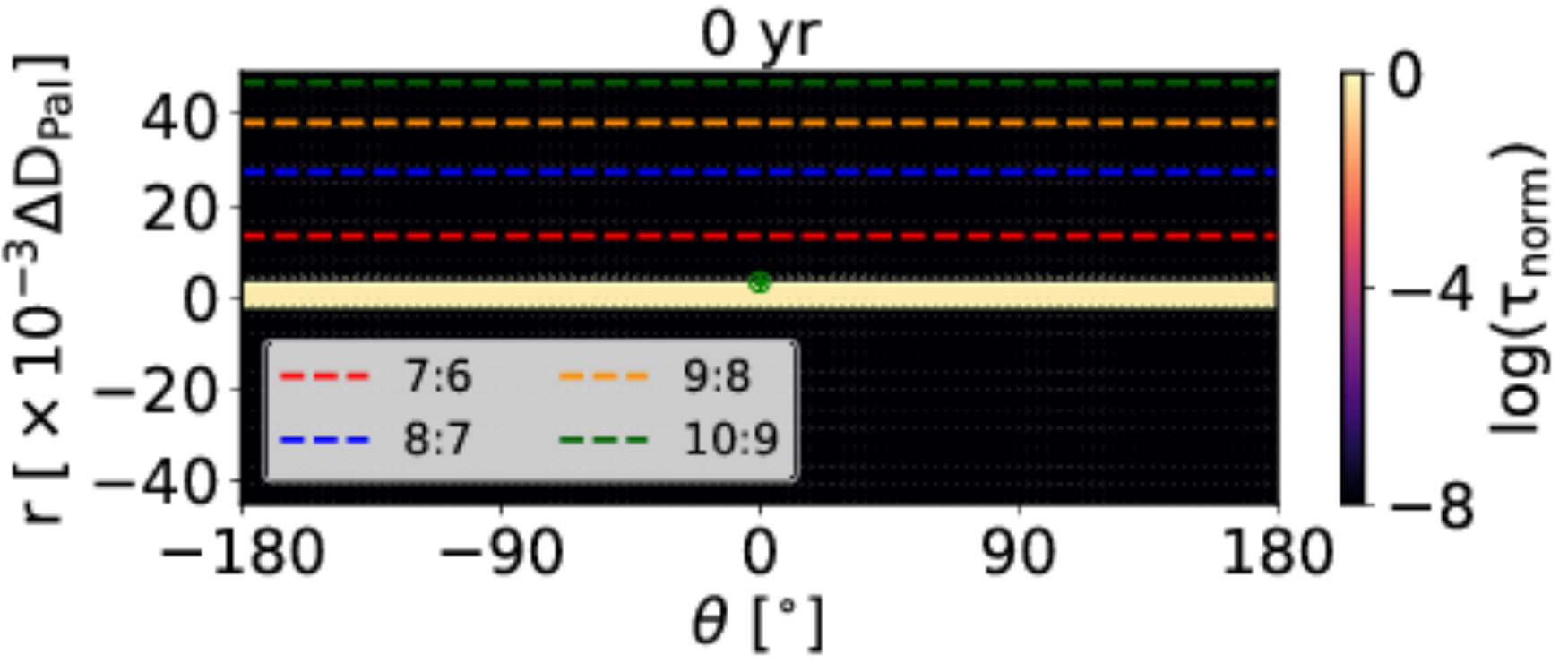} \hfill
\includegraphics[height=2.8cm,trim=25 0 74 0,clip=True]{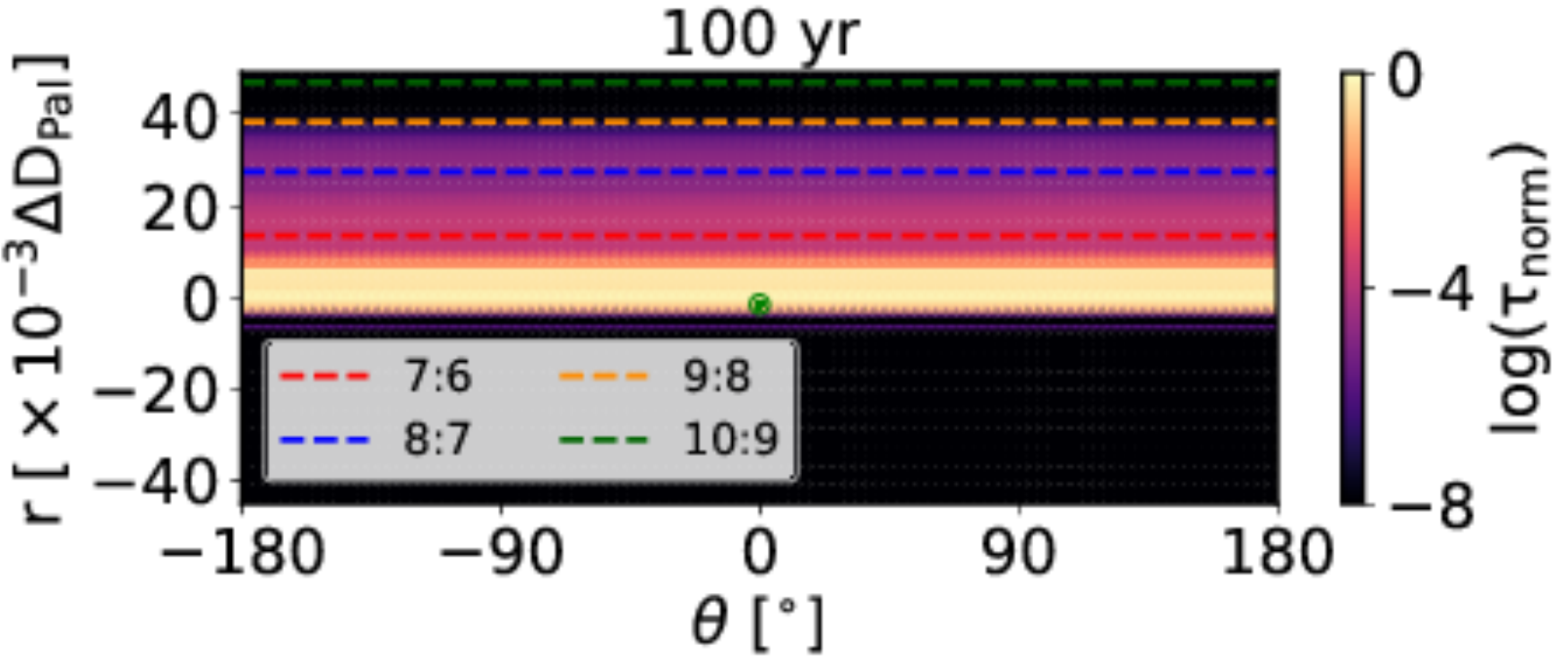} \hfill
\includegraphics[height=2.8cm,trim=25 0 0 0,clip=True]{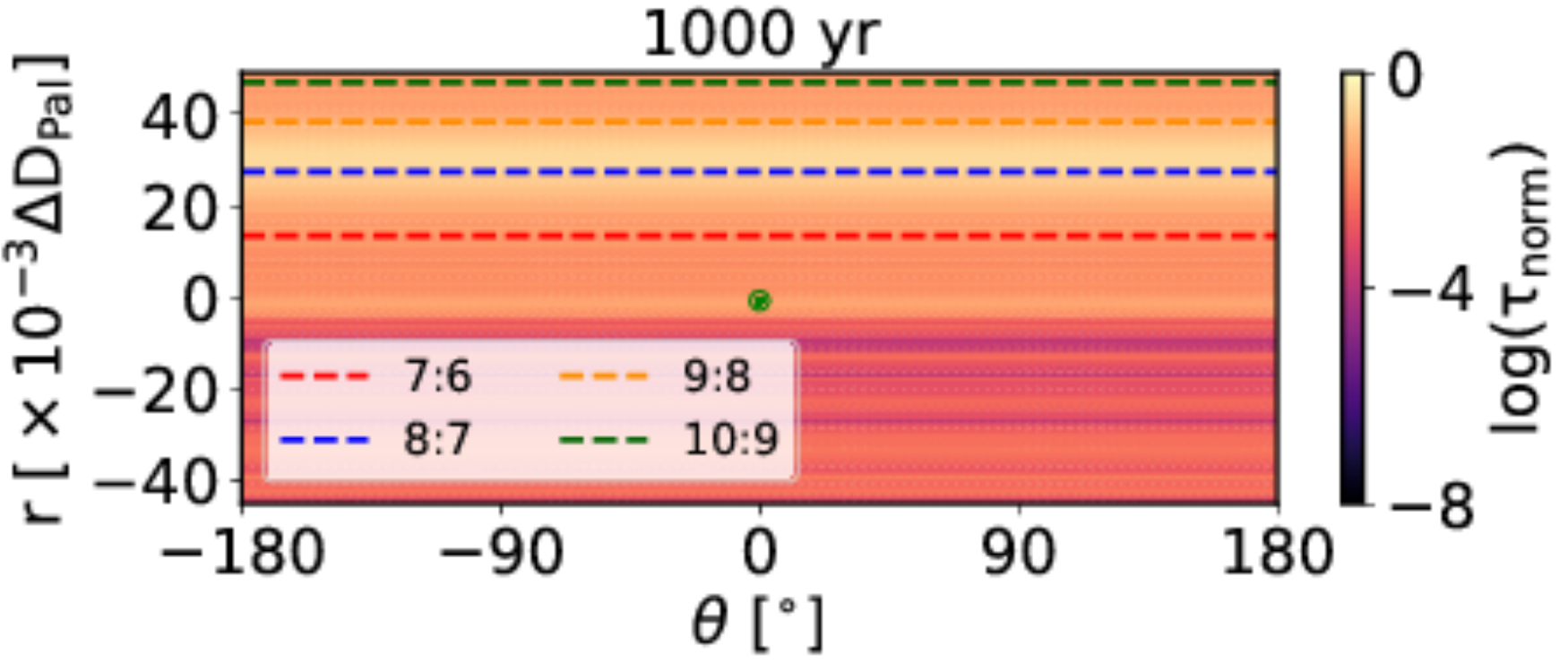} \\[2.5pt]
\includegraphics[height=2.8cm,trim=0 0 74 0,clip=True]{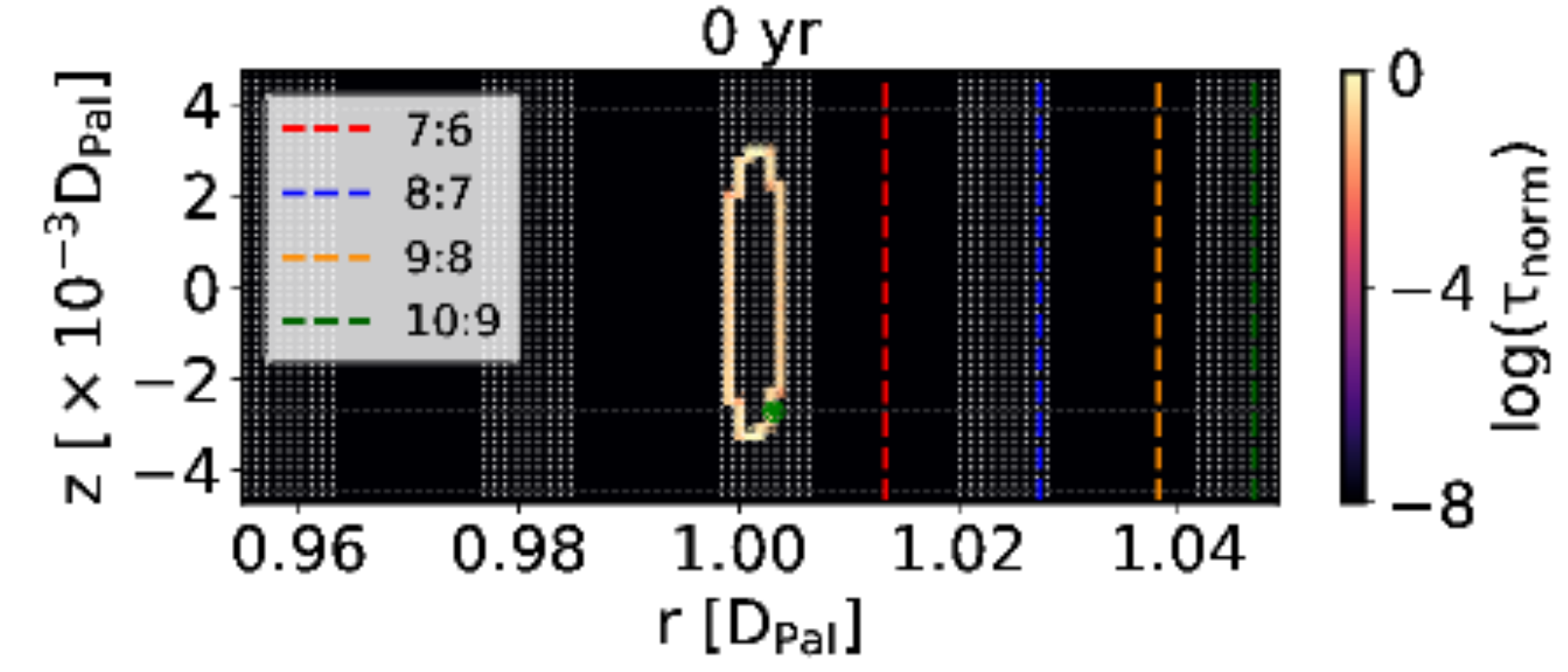} \hfill
\includegraphics[height=2.8cm,trim=36 0 74 0,clip=True]{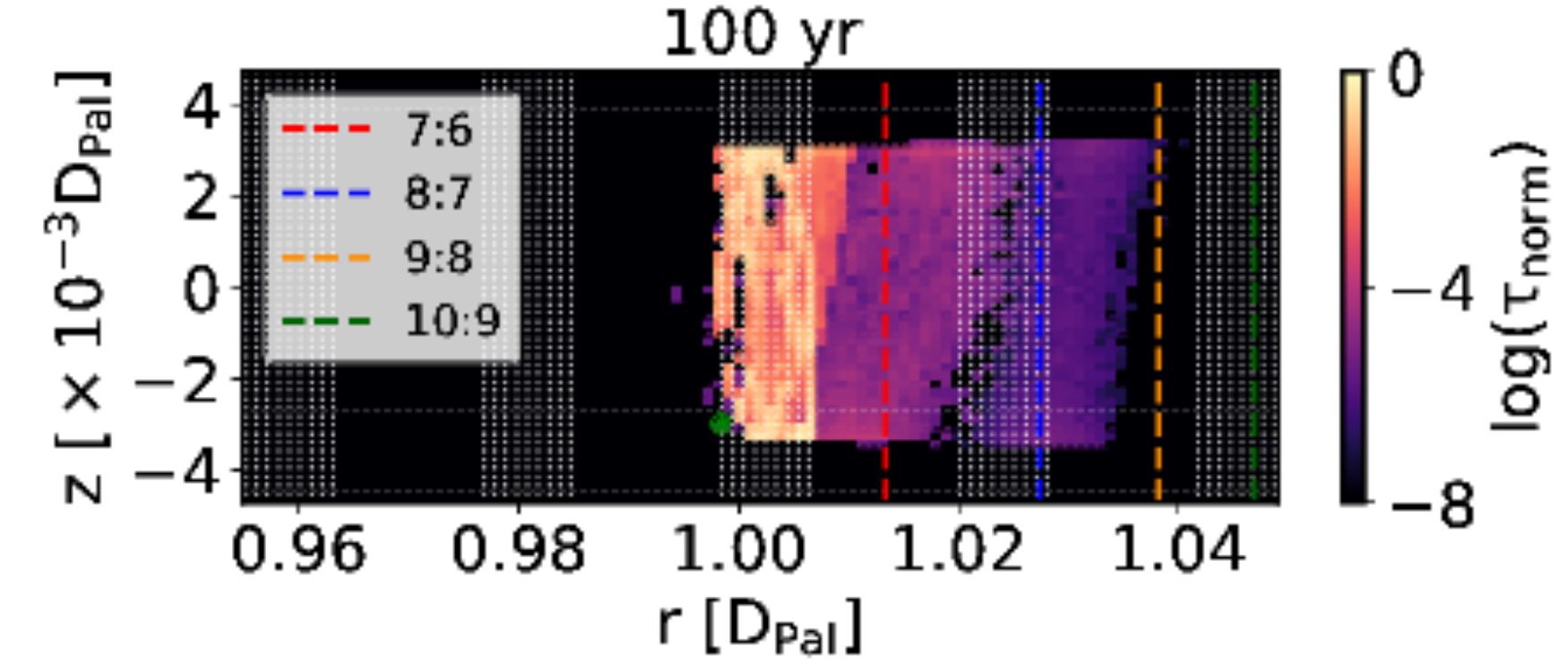} \hfill
\includegraphics[height=2.8cm,trim=36 0 0 0,clip=True]{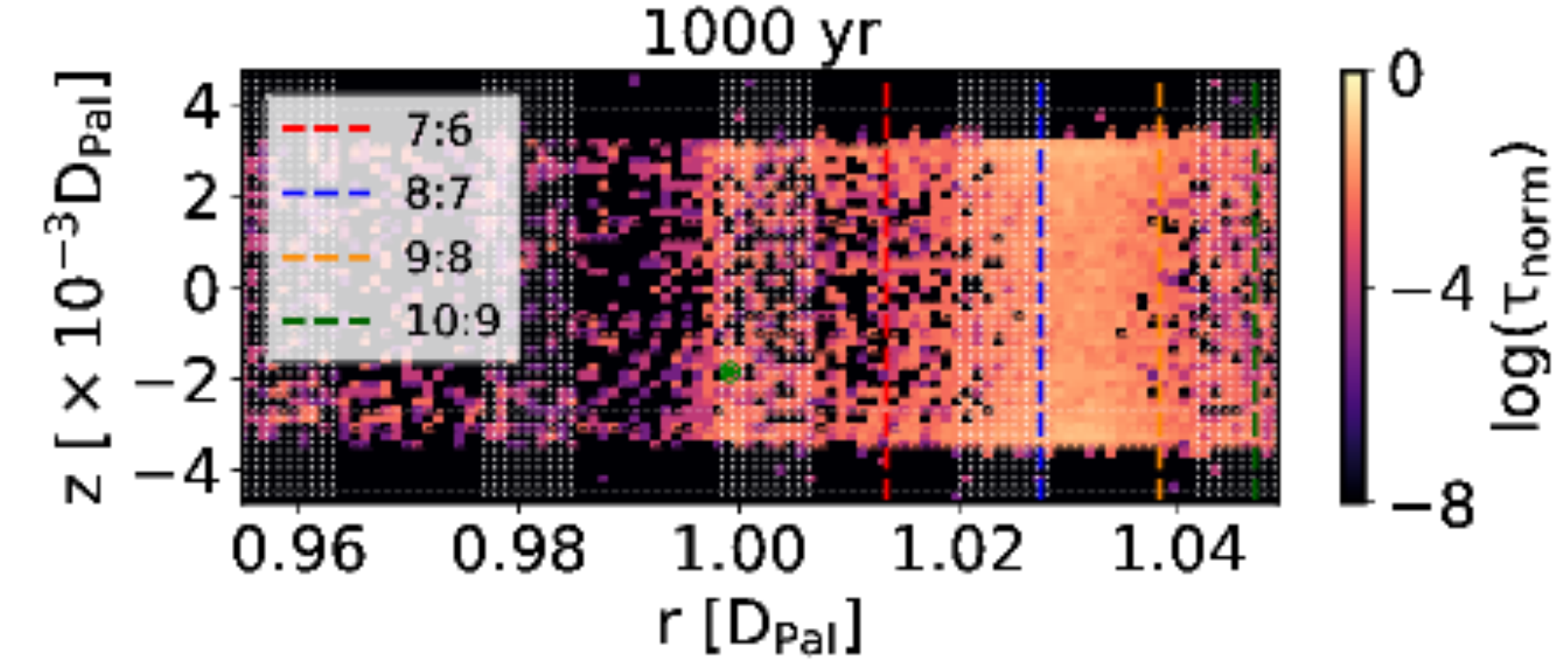} \\[2.5pt]
\caption{Animations showing the normalised optical depth ${\rm \tau_{norm}}$ in the $\theta$-$r$ (top panels) and $r$-$z$ (bottom panels) planes in the rotating frame for co-orbital particles. The green dot gives Pallene's position and the dashed lines indicate the MMRs with Enceladus. The upper limit of the radius in the panels corresponds to the limit ${\rm 1.05~D_{Pal}}$. Adobe Reader version $\geq$9 or similar is required.}
\label{profilerz_coor}
\end{figure*}

\Cref{profilerz_coor} shows animations of the co-orbital particle profiles in the planes $\theta$-$r$ (top panels) and $r$-$z$ (bottom panels). The colour of each pixel gives the normalised optical depth of that pixel, assuming a particle distribution with slope $q=2.5$. The particles are initially distributed along the orbit of Pallene. In 10~yr, we can identify ring-like structures in the $r$-$z$ plane, produced by the precession of the longitude of pericentre (\cref{pericentre}), where each structure is composed of particles with different radii. After 100~yr, the ring shows an asymmetrical profile, with the brightest part close to Pallene's orbit, and structures with lower brightness outside the satellite's orbit. We do not see any bright regions inside the orbit of Pallene, since outward migration is dominant for all particles.

At 400~yr, the torus structure is completely formed, and the ring has an asymmetric structure. The brightest part of the ring is in the region of the 7:6 MMR with Enceladus, but we see dimmer structures inside and outside this location, as an effect of the increased eccentricity of resonant particles. After 1000~yr, the complete structure of the ring has moved outward and the brightest region is located in the 8:7 MMR. After 4000~yr, the structure has moved further away and only a few particles have remained in the ring region. 

\subsection{Particles Ejected from Pallene} \label{ejectedmaterial}

In the numerical simulations presented in this section, 5000~particles were randomly and uniformly distributed in a spherical shell within the Hill radius of Pallene. Particles are ejected radially with random velocities that follow the normalised distribution \citep{Hartmann85,Krivov03,Sun17}:
\begin{equation}
f_v=\frac{1}{v_0}\left(\frac{v}{v_0}\right)^{-2}\Theta[v-v_0],
\end{equation}
where $\Theta(x)$ denotes the Heaviside function. The minimum ejecta speed, $v_0$, is obtained from the transcendental equation \citep{Kruger00}
\begin{equation}
\frac{K_e}{K_i}=Y\left(\frac{v_0}{v_{\rm imp}}\right)^2\left[\left(\frac{v_0}{v_{\rm max}}\right)^{-1}-1\right],
\end{equation}
where $v_{\rm max}$ is the maximum ejecta speed and $K_e/K_i$ is the ratio between the kinetic energy partitioned to the ejecta and the impactor’s kinetic energy, assumed as $K_e/K_i=0.1$ \citep{Sun17}.

\begin{figure*}
\centering
\includegraphics[height=2.8cm,trim=0 0 74 0,clip=True]{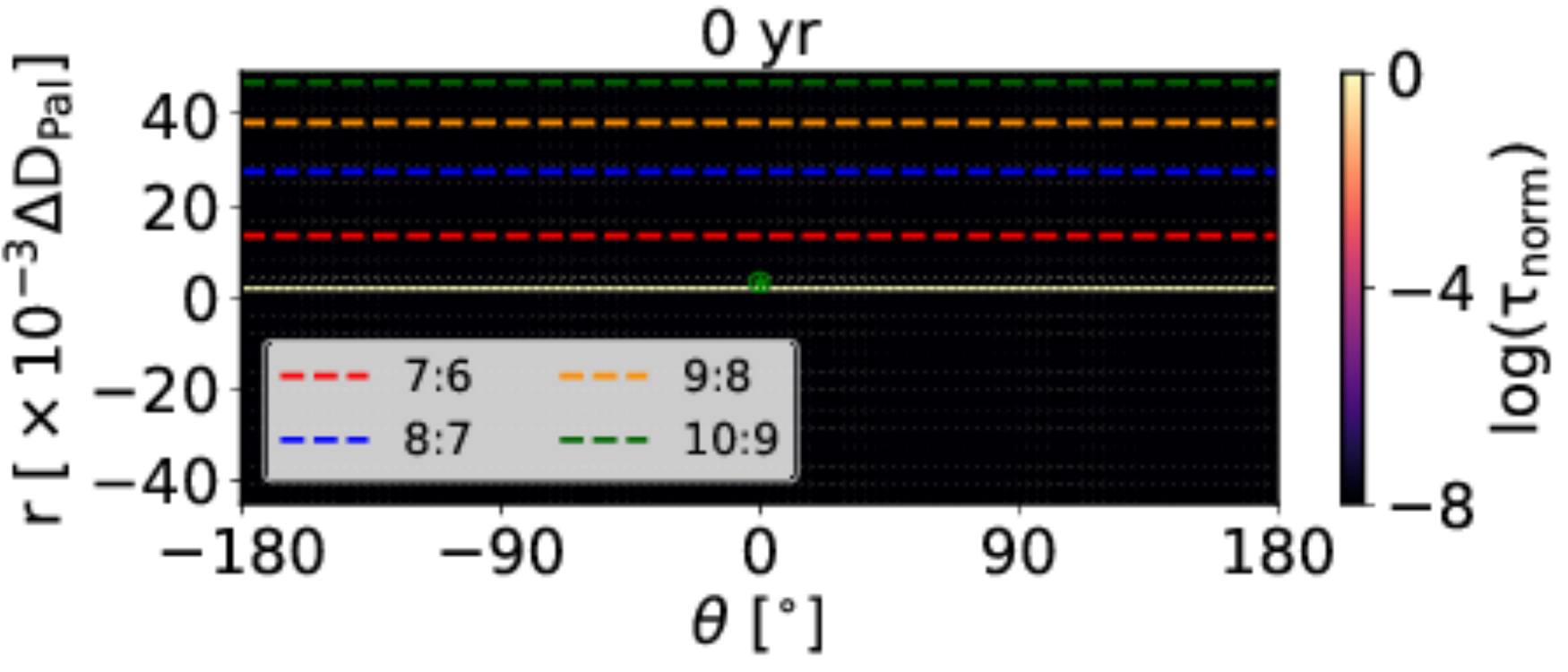} \hfill
\includegraphics[height=2.8cm,trim=25 0 74 0,clip=True]{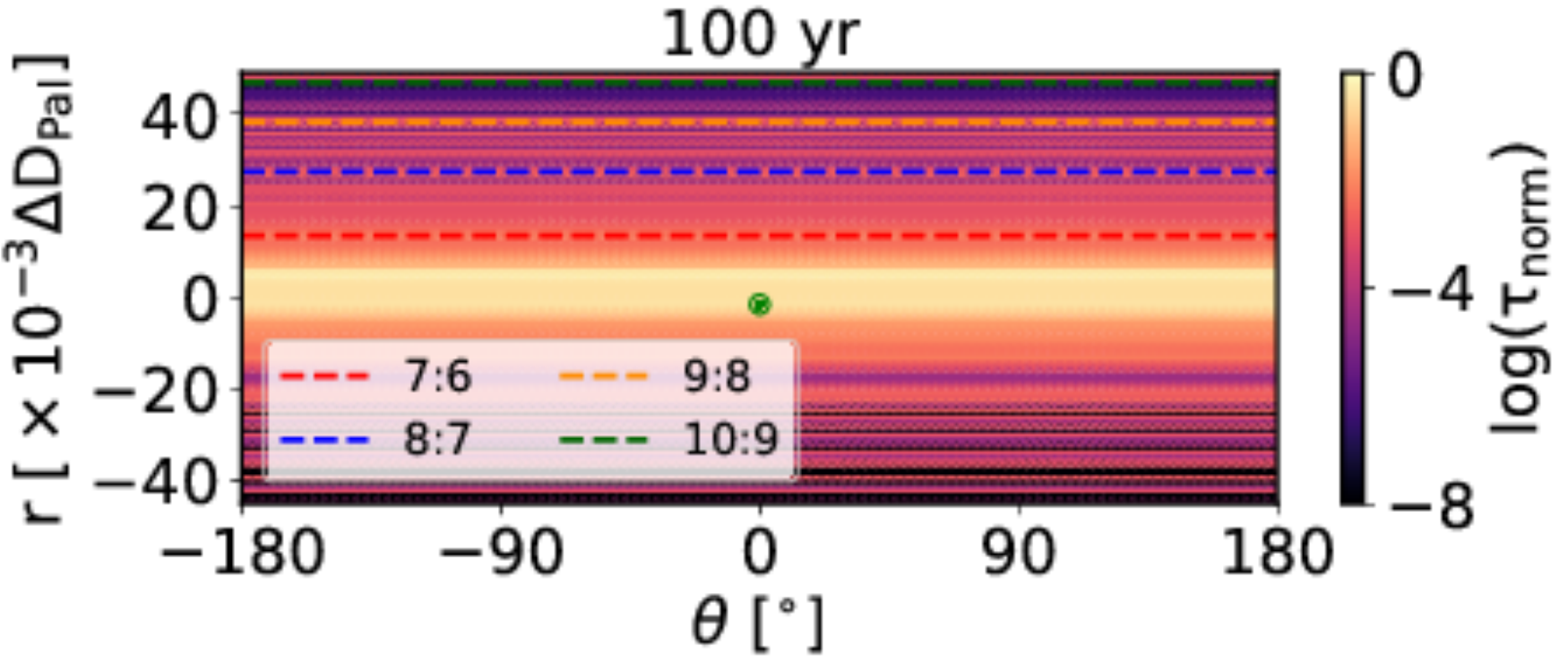} \hfill
\includegraphics[height=2.8cm,trim=25 0 0 0,clip=True]{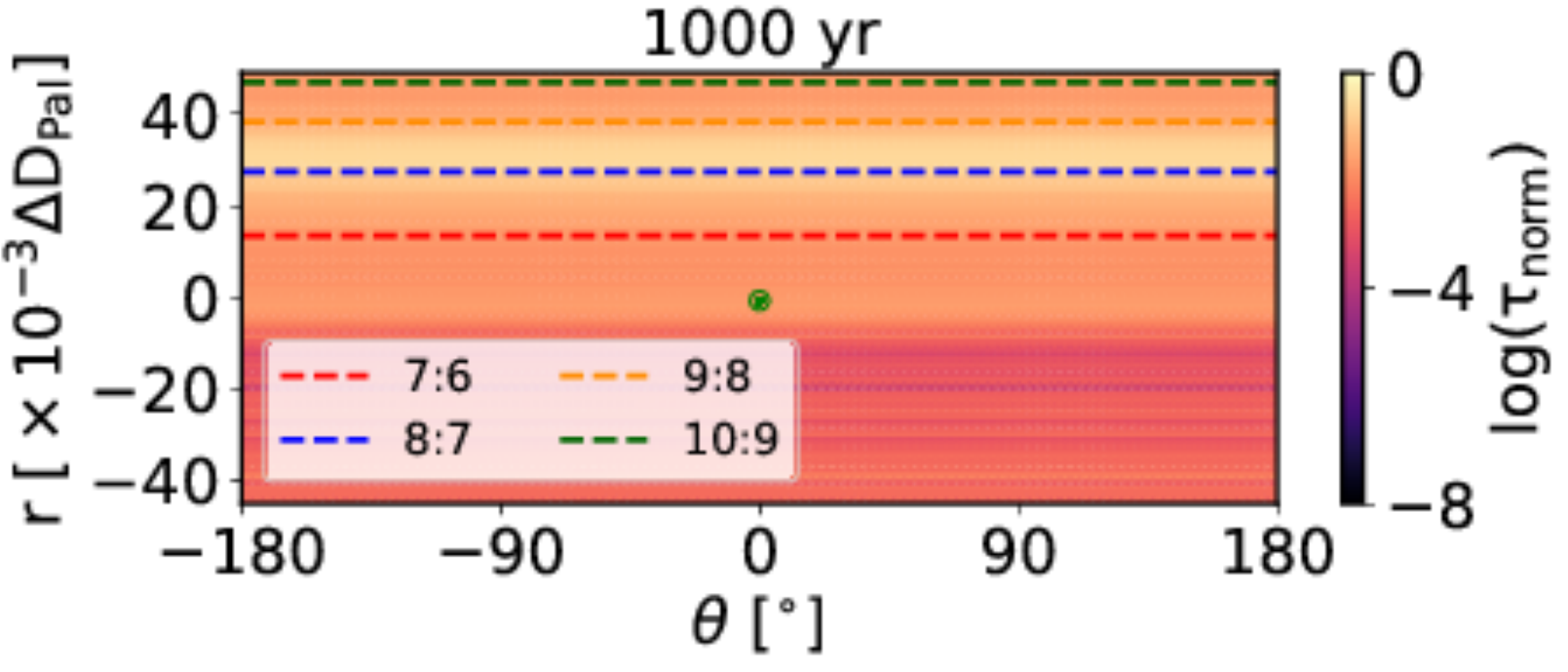} \\[2.5pt]
\includegraphics[height=2.8cm,trim=0 0 74 0,clip=True]{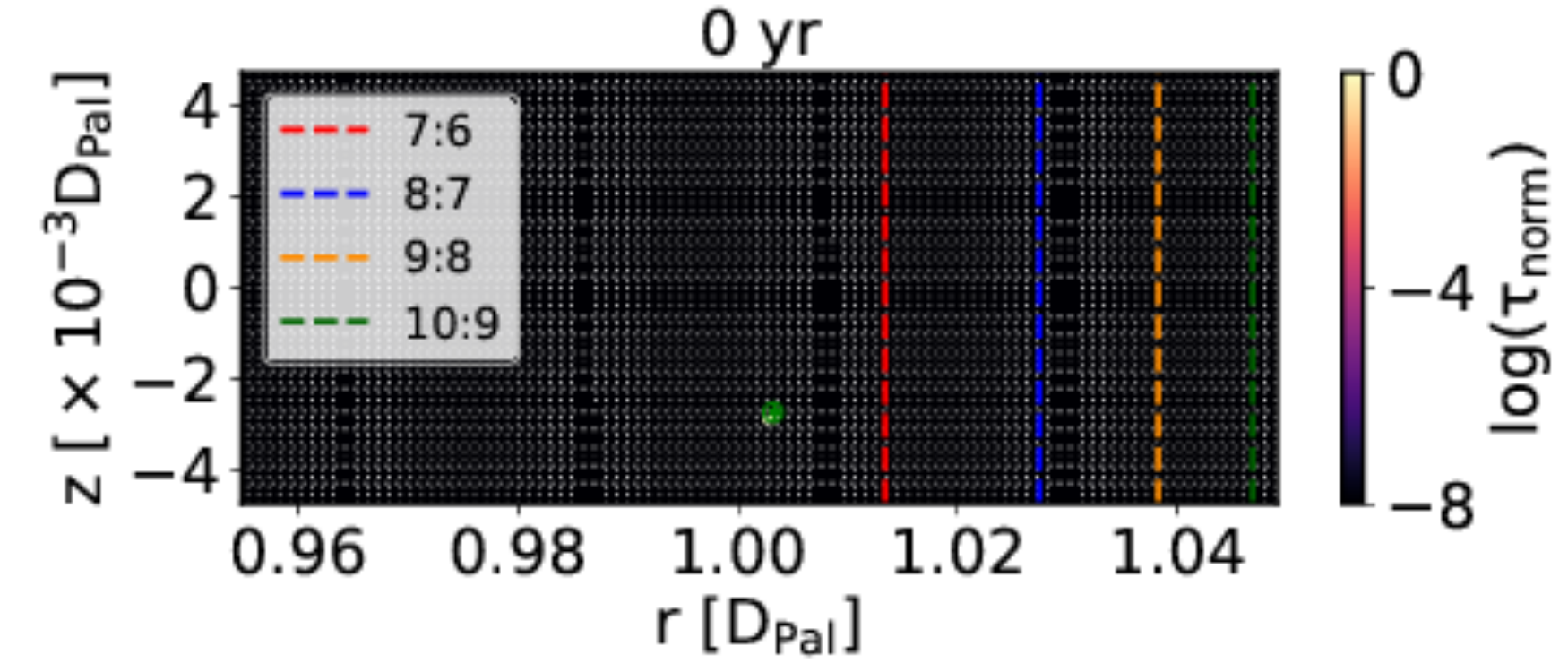} \hfill
\includegraphics[height=2.8cm,trim=36 0 74 0,clip=True]{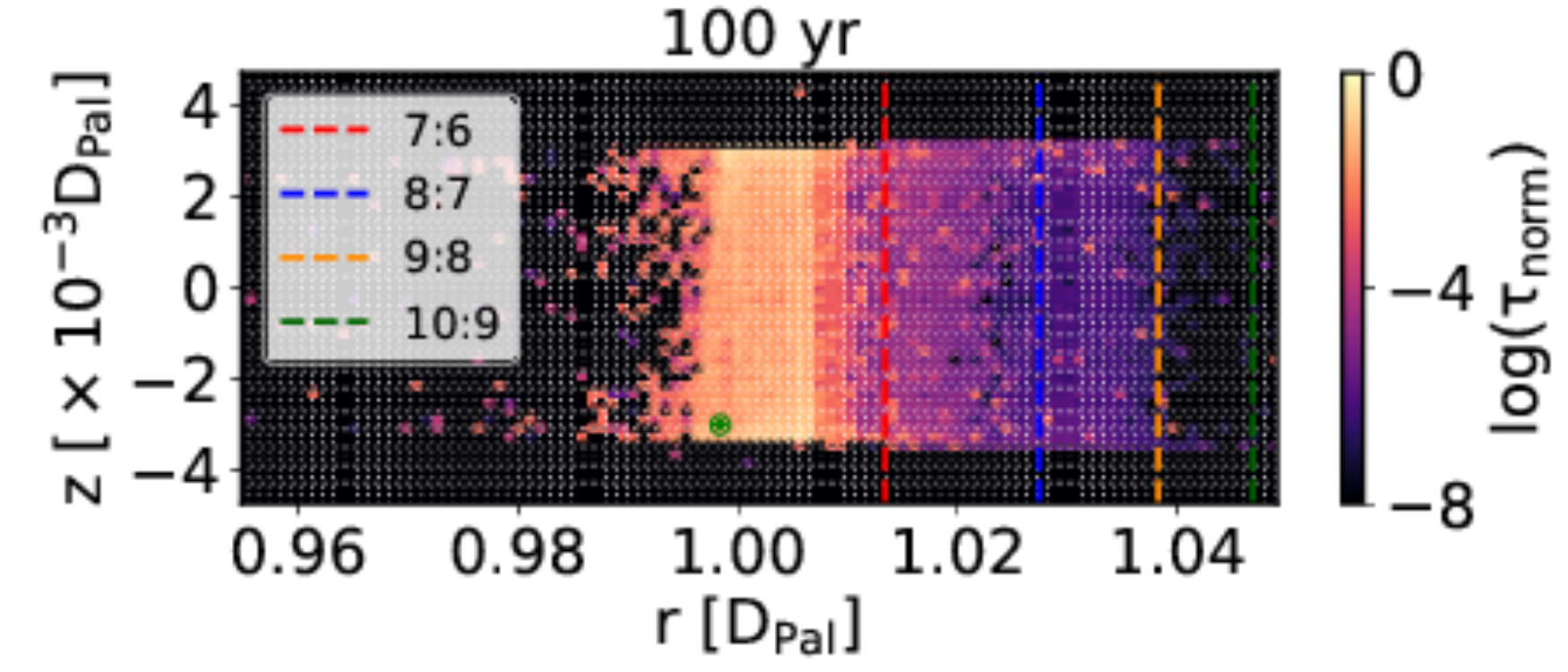} \hfill
\includegraphics[height=2.8cm,trim=36 0 0 0,clip=True]{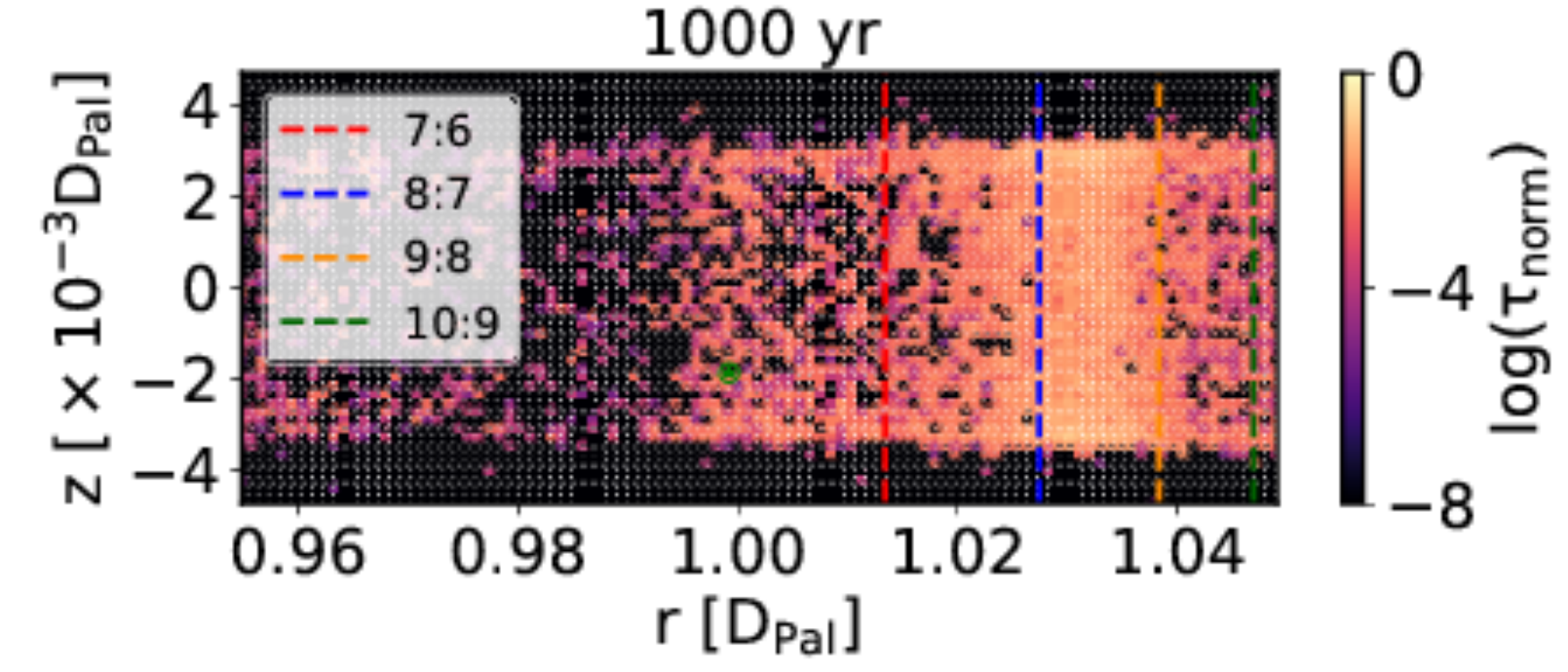} \\[2.5pt]
\caption{
    Normalised optical depth ${\rm \tau_{norm}}$ for the ejected particles. Similarly to \cref{profilerz_coor}, we present a cut in the $\theta$-$r$ and $r$-$z$ planes in the rotating frame. The green dot gives Pallene's position and the vertical dashed lines are MMRs with Enceladus. Adobe Reader version $\geq$9 or similar is required.
} 
\label{profilerz_ejec}
\end{figure*}

\begin{figure}
\centering
\subfloat[\label{r_e-a}]{\includegraphics[width=\columnwidth]{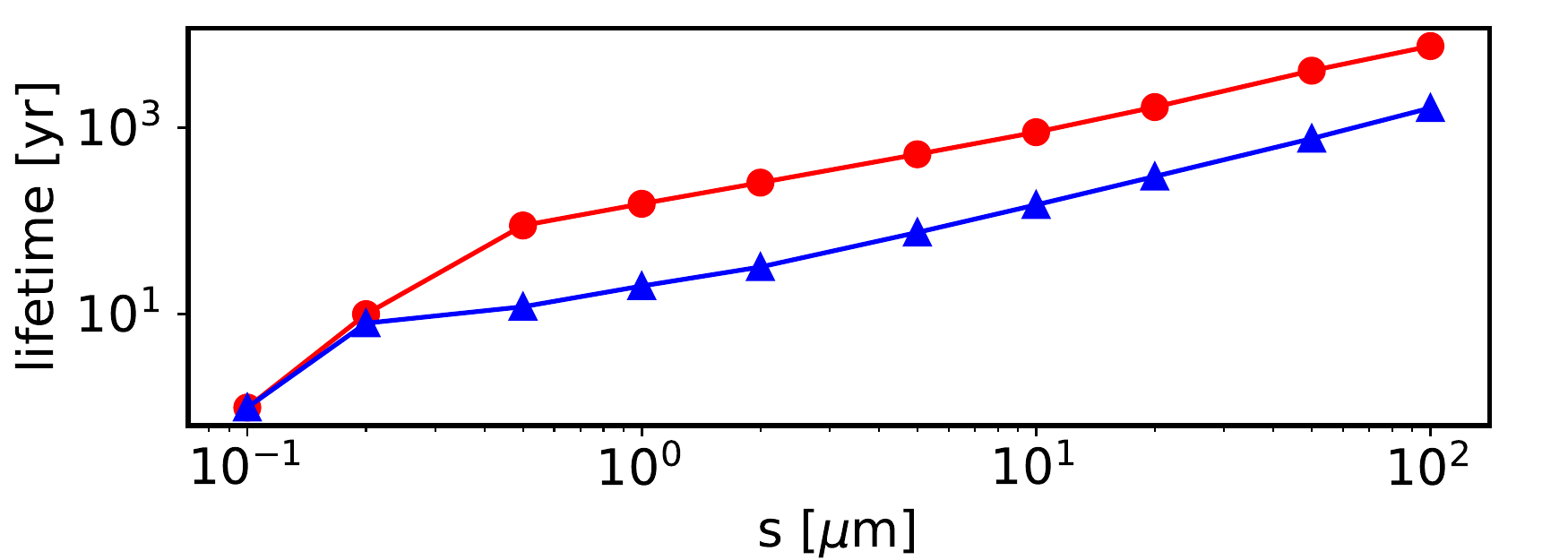}}
\quad
\subfloat[\label{r_e-b}]{\includegraphics[width=\columnwidth]{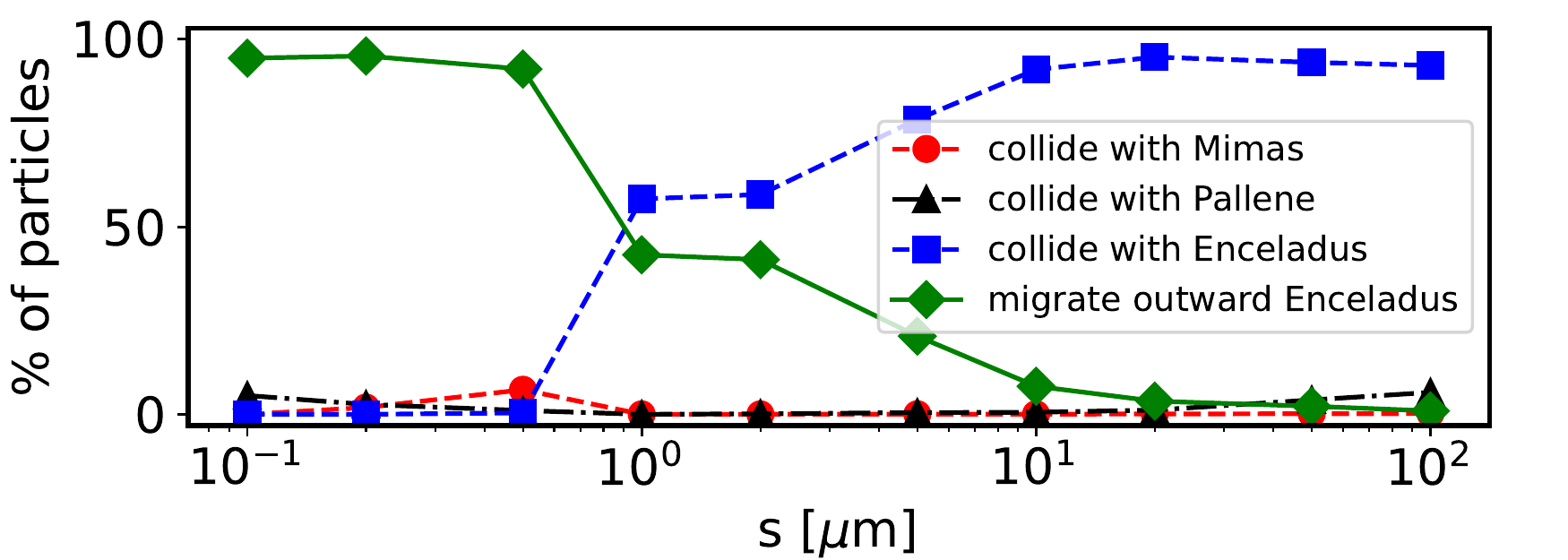}}
\quad
\subfloat[\label{r_e-c}]{\includegraphics[width=\columnwidth]{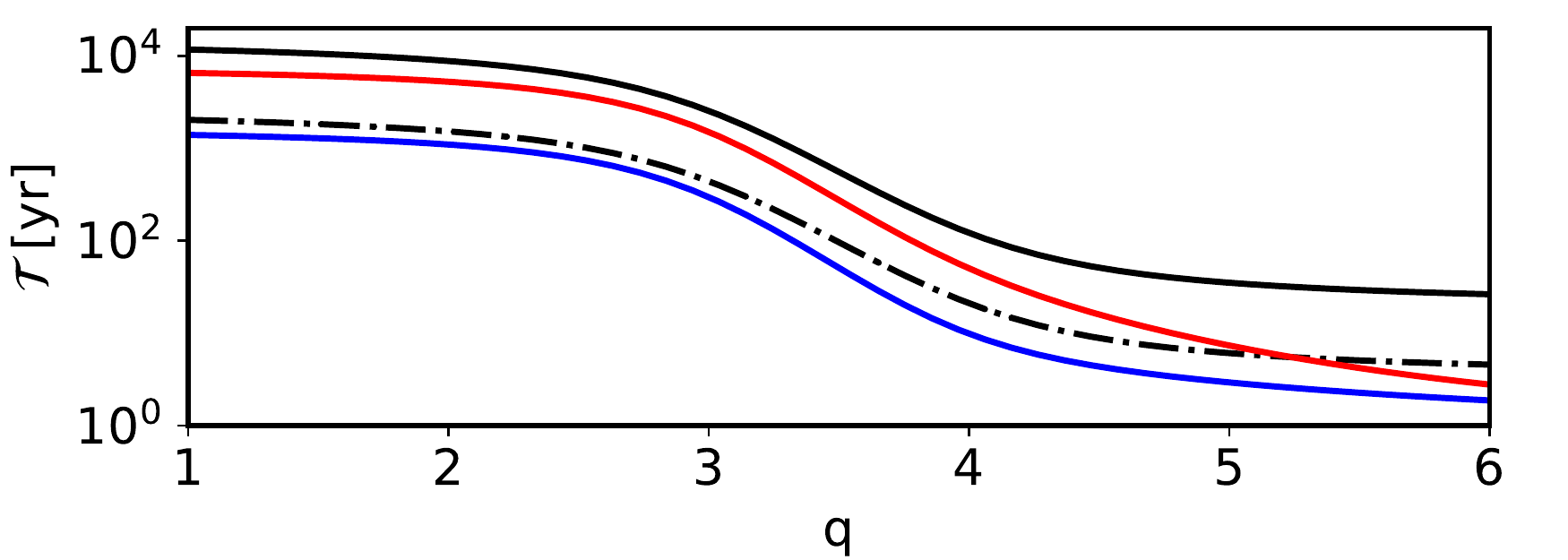}}
\caption{
    a) The solid lines in blue and red show the time for 50\% and 100\% of the ejected particles to be removed from Pallene ring, respectively. b) The coloured lines show the fraction of particles that collide with Mimas (in red), Pallene (in black), and Enceladus (in blue), and the fraction that migrates outside the orbit of Enceladus (in green). c) The time for Pallene to produce the ring material is given by the black lines, in the non-porous (solid) and porous (dot-dashed) cases, while the ring lifetime and half-life are given by the red and blue lines, respectively.
}
\label{rate_eject}
\end{figure}

\Cref{profilerz_ejec} is similar to \cref{profilerz_coor} but for the ejected particles. The temporal evolution of the ejected particles is similar to the co-orbital particles scenario. The same is true for the ring profiles, with greater distinctions only in the first years of the simulation, due to the different initial conditions. \Cref{rate_eject} shows the half-lifetime and lifetime of the ring (top panel), the particle sinks (middle panel), the times required for Pallene to produce the ring material, as well as the lifetimes as a function of the slope of the size distribution (bottom panel). Our results are similar to those discussed in \cref{coorbitalmaterial}. In both scenarios, Pallene could produce the material to keep the ring in a steady-state if the distribution of the particles in the ring is given by $q\lesssim 3$.

\subsection{Comments on ring sources}
Similar to \citet{Madeira18} and \citet{Madeira20}, we only computed the production due to external projectile impacts with the immersed moon. Therefore, we are analysing whether the satellite can produce the amount of material needed to keep the systems in steady-state, not whether they are in steady-state. In fact, the most likely case is that all the mentioned dusty arcs/rings are in a quasi-steady state, demonstrating that more sophisticated models are needed to understand their stability.

As we pointed out in this section, satellite porosity can be a factor influencing material production; however, the systems also have other sources. For example, ring particles are also impacted by external projectiles and therefore also produce material. However, following the prescription given in \citet{Dikarev05}, we obtained that such source is at least three orders of magnitude less efficient than the satellite for the systems analysed here.

The mentioned arcs/rings have the similarity of having a population of larger particles \citep[$\sim$ cm-m,][]{Hedman09,Hedman10,Spahn19}, which lead us to speculate whether the mutual collision of these objects or their impacts with the moon would be the main source of these systems \citep{colwell1990model,colwell1990numerical}. Just as a proof of concept, we will assume that in the Pallene ring is immersed a family of moonlets with radii ranging from $1$ m to $100$ m, following a size distribution $N\sim s^{-3.5}$ and total optical depth $\tau_{\rm mlets}=10^{-8}$. Production due to impacts between the moonlets can be roughly estimated as \citep{Sun15}
\begin{equation}
\dot{M}_{\rm mlets}=3\tau_{\rm mlets}NM_{\rm col}   
\end{equation}
where $M_{\rm col}$ is the amount of dust released per collision, assumed as $0.12M_{\rm mlet}$ \citep{canup1995accretion}, and $M_{\rm mlet}$ is the total mass of the moonlet population.

As a result, we get $\dot{M}_{\rm mlets}\sim10^{-2}~{\rm kg/s}$ corresponding to a value more than one order of magnitude higher than the production due to the non-porous Pallene. This shows that impacts between larger particles are an appealing possibility to keep the arcs/rings in steady-state. However, production due to impacts between centimetric-metric bodies is a very intricate problem, and is beyond the scope of this work.

\section{Summary and Conclusions}
\label{sec:summary}

In this work, we performed an exhaustive numerical exploration of the evolution of the small Saturnian moon Pallene, as well as of the diffuse dusty ring sharing its orbit. We used both short- and long-term numerical simulations, spanning a wide range of timescales to cover in detail the evolution of Pallene and its ring. 

By using the frequency map analysis technique, we produced a diffusion map to characterise the current dynamical state of a wide region of phase-space surrounding Pallene. We identified all the MMRs of relevance in the region, among Pallene and any of the six major moons considered in this study, up to fourth order. We used a simple tidal evolution calculation for Mimas, Pallene, and Enceladus in order to set the context for our longer-term simulations. We made note that the most recent resonance Pallene may have escaped from is the 4:5 resonance with Mimas. Pallene's current eccentricity or inclination could be signs of this or another past resonance crossing.

From the short- and long-term N-body simulations, we analysed all the direct and indirect arguments of the disturbing function identified in the diffusion map in the vicinity of Pallene. These arguments included zeroth-order arguments, with degrees $j\leq$~15, and first- to fourth-order arguments with degrees $j \leq 30$. In brief, we found that some arguments displayed interesting behaviour by temporally librating at various timescales. In particular, the direct argument $\phi_\mathrm{tP}=8\lambda' - 5\lambda - \varpi' - 2\varpi$ of Pallene with Tethys that librates for $\sim 10$~kyr and the zeroth-order argument $\Phi = \varpi' - \varpi + \Omega' - \Omega$ of Pallene with Tethys, Dione and Titan, which coincides with the angle combination suggested for Pallene with Mimas by \citet{Callegari10}. The recurrence of this zeroth-order combination suggests a possible secular alignment of the lines of apsides and nodes among Pallene, Dione, Rhea, and Titan in timescales $\sim 800$~yr. 

Furthermore, after a thorough search of possible (two-body) resonant arguments for Pallene, we conclude that the small moon is not currently in resonance with either Mimas, Enceladus, Tethys, Dione, Rhea, or Titan. It is unlikely that Pallene would be in a higher-order MMR, i.e., $\geq$ 5th order, with any of these satellites, due to their small eccentricity/inclination, and the corresponding $e$-$I$ coefficients of the disturbing function. Nevertheless, the lack of two-body MMRs for Pallene does not exclude the hypothesis that Pallene might be part of a three-body resonance. Moreover, under the present considerations and without accounting for Saturn's tidal forces in the numerical simulations, we cannot dismiss either the past escape of Pallene from a resonance or its future trapping, particularly at times longer than 5~Myr.

We analysed the dynamical evolution of the Pallene ring assuming a scenario where particles are ejected from the satellite's surface, as well as a scenario where the material is originally co-orbital to Pallene. We found that non-gravitational forces dynamically dominate the system and the material experiences a similar dynamical evolution in both scenarios.  

The outward migration due to plasma drag causes the loss of particles with radius of a few micrometres in just tens of years, while larger particles ($\gtrsim 10~\mu$m) can survive for a few hundred years in the ring. \cite{Spahn19} measured the radial mean position of the ring to be more than $1000$~km beyond the satellite's orbit; this is likely caused by plasma drag. Our ring profiles clearly show the formation of particle clusters beyond Pallene's orbit. Furthermore, the profiles show that the ring evolves into structures that are radially asymmetrical in relation to the satellite's orbit.

The precession of the longitude of pericentre due to non-gravitational forces produces vertical excursions of the particles in relation to Pallene's orbital plane. This could be the mechanism responsible for vertical excursions discussed in \cite{Hedman09}.

\emph{Cassini} data indicate a concentration of larger particles around Pallene's orbit, which is in line with the significantly longer lifetime of the larger particles that we found. In fact, when calculating the mass production rate due to IDPs and ERPs, we find that Pallene can keep the ring in a steady-state only if it is predominantly composed of larger micrometre-sized particles ($q\lesssim 3$).

If we assume Pallene as the only source of material for the rings, we conclude that the ring would spread for $q\lesssim 4$. This corresponds to the slope range given by \citet{Kempf08a,Ye14a} for the E~ring, in which Pallene is immersed. In this scenario, our profiles show that the ring will evolve into a toroidal structure similar to the gossamer rings of Jupiter, and then it will  continuously spread out, both radially and vertically, until it finally disappears. From our numerical results, we cannot constrain whether the ring originated from the material ejected from the satellite or from the disruption of an ancient proto-Pallene.

We must point out that our dynamical model is not complete; if the ring has a high concentration of larger particles, additional effects such as collisions between the  particles, self-gravity, and local viscosity may be significant to the system. However, even in this case, plasma drag may dominate, and our main results would still hold valid.

\section*{Acknowledgements}

We thank the anonymous referee for a detailed and careful report that helped to greatly improve the quality of this paper. G.~Madeira thanks FAPESP for financial support via grant 2018/23568-6. J.~A'Hearn thanks M.~Hedman, M.~Tiscareno, and M.~Showalter for useful discussions; and also thanks NASA for partial support through the \textit{Cassini} Data Analysis and Participating Scientist Program grant NNX15AQ67G. S.~M.~Giuliatti Winter thanks FAPESP (2016/24561-0), CNPq (313043/2020-5) and Capes for the financial support.

\section*{Data Availability}

The data underlying this article will be shared on reasonable request to the corresponding author.




\bibliographystyle{mnras}
\bibliography{pallene} 





\bsp	
\label{lastpage}

\end{document}